\documentclass[twocolumn,letterpaper,pra,showpacs,superscriptaddress]{revtex4}
\usepackage{graphicx}
\usepackage{bm}
\usepackage{amsfonts}
\usepackage[latin1]{inputenc} 
\usepackage[intlimits]{amsmath}
\usepackage{amsxtra}
\usepackage{amssymb}
\usepackage{psfrag}
\usepackage{amstext}
\newlength{\twocolumnwidth}

\newlength{\auxlv}

\newlength{\DL}

\newlength{\pwd} 
\begin{document} 
\title{Causal signal transmission by quantum fields.%
\\ Electromagnetic interaction of distinguishable devices 
and the optical equivalence theorem.%
} 
\author{L.I.\ Plimak} 
\affiliation{Institut f\"ur Quantenphysik, Universit\"at Ulm, 89069 Ulm, Germany.} 
\author{S.\ Stenholm} 
\affiliation{Institut f\"ur Quantenphysik, Universit\"at Ulm, 89069 Ulm, Germany.} 
\affiliation{Physics Department, Royal Institute of Technology, KTH, Stockholm, Sweden.} 
\affiliation{Laboratory of Computational Engineering, HUT, Espoo, Finland.} 
\date{\today} 
\begin{abstract} 
Generalised phase-space techniques for electromagnetic\ interactions beyond the rotating wave approximation\ [L.P.\ and S.S., arXive:1104.3825\ (2011)] is applied to interactions of distinguishable devices. The paper is built around the concept of ``doing quantum electrodynamics\ while thinking classically,'' which is a generalisation of Sudarshan's renowned optical equivalence theorem\ [E.C.G.\ Sudarshan, Phys.\ Rev.\ Lett.\ \textbf{10}, 277 (1963)] to the interacting electromagnetic\ field. This concept allows one to reduce inherently quantum problems to semiclasical considerations. 
\end{abstract}
\pacs{XXZ}
\maketitle 
\section{Introduction}%
\label{ch:IntVII}
This article concludes the subseries of this series of papers, concerned with quantum electrodynamics\ under macroscopic (often termed mesoscopic, cf.\ endnote 21 in paper \cite{WickCaus}) 
conditions. In paper \cite{QDynResp}, we applied {\em response transformations\/} \cite{API,APII,APIII} and the {\em causal Wick theorem\/} \cite{WickCaus} to the standard perturbative approach of quantum field theory \cite{Schweber}. 
An astonishing feature of formulae thus found is that they lack Planck's constant. Such formulae survive the classical limit \mbox{$
\hbar \to 0
$} unchanged, and must therefore equally make sense in classical stochastic electrodynamics. This quantum-classical correspondence was the subject of paper \cite{PFunc}. It was demonstrated that formulae of \mbox{Ref.\ \cite{QDynResp}} are naturally written in phase-space terms. We introduced the concept of {\em conditional functional quasi-probability distribution\/}, or {\em conditional P-functional\/}, where ``conditional'' means dependent on external c-number sources. The conditional P-functional generalises the conventional P-function \cite{MandelWolf} in two ways: to general nonlinear non-Markovian quantum systems, and to {\em response properties\/} of quantum systems. It also establishes a natural relation to classical stochastic electrodynamics. If a P-functional is nonnegative, it may be interpreted as a functional probability distribution, thus mapping a quantum system {\em exactly\/} on a classical stochastic system (recall that corresponding formulae lack Planck's constant). 

The subject of \mbox{Refs.\ \cite{QDynResp,PFunc}} was a solitary quantum device. Formally, the latter is just a placeholder for a quantum model of matter. There are no restrictions on this model, so that results of \mbox{Refs.\ \cite{QDynResp,PFunc}} apply to any case of electromagnetic\ interactions, including relativistic quantum electrodynamics. The macroscopic (mesoscopic) approximation is introduced in this paper, by regarding the matter as a collection of {\em distinguishable devices\/}. Distinguishability is an approximation: after all, electrons in all devices are identical. Under macroscopic conditions, such effects are negligible for all practical purposes. The approximation of there being distinguishable devices may be regarded a definition of macroscopic conditions in quantum mechanics. 

The notion this paper is built around is ``doing quantum electrodynamics\ while thinking classically.'' Any relation for P-functionals is identical with some relation for probability distributions, and may be obtained as formal generalisation of the latter. One may therefore derive quantum relations by ``demoting'' problems to classical mechanics, doing the classical theory, then ``promoting'' the results to quantum mechanics. This recipe must be applied with care, because there exist classical models incompatible with quantum mechanics, such as, for instance, noiseless photodetector and noiseless coherent amplifier. The necessary reservations constitute an important part of our analyses. 

As in our previous papers \cite{WickCaus,QDynResp,PFunc} we distinguish the {\em narrow-band\/} and {\em broad-band\/} cases, which differ in whether the rotating wave approximation (RWA) is or is not made in dynamics. The broad-band\ case\ is most general, but is also very much disconnected from the quantum-optical paradigm where RWA is quite common. For instance, in Glauber-Kelley-Kleiner's photodetection theory \cite{GlauberPhDet,KelleyKleiner,GlauberTN}, the optical field is commonly treated under the RWA. However, the latter cannot be extended to the photocurrent and photovoltage. These are broad-band processes which are not subject to any kind of resonance approximation. 

The paper is structured as follows. In Sec.\ \ref{ch:S}, we summarise the key results of papers \cite{QDynResp,PFunc}. In Sec.\ \ref{ch:E}, we demonstrate that the concept of ``doing quantum electrodynamics\ while thinking classically'' naturally extends to interactions of distinguishable devices. As an illustration, in Sec.\ \ref{ch:T}, we apply this concept to the so-called {\em cascaded systems\/} \cite{CascadeC,CascadeH}. We formulate a photodetection theory without the rotating wave approximation\ (Sec.\ \ref{ch:TF}) and a generalisation of Sudarshan's optical equivalence theorem\ to interacting electromagnetic field (Sec.\ \ref{ch:SGO}). In Sec.\ \ref{ch:TC}, we apply ``doing quantum electrodynamics\ while thinking classically'' to photodetection of the electromagnetic\ field in a quantum state preceded by coherent quantum amplification. This example allows us to illustrate the reservations one has to make when applying this concept to quantum systems. In appendices, we concern ourselves with formal particulars neglected in the main body of the paper. 
\section{Solitary device revisited}%
\label{ch:S}
\subsection{The model}%
\label{ch:SM}
We start our analyses from the {\em broad-band\ case\/} \cite{WickCaus,QDynResp,PFunc}. We employ a somewhat simplified version of the structural model of electromagnetic interaction used in \mbox{Refs.\ \cite{QDynResp,PFunc}}. For convenience of the reader, we copy here the key definitions from \mbox{Ref.\ \cite{PFunc}}. We consider a quantum device interacting with a collection of oscillator modes, with the Hamiltonian in the interaction picture being, 
\begin{align} 
\begin{aligned} 
 &\hat H(t) = \hbar 
\sum_{\kappa =1}^{N}\omega_{\kappa}\hat a_{\kappa}^{\dag}\hat a_{\kappa} 
+ \hat H_{\mathrm{dev}}(t) + \hat H_{\text{I}}(t) 
. 
\end{aligned} 
\label{eq:86FB} 
\end{align}%
The oscillators, represented by the standard creation and annihilation\ operators, 
\begin{align} 
\begin{aligned} 
 &\big [ 
\hat a_{\kappa},\hat a_{\kappa '}^{\dag}
 \big ] = \delta_{k\kappa '} , & &\kappa ,\kappa'=1,\cdots,N.
\end{aligned} 
\label{eq:87FC} 
\end{align}%
are organised in a quantised field, 
\begin{align} 
\begin{aligned} 
\hat A(x,t) &= \sum_{\kappa =1}^{N}
\sqrt{\frac{\hbar}{2\omega_{\kappa}}}
 u_{\kappa}(x) \hat a_{\kappa}\mathrm{e}^{-i\omega_{\kappa}t} + \mathrm{H.c.}\, . 
\end{aligned} 
\label{eq:45LB} 
\end{align}%
where $ u_k(x)$ are complex mode functions. Variable $x$ comprises all field arguments except time. The field interacts with the device according to the nonresonant\ (broad-band) Hamiltonian, 
\begin{align} 
\begin{aligned} 
\hat H_{\text{I}}(t) &= 
- 
\int dx \big [ 
\hat A(x,t)+A_{\mathrm{e}}(x,t)
 \big ]\hat J(x,t) 
 . 
\end{aligned} 
\label{eq:81AP} 
\end{align}%
The Hamiltonian $\hat H_{\mathrm{dev}}(t)$ and the current operator $\hat J(x,t)$ describe the device. They commute with all \mbox{$
\hat a_{\kappa },\hat a_{\kappa }^{\dag}
$} and otherwise remain arbitrary. The {Heisenberg}\ density matrix factorises into the vacuum state of all oscillators and an arbitrary state of the device, 
\begin{align} 
\begin{aligned} 
\hat \rho = 
\big | 
0
 \big \rangle 
\big \langle 
0
 \big | \otimes \hat \rho_{\mathrm{dev}}.
\end{aligned} 
\label{eq:4YP} 
\end{align}%
The c-number external source $ A_{\mathrm{e}}(x,t)$ is added for formal purposes. The aforementioned simplification is the absence of an external c-number current in the interaction which is of no use in this paper. 
\subsection{Retarded Green function of the field}%
\label{ch:DR}
The electromagnetic\ field enters the theory through its {\em retarded Green function\/}
\begin{align} 
\begin{aligned} 
 &G_{\text{R}}(x,x',t-t') = \frac{i}{\hbar }\theta(t-t') 
\big [ 
\hat A(x,t),\hat A(x',t')
 \big ] . 
\end{aligned} 
\label{eq:83LF} 
\end{align}%
This definition is Kubo's formula for a linear response function \cite{Kubo}; for more details see \mbox{Ref.\ \cite{API}}. The commutator in (\ref{eq:83LF}) is a c-number so that quantum averaging present in Kubo's formula could be omitted (i.e., response of a linear system does not depend on its state). The explicit expression for $G_{\text{R}}$ follows by combining (\ref{eq:83LF}) with (\ref{eq:45LB}), see \mbox{Ref.\ \cite{QDynResp}}. 
\subsection{Condensed notation}%
\label{ch:DN}
To keep the bulk of formulae under the lid and make their structure more transparent, we make extensive use of condensed notation, 
\begin{align} 
fg &= \int dx dt f(x,t) g(x,t) , 
\label{eq:3VS} 
\\ 
fKg &= \int dx dx' dt dt' f(x,t)
\nonumber\\ &\qquad\times 
K(x,x',t-t') g(x',t') , 
\label{eq:76NU} 
\\ 
(Kf)(x,t) &= \int dx' dt' K(x,x',t-t')f(x',t') , 
\label{eq:77NV} 
\\ 
(fK)(x,t) &= \int dx' dt' g(x',t')K(x',x,t'-t) , 
\label{eq:78NW} 
\end{align}%
where $f(x,t)$ and $g(x,t)$ are c-number or q-number functions, and $K(x,x',t-t')$ is a c-number kernel. The ``products'' $fg$ and $fKg$ denote scalars, while $Kg$ and $fK$ --- functions (fields). 
\subsection{Conditional time-normal averages and conditional P-functionals}%
\label{ch:ST}
A formal solution to the problem of Sec.\ \ref{ch:SM} is constructed applying the standard perturbative techniques of quantum field theory \cite{Schweber}. Of interest is however not this solution as such, but the fact that there exists a ``language'' in which it looks essentially classical. Referring the reader for details to \mbox{Refs.\ \cite{QDynResp,PFunc}}, here we only reiterate the key points. 

As was demonstrated in \mbox{Ref.\ \cite{PFunc}}, full quantum description of an electromagnetic device reduces to time-normal averages \cite{KelleyKleiner,GlauberTN,APII,APIII,PFunc} of the quantum fields and currents, conditional on the sources. For the broad-band\ field and current, one should use the amended definition of the time-normal ordering introduced in \mbox{Ref.\ \cite{APII}}. The conventional as well as amended definitions of the time-normal ordering are reiterated in appendix \ref{ch:DA}. For a general discussion of the time-normal operator ordering see \mbox{Ref.\ \cite{PFunc}}, section 
II, 
and references therein. 

The said averages enter the theory through their generating functional, 
\begin{align} 
\begin{aligned} 
\Phi{\big( 
\eta , \zeta 
 \big| 
A_{\mathrm{e}} 
 \big)} = 
\big \langle {\mathcal T}{\mbox{\rm\boldmath$:$}}\exp\big(
i\eta{\hat{\mathcal A}}+i\zeta{\hat{\mathcal J}}
 \big) 
{\mbox{\rm\boldmath$:$}}
 \big \rangle 
 , 
\end{aligned} 
\label{eq:76FZ} 
\end{align}%
where $\eta (x,t)$ and $\zeta (x,t)$ are auxiliary c-number functions. Equation (\ref{eq:76FZ}) implies condensed notation (\ref{eq:3VS}). The operators ${\hat{\mathcal A}}(x,t),{\hat{\mathcal J}}(x,t)$ are $\hat A(x,t),\hat J(x,t)$ in the {Heisenberg}\ picture. These operators are by construction dependent (conditional) on the c-number source in the Hamiltonian. The averaging in (\ref{eq:76FZ}) is over the initial (Heisenberg) state of the system (\ref{eq:4YP}), 
\begin{align} 
\begin{aligned} 
\big \langle 
\cdots
 \big \rangle = \text{Tr}\hat \rho (\cdots), 
\end{aligned} 
\label{eq:77HA} 
\end{align}%
where the ellipsis stands for an arbitrary operator. 

Formulae relating functional (\ref{eq:76FZ}) to conventional quantum averages of the Heisenberg operators (termed {\em response transformations\/}) are summarised in paper \cite{PFunc}, section IIIB. In the terminology of \mbox{Refs.\ \cite{APII,APIII,QDynResp,PFunc}}, that functional (\ref{eq:76FZ}) contains full information on the quantum device is a {\em consistency condition\/}. For a summary of consistency conditions see \mbox{Ref.\ \cite{PFunc}}, section 
IIID. Some examples may be found in Sec.\ \ref{ch:MW} below. 
\subsection{Reduction to quantum current}%
\label{ch:SR}
In classical stochastic electrodynamics, a device may be seen as a random current $J(x,t)$. This current radiates the random field, 
\begin{align} 
\begin{aligned} 
A(x,t) = \int dx' dt' G_{\text{R}}(x,x',t-t')J(x',t'), 
\end{aligned} 
\label{eq:96YF} 
\end{align}%
where \mbox{$G_{\text{R}}(x,x',t-t')$} is the retarded Green function (also known as transfer function, or linear response function) characteristic of the linear media, or vacuum, in which the device is submerged. Linear response functions of a classical field and of the corresponding quantised field coincide \cite{API,Corresp}, so that $G_{\text{R}}$ in (\ref{eq:96YF}) is in fact given by Kubo's quantum formula (\ref{eq:83LF}). 

If the radiating current is stochastic, joint statistical averages of the field and current are given by the formula, 
\begin{widetext} 
\begin{align} 
 &\overline{\hspace{0.1ex}{
 A(x_1,t_1)\cdots A(x_m,t_m) 
 J(x_{m+1},t_{m+1})\cdots J(x_{m+n},t_{m+n})
}\hspace{0.1ex}} \nonumber\\ &\quad
= \int dx_1'dt_1'\cdots dx_m'dt_m'
G_{\text{R}}(x_1,x_1',t_1-t_1')\cdots G_{\text{R}}(x_m,x_m',t_m-t_m')
\nonumber\\ &\qquad\times 
\overline{\hspace{0.1ex}{
 J(x_1',t_1')\cdots J(x_m',t_m') 
 J(x_{m+1},t_{m+1})\cdots J(x_{m+n},t_{m+n})
}\hspace{0.1ex}} 
 , 
\label{eq:21ZH} 
\end{align}%
As was shown in \mbox{Ref.\ \cite{PFunc}}, the corresponding quantum formula for joint time-normal averages of the field and current operators generated by functional (\ref{eq:76FZ}) is found replacing statistical averages by time-normal averages, 
\begin{align} 
 &\big \langle {\mathcal T}{\mbox{\rm\boldmath$:$}}
{\hat{\mathcal A}}(x_1,t_1)\cdots{\hat{\mathcal A}}(x_m,t_m) 
{\hat{\mathcal J}}(x_{m+1},t_{m+1})\cdots{\hat{\mathcal J}}(x_{m+n},t_{m+n})
{\mbox{\rm\boldmath$:$}} \big \rangle \nonumber\\ &\quad
= \int dx_1'dt_1'\cdots dx_m'dt_m'
G_{\text{R}}(x_1,x_1',t_1-t_1')\cdots G_{\text{R}}(x_m,x_m',t_m-t_m')
\nonumber\\ &\qquad\times 
\big \langle {\mathcal T}{\mbox{\rm\boldmath$:$}}
{\hat{\mathcal J}}(x_1',t_1')\cdots{\hat{\mathcal J}}(x_m',t_m') 
{\hat{\mathcal J}}(x_{m+1},t_{m+1})\cdots{\hat{\mathcal J}}(x_{m+n},t_{m+n})
{\mbox{\rm\boldmath$:$}} \big \rangle 
. 
\label{eq:97YH} 
\end{align}%
\end{widetext}%
One may say that, under the time-normal ordering, Eq.\ (\ref{eq:96YF}) applies directly to {Heisenberg}\ operators. 
\subsection{Response and phase-space characterisation of solitary devices}%
\label{ch:MW}
\subsubsection{``Dressed'' device}%
\label{ch:MWD}
We remind that, by definition, $\hat J(x,t)$ is the interaction-picture operator (``bare'' current). Its Heisenberg counterpart (``dressed'' current) is denoted as ${\hat{\mathcal J}}(x,t)$; it is by construction dependent (conditional) on the external field \mbox{$
A_{\mathrm{e}}(x,t)
$} present in (\ref{eq:81AP}). 

Equation (\ref{eq:97YH}) has an obvious implication: it suffices to calculate time-normal averages of the quantum current. Those of the field are recovered applying the classical radiation law (\ref{eq:96YF}). 
Using Eq.\ (\ref{eq:97YH}), for the functional (\ref{eq:76FZ}) we obtain, 
\begin{align} 
\begin{aligned} 
\Phi{\big( \eta ,\zeta \big| A_{\mathrm{e}} \big)} = 
\Phi{\big( 0,\zeta+\eta R_{\text{R}}\big| A_{\mathrm{e}} \big)} \equiv 
\Phi_{\mathrm{dev}}{\big( \zeta+\eta R_{\text{R}}\big| A_{\mathrm{e}} \big)} . 
\end{aligned} 
\label{eq:24KA} 
\end{align}%
The ``dressed'' device is completely characterised by the time-normal averages of the Heisenberg current operator, generated by the functional, 
\begin{align} 
\Phi_{\mathrm{dev}}\big(
\zeta 
\big | 
A_{\mathrm{e}} 
 \big) 
 &= 
\Big \langle {\mathcal T}{\mbox{\rm\boldmath$:$}}\exp\big(
i\zeta{\hat{\mathcal J}}
 \big) 
{\mbox{\rm\boldmath$:$}}
 \Big \rangle 
\nonumber\\ &
= \prod_{x,t}\bigg\{\int d J(x,t)\bigg\} 
p\big(
J \big| A_{\mathrm{e}}
 \big) 
\exp\big(
i\zeta J
 \big)
, 
\label{eq:53HM} 
\end{align}%
We also took this opportunity to introduce the phase-space characterisation of the ``dressed'' current by the {\em conditional P-functional\/}, or {\em conditional functional quasiprobability distribution\/}, \mbox{$
p\big(
J \big| A_{\mathrm{e}}
 \big)
$} \cite{PFunc}. 
\subsubsection{Consistency condition}%
\label{ch:MWC}
An alternatively way of defining functional (\ref{eq:53HM}) is applying {\em response transformation\/} \cite{API,APII,APIII} to the closed-time-loop (Schwinger-Perel-Keldysh \cite{SchwingerC,Perel,Keldysh}) averages of ${\hat{\mathcal J}}(x,t)$ defined {\em without the source\/}. Namely, 
\begin{align} 
\begin{aligned} 
\Phi_{\mathrm{dev}}\big(
\zeta \big| a_{\mathrm{e}}
 \big) 
= \big [ 
\big \langle 
T_C\exp\big(
i{\zeta}_+{\hat{\mathcal J}}_+
-i{\zeta}_-{\hat{\mathcal J}}_-
 \big) 
 \big \rangle \settoheight{\auxlv}{$|$}%
\raisebox{-0.3\auxlv}{$|_{A_{\mathrm{e}}=0}$}
 \big ] \settoheight{\auxlv}{$|$}%
\raisebox{-0.3\auxlv}{$|_{\mathrm{c.v.}}$} , 
\end{aligned} 
\label{eq:25KB} 
\end{align}%
where c.v.\ refers to the {\em response substitution\/}, 
\begin{align} 
\begin{aligned} 
 &\zeta_{\pm}(x,t) = \frac{ a_{\mathrm{e}}(x,t)}{\hbar }\pm \zeta ^{(\mp)}(x,t), 
\end{aligned} 
\label{eq:12YX} 
\end{align}%
with ${^{(\pm)}}$ standing for separation of the frequency-positive and negative\ parts of a function. Definition of the $T_C$-ordering is reiterated in appendix \ref{ch:DO} and that of the said separation --- in appendix \ref{ch:DF}. 

Equivalence of definitions (\ref{eq:53HM}) and (\ref{eq:25KB}) is, in terminology of \mbox{Refs.\ \cite{APII,APIII,QDynResp,PFunc}}, a {\em consistency condition\/}. More precisely, the latter is expressed by the relation \cite{APII,QDynResp}, 
\begin{align} 
\begin{aligned} 
\Phi_{\mathrm{dev}}\big(
\zeta\big | a_{\mathrm{e}} +A_{\mathrm{e}}
 \big) 
= \big \langle 
T_C\exp\big(
i{\zeta}_+{\hat{\mathcal J}}_+
-i{\zeta}_-{\hat{\mathcal J}}_-
 \big) 
 \big \rangle \settoheight{\auxlv}{$|$}%
\raisebox{-0.3\auxlv}{$|_{\mathrm{c.v.}}$} 
. 
\end{aligned} 
\label{eq:14YZ} 
\end{align}%
showing that the auxiliary variable \mbox{$a_{\mathrm{e}}(x,t)$} and the external source \mbox{$A_{\mathrm{e}}(x,t)$} occur in the theory as a sum. With \mbox{$
a_{\mathrm{e}}(x,t)=0
$} we recover Eq.\ (\ref{eq:53HM}), while with \mbox{$
A_{\mathrm{e}}(x,t)=0
$} --- Eq.\ (\ref{eq:25KB}).
A summary of consistency conditions may be found in \mbox{Ref.\ \cite{PFunc}}, section IIID. 
For details see \mbox{Refs.\ \cite{APII,APIII,QDynResp,PFunc}} 

Equation (\ref{eq:20ZF}) gives a fair idea of how the language of conditional time-normal averages is related to the conventional closed-time-loop formalism \cite{SchwingerC,Perel,Keldysh}. In particular, it makes it evident that the time-normal averages (\ref{eq:53HM}) indeed provide {\em complete\/} quantum characterisation of the device. 
\subsubsection{``Bare'' device}%
\label{ch:MWB}
Similar to the ``dressed'' current, 
the ``bare'' current may be characterised in two equivalent ways. One is in terms of the time-normal averages of the Heisenberg current operator in the presence of a given c-number source with the quantised electromagnetic\ field ``switched off.'' These are conveniently accessed through their generating functional, 
\begin{align} 
\Phi^{\mathrm{I}}_{\mathrm{dev}}\big(
\zeta 
\big | 
A_{\mathrm{e}} 
 \big) 
 &= 
\text{Tr}\hat\rho _{\mathrm{dev}}{\mathcal T}{\mbox{\rm\boldmath$:$}}\exp\big(
i\zeta\hat J'
 \big) 
{\mbox{\rm\boldmath$:$}} \nonumber\\ &
= 
\prod_{x,t}\bigg\{\int d J(x,t)\bigg\} 
p^{\mathrm{I}}\big(
J \big| A_{\mathrm{e}}
 \big) 
\exp\big(
i\zeta J
 \big)
, 
\label{eq:96KJ} 
\end{align}%
where $\hat J'(x,t)$ is defined as a {Heisenberg}\ operator for the Hamitonian, 
\begin{align} 
\begin{aligned} 
\hat H'(t) = \hat H_{\mathrm{dev}}(t)-\int dx \hat J(x,t)A_{\mathrm{e}}(x,t), 
\end{aligned} 
\label{eq:97KK} 
\end{align}%
and \mbox{$
p^{\mathrm{I}}\big(
J \big| A_{\mathrm{e}}
 \big)
$} is the corresponding ``bare'' P-functional. 
Alternatively, 
\begin{align} 
\begin{aligned} 
\Phi^{\mathrm{I}}_{\mathrm{dev}}\big(
\zeta \big| a_{\mathrm{e}}
 \big) 
= \big \langle 
T_C\exp\big(
i{\zeta}_+\hat J_+
-i{\zeta}_-\hat J_-
 \big) 
 \big \rangle\settoheight{\auxlv}{$|$}%
\raisebox{-0.3\auxlv}{$|_{\mathrm{c.v.}}$} . 
\end{aligned} 
\label{eq:98KL} 
\end{align}%
Equation (\ref{eq:98KL}) makes it evident that \mbox{$
\Phi^{\mathrm{I}}_{\mathrm{dev}}
$} and \mbox{$
p^{\mathrm{I}}_{\mathrm{dev}}
$} are determined solely by the free current operator \mbox{$\hat J(x,t)$}. Equations (\ref{eq:96KJ}) and (\ref{eq:98KL}) are particular cases of yet another instance of consistency condition, 
\begin{align} 
\begin{aligned} 
\Phi_{\mathrm{dev}}^{\mathrm{I}}\big(
\zeta\big | a_{\mathrm{e}} +A_{\mathrm{e}}
 \big) 
= \big \langle 
T_C\exp\big(
i{\zeta}_+\hat J_+'
-i{\zeta}_-\hat J_-'
 \big) 
 \big \rangle \settoheight{\auxlv}{$\big|$}%
\raisebox{-0.3\auxlv}{$\big|_{\mathrm{c.v.}}$} , 
\end{aligned} 
\label{eq:20ZF} 
\end{align}%
In fact, Eq.\ (\ref{eq:14YZ}) reduces to (\ref{eq:20ZF}) if formally regarding the interaction with the quantised field as part of $\hat H_{\mathrm{dev}}(t)$. 
\subsection{How to do quantum electrodynamics\ while thinking classically}%
\label{ch:MA}
The main advantage of the {quasiprobability distribution}s is that they may be manipulated to a large extent as if they were classical probability distributions. For example, one may introduce a joint {quasiprobability distribution}\ of the quantum field and current by the formula, 
\begin{align} 
 &\Big \langle {\mathcal T}{\mbox{\rm\boldmath$:$}}\exp\big(
i\eta{\hat{\mathcal A}}+i\zeta{\hat{\mathcal J}}
 \big) 
{\mbox{\rm\boldmath$:$}}
 \Big \rangle 
\nonumber\\ &
= 
\prod_{x,t}\bigg\{\int dJ(x,t)dA(x,t)\bigg\}
p{\big( 
A,J \big| A_{\mathrm{e}}
 \big)} 
\exp\big(
i\eta A+i\zeta J
 \big). 
\label{eq:79HC} 
\end{align}%
In classical stochastic electrodynamics, we would apply Eq.\ (\ref{eq:96YF}), resulting in, 
\begin{align} 
\begin{aligned} 
p{\big( 
A,J \big| A_{\mathrm{e}}
 \big)} = p{\big( 
J \big| A_{\mathrm{e}}
 \big)}\prod_{x,t}\delta\big(
A(x,t) - (
G_{\text{R}}J
 ) (x,t)
 \big) , 
\end{aligned} 
\label{eq:80HD} 
\end{align}%
where we use notation (\ref{eq:77NV}). Substituting this relation into Eq.\ (\ref{eq:79HC}) we find, 
\begin{align} 
 &\Big \langle {\mathcal T}{\mbox{\rm\boldmath$:$}}\exp\big(
i\eta{\hat{\mathcal A}}+i\zeta{\hat{\mathcal J}}
 \big) 
{\mbox{\rm\boldmath$:$}}
 \Big \rangle \nonumber\\ &
= \prod_{x,t}\bigg\{\int d J(x,t)\bigg\}p{\big( 
J \big| A_{\mathrm{e}}
 \big)}\exp\big(
i\zeta J + i \eta G_{\text{R}}J
 \big) . 
\label{eq:81HE} 
\end{align}%
We use here notation (\ref{eq:76NU}). In view of Eq.\ (\ref{eq:53HM}), we have recovered Eq.\ (\ref{eq:24KA}), proving that Eq.\ (\ref{eq:80HD}) is in fact a genuine quantum formula. 

Furthermore, the solution to the self-action, or electromagnetic dressing, problem in terms of the {quasidistribution}s reads, 
\begin{align} 
\begin{aligned} 
p\big(
J\big| A_{\mathrm{e}}
 \big) = p^{\mathrm{I}}\big(
J \big| A_{\mathrm{e}} + G_{\text{R}}J
 \big) , 
\end{aligned} 
\label{eq:87ZJ} 
\end{align}%
where we use notation (\ref{eq:77NV}). The classical content of this relation is obvious. Bare devices are characterised by statistics of the current conditional on the local field, 
\begin{align} 
A_{\mathrm{l}}(x,t) 
 &= A_{\mathrm{e}}(x,t) 
\nonumber\\ &\quad
+ \int dx'dt' G_{\text{R}}(x,x',t-t')J(x',t') . 
\label{eq:1YL} 
\end{align}%
Equation (\ref{eq:87ZJ}) states that, 
\begin{align} 
\begin{aligned} 
p\big(
J\big| A_{\mathrm{e}}
 \big) = p^{\mathrm{I}}\big(
J \big| A_{\mathrm{l}} \big) . 
\end{aligned} 
\label{eq:22ZJ} 
\end{align}%
As was shown in \cite{PFunc}, Eq.\ (\ref{eq:87ZJ}) is equivalent to the operator {\em dressing formula\/}, 
\begin{align} 
\begin{aligned} 
\Phi_{\mathrm{dev}}\big(
\zeta\big|a_{\mathrm{e}} \big) 
= 
\exp\bigg(
-i\frac{\delta }{\delta a_{\mathrm{e}}} G_{\text{R}}
\frac{\delta }{\delta \zeta }
 \bigg) 
\Phi_{\mathrm{dev}}^{\mathrm{I}}\big(
\zeta\big|a_{\mathrm{e}} \big) , 
\end{aligned} 
\label{eq:23YA} 
\end{align}%
found in \mbox{Ref.\ \cite{QDynResp}}. 
\section{Electromagnetic interaction of a pair of quantum devices}%
\label{ch:E}
\begin{figure}
\begin{center}
\includegraphics[width=220pt]{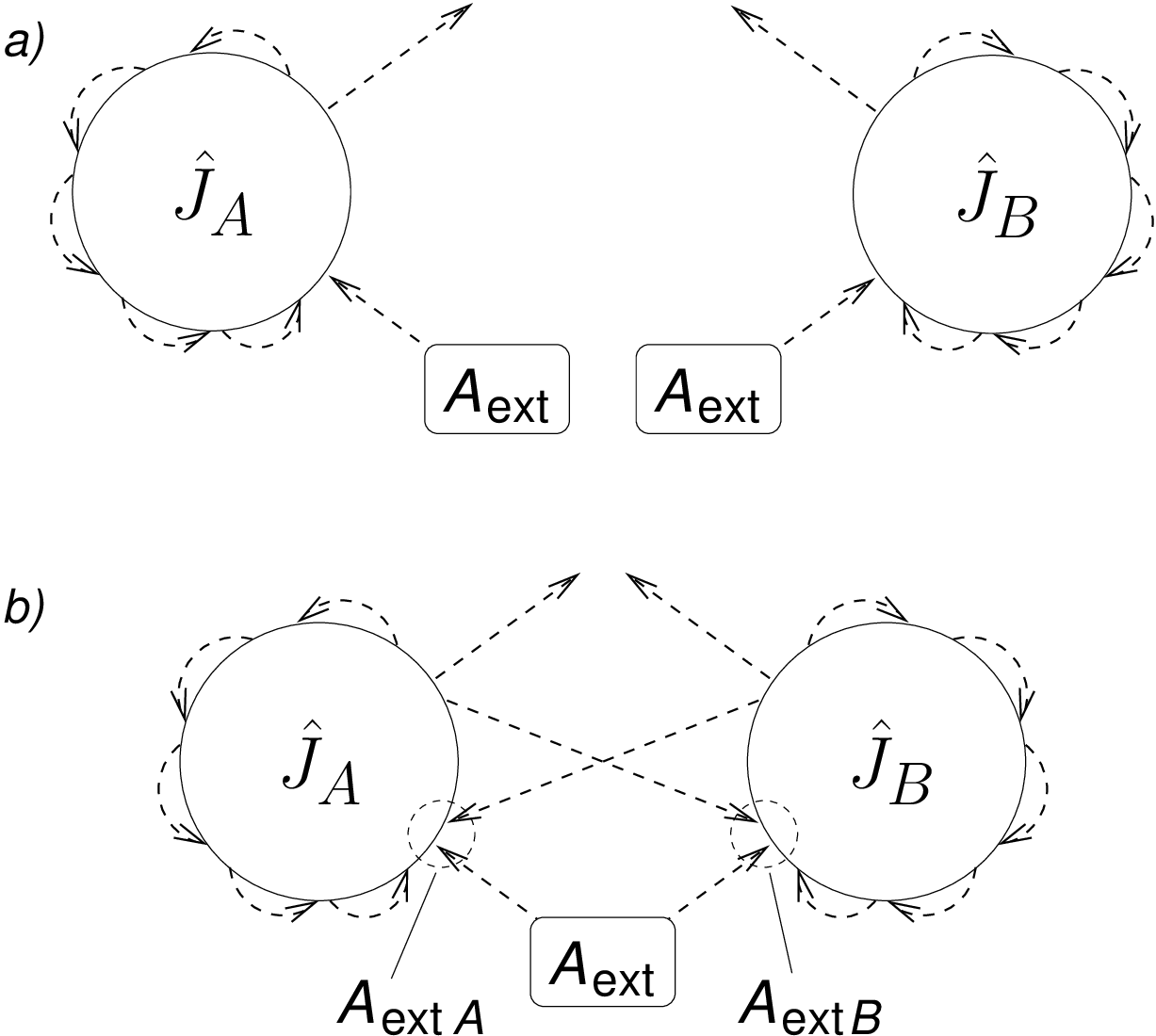}
\end{center}
\caption{Schematics of electromagnetic interaction of distinguishable devices. a) Two solitary devices in response representation. b) The same devices interacting.}\label{fig:BirdsEye6}
\end{figure}
\subsection{Statement of the problem}%
\label{ch:MB}
The crucial step to physics is from a solitary device to a pair of interacting distinguishable devices. Assume that the quantum device consists of two components, $A$ and $B$. Of interest to us is the connection among three quantum problems: those of solitary devices $A$ and $B$ (fig.\ \ref{fig:BirdsEye6}a) , and that of the composite device (fig.\ \ref{fig:BirdsEye6}b). All three are governed by the Hamiltonian (\ref{eq:86FB}).
The {\em problem of composite device\/} emerges by postulating, 
\begin{align} 
\hat H_{\mathrm{dev}}(t) &= \hat H_{\mathrm{dev}A}(t)
+\hat H_{\mathrm{dev}B}(t) , 
\label{eq:1FS} 
\\ 
\hat J(x,t) &= \hat J_A(x,t) + \hat J_B(x,t), 
\label{eq:60BC} 
\\ 
\hat \rho_{\mathrm{dev}} &= 
\hat \rho_{\mathrm{dev}A} \otimes 
\hat \rho_{\mathrm{dev}B} . 
\label{eq:62BE} 
\end{align}%
This implies factorisation of the matter subspace of the Hilbert space, so that $\hat H_{\mathrm{dev}A}(t)$, $\hat J_A(x,t)$ and $\hat \rho_{\mathrm{dev}A}$ commute with $\hat H_{\mathrm{dev}B}(t)$, $\hat J_B(x,t)$ and $\hat \rho_{\mathrm{dev}B}$. Furthermore, the {\em problem of device $A$\/} is found specifying, 
\begin{align} 
\hat H_{\mathrm{dev}}(t) &= \hat H_{\mathrm{dev}A}(t) , 
\label{eq:5YQ} 
\\ 
\hat J(x,t) &= \hat J_A(x,t), 
\label{eq:6YR} 
\\ 
\hat \rho_{\mathrm{dev}} &= 
\hat \rho_{\mathrm{dev}A} . 
\label{eq:7YS} 
\end{align}%
while the {\em problem of device $B$\/} implies that, 
\begin{align} 
\hat H_{\mathrm{dev}}(t) &= \hat H_{\mathrm{dev}B}(t) , 
\label{eq:8YT} 
\\ 
\hat J(x,t) &= \hat J_B(x,t), 
\label{eq:9YU} 
\\ 
\hat \rho_{\mathrm{dev}} &= 
\hat \rho_{\mathrm{dev}B} . 
\label{eq:10YV} 
\end{align}%
Quantum averages defined under conditions (\ref{eq:1FS})--(\ref{eq:62BE}), (\ref{eq:5YQ})--(\ref{eq:7YS})\ and (\ref{eq:8YT})--(\ref{eq:10YV})\ will be denoted, respectively, as \mbox{$
 \langle 
\cdots
 \rangle 
$}, \mbox{$
 \langle 
\cdots
 \rangle_A 
$} and \mbox{$
 \langle 
\cdots
 \rangle_B 
$}. 

Definitions of Sec.\ \ref{ch:MW} are extended to devices $A$ and $B$ by assigning subscripts $A$ and $B$ to functionals, averages and {quasidistribution}s. So, 
\begin{multline} 
\hspace{0.4\columnwidth}\hspace{-0.4\twocolumnwidth}
\Phi^{\mathrm{I}}_{\mathrm{dev}A,B}\big(
\zeta 
\big | 
A_{\mathrm{e}} 
 \big) 
= 
\text{Tr}\hat\rho _{\mathrm{dev}A,B}{\mathcal T}{\mbox{\rm\boldmath$:$}}\exp\big(
i\zeta\hat J'_{A,B}
 \big) 
{\mbox{\rm\boldmath$:$}} \\ 
= 
\prod_{x,t}\bigg\{\int d J_{A,B}(x,t)\bigg\} 
p^{\mathrm{I}}_{A,B}\big(
J_{A,B} \big| A_{\mathrm{e}}
 \big) 
\exp\big(
i\zeta J_{A,B}
 \big)
, 
\hspace{0.4\columnwidth}\hspace{-0.4\twocolumnwidth}%
\label{eq:18ZD} 
\end{multline}%
where \mbox{$
\hat J'_{A,B}(x,t)
$} are the Heisenberg currents with respect to the Hamiltonians, 
\begin{align} 
\begin{aligned} 
\hat H(t) = \hat H_{\mathrm{dev}A,B}(t)-\int dx \hat J_{A,B}(x,t)A_{\mathrm{e}}(x,t). 
\end{aligned} 
\label{eq:19ZE} 
\end{align}%
Similarly, for dressed components, 
\begin{multline} 
\hspace{0.4\columnwidth}\hspace{-0.4\twocolumnwidth}
\Phi_{\mathrm{dev}A,B}\big(
\zeta 
\big | 
A_{\mathrm{e}} 
 \big) 
= 
\Big \langle {\mathcal T}{\mbox{\rm\boldmath$:$}}\exp\big(
i\zeta{\hat{\mathcal J}}
 \big) 
{\mbox{\rm\boldmath$:$}}
 \Big \rangle _{A,B}
\\ 
= \prod_{x,t}\bigg\{\int d J_{A,B}(x,t)\bigg\} 
p _{A,B}\big(
J_{A,B} \big| A_{\mathrm{e}}
 \big) 
\exp\big(
i\zeta J_{A,B}
 \big)
. 
\hspace{0.4\columnwidth}\hspace{-0.4\twocolumnwidth}%
\label{eq:11YW} 
\end{multline}%
Finally, eq.\ (\ref{eq:87ZJ}) turns into a pair of relations, 
\begin{align} 
\begin{aligned} 
p_A\big(
 J_A\big | A_{\mathrm{e}}
 \big) 
 &= p^{\mathrm{I}}_A\big(
 J_A\big | A_{\mathrm{e}} + G_{\text{R}}J_A
 \big) , \\ 
p_B\big(
 J_B\big | A_{\mathrm{e}}
 \big) 
 &= p^{\mathrm{I}}_B\big(
 J_B\big | A_{\mathrm{e}} + G_{\text{R}}J_B
 \big) . 
\end{aligned} 
\label{eq:3YN} 
\end{align}%
Assigning indices to phase-space variables in (\ref{eq:18ZD})--(\ref{eq:3YN}) is a matter of physical clarity rather than mathematical necessity. 
\subsection{Phase-space approach to interacting quantum devices}%
\label{ch:MS}
It is not difficult to derive the formula relating $\Phi_{\mathrm{dev}}$ to $\Phi_{\mathrm{dev}A,B}$ directly, using Eqs.\ (\ref{eq:23YA}) and the obvious relation, 
\begin{align} 
\begin{aligned} 
\Phi_{\mathrm{dev}} ^{\mathrm{I}}\big(
\zeta \big | A_{\mathrm{l}}
 \big) = 
\Phi_{\mathrm{dev}A} ^{\mathrm{I}}\big(
\zeta \big | A_{\mathrm{l}}
 \big) 
\Phi_{\mathrm{dev}B} ^{\mathrm{I}}\big(
\zeta \big | A_{\mathrm{l}}
 \big) . 
\end{aligned} 
\label{eq:63BF} 
\end{align}%
Such derivation is outlined in Sec.\ \ref{ch:MT} below. Our goal is, however, not to construct a formal solution to the interaction problem --- which at this level of abstraction is more or less trivial --- but to show that this solution is consistent with ``doing quantum electrodynamics\ while thinking classically.'' We therefore start from constructing the said solution in terms of the {quasiprobability distribution}s, manipulating them as if they were classical probability distributions. In Secs \ref{ch:MG} and \ref{ch:MT} we demonstrate that the formula thus found is equivalent to the exact q-number formula. 

The total random current characterising the composite device is a sum of two components, 
\begin{align} 
\begin{aligned} 
J(x,t) = J_A(x,t)+J_B(x,t). 
\end{aligned} 
\label{eq:3KQ} 
\end{align}%
Device $A$ is formally described by random current $J_A$, which is characterised by two conditional {quasiprobability distribution}s, $p^{\mathrm{I}}_{A}\big(
J_{A} \big| A_{\mathrm{l}}
 \big)$ and $p_{A}\big(
J_{A} \big| A_{\mathrm{e}}
 \big)$. For device $B$, we have current $J_B$, characterised by $p^{\mathrm{I}}_{B}\big(
J_{B} \big| A_{\mathrm{l}}
 \big)$ and $p_{B}\big(
J_{B} \big| A_{\mathrm{e}}
 \big)$. The {quasiprobability distribution}s are pairwise connected by Eqs.\ (\ref{eq:3YN}). 

Solution to the interaction problem in terms of the ``bare'' distributions is trivial. The currents $J_A$ and $J_B$ are independent if considered conditional on the local field. 
The corresponding joint {quasiprobability distribution}\ factorises, 
\begin{align} 
\begin{aligned} 
p^{\mathrm{I}}\big(
J_{A}, J_{B} \big| A_{\mathrm{l}}
 \big) = p^{\mathrm{I}}_{A}\big(
J_{A} \big| A_{\mathrm{l}}
 \big)p^{\mathrm{I}}_{B}\big(
J_{B} \big| A_{\mathrm{l}}
 \big) . 
\end{aligned} 
\label{eq:4KR} 
\end{align}%
Integrating over the redundant information we find the {quasiprobability distribution}\ of the full current, 
\begin{align} 
p^{\mathrm{I}}\big(
J \big| A_{\mathrm{l}}
 \big) &= \prod_{x,t}\bigg\{\int dJ_A(x,t)\bigg\} 
p^{\mathrm{I}}_{A}\big(
J_{A} \big| A_{\mathrm{l}}
 \big)p^{\mathrm{I}}_{B}\big(
J_{B} \big| A_{\mathrm{l}}
 \big) , 
\label{eq:5KS} 
\end{align}%
where here and hereafter till the end of the paragraph \mbox{$
J_B(x,t)
$} stands for the current difference, 
\begin{align} 
\begin{aligned} 
J_B(x,t)
= J(x,t)-J_A(x,t)
. 
\end{aligned} 
\label{eq:26KC} 
\end{align}%
Based on this relation it is also easy to construct a solution to the interaction problem in terms of the ``dressed'' distributions. 
Applying the dressing formula (\ref{eq:87ZJ}) to (\ref{eq:5KS}) we obtain, 
\begin{align} 
p\big(
J \big| A_{\mathrm{e}}
 \big) &= \prod_{x,t}\bigg\{\int dJ_A(x,t)\bigg\} 
p^{\mathrm{I}}_{A}\big(
J_{A} \big| A_{\mathrm{e}} + G_{\text{R}}J
 \big) 
\nonumber\\ &\qquad\times 
p^{\mathrm{I}}_{B}\big(
J_{B} \big| A_{\mathrm{e}} + G_{\text{R}}J
 \big) . 
\label{eq:6KT} 
\end{align}%
We use notation (\ref{eq:77NV}). We then dress the components of the device by taking notice of Eqs.\ (\ref{eq:3YN}) and (\ref{eq:3KQ}), 
\begin{align} 
\begin{aligned} 
 &p^{\mathrm{I}}_{A}\big(
J_{A} \big| A_{\mathrm{e}} + G_{\text{R}}J
 \big) 
\\ &\quad
= p^{\mathrm{I}}_{A}\big(
J_{A} \big| A_{\mathrm{e}} + G_{\text{R}}\big [ 
J_A+J_B
 \big ] 
 \big) 
= p_{A}\big(
J_{A} \big| A_{\mathrm{e}A} 
 \big) , \\ 
 &p^{\mathrm{I}}_{B}\big(
J_{B} \big| A_{\mathrm{e}} + G_{\text{R}}J
 \big) 
\\ &\quad
= p^{\mathrm{I}}_{B}\big(
J_{B} \big| A_{\mathrm{e}} + G_{\text{R}}\big [ 
J_A+J_B
 \big ] 
 \big) 
= p_{B}\big(
J_{B} \big| A_{\mathrm{e}B}
 \big) , 
\end{aligned} 
\label{eq:7KU} 
\end{align}%
where 
\begin{align} 
\begin{aligned} 
A_{\mathrm{e}A}(x,t) &= 
A_{\mathrm{e}}(x,t) 
+ \int dx'dt' G_{\text{R}}(x,x',t-t') J_B(x,t') , \\ 
A_{\mathrm{e}B}(x,t) &= 
A_{\mathrm{e}}(x,t) 
+ \int dx'dt' G_{\text{R}}(x,x',t-t') J_A(x,t') . 
\end{aligned} 
\label{eq:2ZY} 
\end{align}%
Finally, 
\begin{align} 
\begin{aligned} 
p\big(
J \big| A_{\mathrm{e}}
 \big) = \prod_{x,t}\bigg\{\int dJ_A(x,t)\bigg\} 
p_{A}\big(
J_{A} \big| A_{\mathrm{e}A} 
 \big)
p_{B}\big(
J_{B} \big| A_{\mathrm{e}B} 
 \big) , 
\end{aligned} 
\label{eq:8KV} 
\end{align}%

The physical content of Eq.\ (\ref{eq:8KV}) is illustrated in fig.\ \ref{fig:BirdsEye6}b. It describes a random current $J(x,t)$ which is a sum of two components, $J_A(x,t)$ and $ J_B(x,t)$ . Statistics of the components depend (are conditional) on the external fields $A_{\mathrm{e}A}(x,t)$ and $A_{\mathrm{e}B}(x,t)$. 
With these fields given, components of the current are statistically independent. Because of the electromagnetic interaction, the external fields themselves become stochastic. Each is a sum of the external field affecting the composite device plus radiation of the other component. From the point of view of the device, radiation of the components is part of the local microscopic field. From the point of view of each of the components, this radiation is part of the external field. Electromagnetic self-action of the components is included in $p_{A,B}$ and is therefore excluded from Eq.\ (\ref{eq:8KV}). In quantum mechanics, these considerations apply with replacement of ``statistical'' by ``quasistatistical.'' The last statement remains subject to independent verification of Eq.\ (\ref{eq:8KV}) in Secs \ref{ch:MG} and \ref{ch:MT} below. 
\subsection{From {quasiprobability distribution}s to generating functionals}%
\label{ch:MG}
To assure a natural connection with quantum electrodynamics\ we reformulate the theory of Sec.\ \ref{ch:MS} in terms of the generating functionals (for all definittions see Sec.\ \ref{ch:MW}). Substituting Eq.\ (\ref{eq:5KS}) into Eq.\ (\ref{eq:96KJ}) we find, 
\begin{multline} 
\hspace{0.4\columnwidth}\hspace{-0.4\twocolumnwidth}
\Phi^{\mathrm{I}}_{\mathrm{dev}}\big(
\zeta 
\big | 
A_{\mathrm{l}} 
 \big) = \prod_{x,t}\bigg\{\int dJ_A(x,t)dJ_B(x,t)\bigg\} 
\\ \times 
p^{\mathrm{I}}_{A}\big(
J_{A} \big| A_{\mathrm{l}}
 \big)p^{\mathrm{I}}_{B}\big(
J_{B} \big| A_{\mathrm{l}}
 \big) \exp\big [ 
i\zeta\big(
J_A + J_B
 \big) 
 \big ] . 
\hspace{0.4\columnwidth}\hspace{-0.4\twocolumnwidth}%
\label{eq:9KW} 
\end{multline}%
The integral factorises; using Eqs.\ (\ref{eq:18ZD}) we arrive at Eq.\ (\ref{eq:63BF}). 

Finding the Hilbert-space counterpart of Eq.\ (\ref{eq:8KV}) takes a bit of ingenuity. Substituting Eq.\ (\ref{eq:8KV}) into (\ref{eq:53HM}) yields, 
\begin{multline} 
\hspace{0.4\columnwidth}\hspace{-0.4\twocolumnwidth}
\Phi_{\mathrm{dev}}\big(
\zeta 
\big | 
A_{\mathrm{e}} 
 \big) = \prod_{x,t}\bigg\{\int dJ_A(x,t)dJ_B(x,t)\bigg\} 
\\ \times 
p_{A}\big(
J_{A} \big| A_{\mathrm{e}} + G_{\text{R}}J_B
 \big) 
p_{B}\big(
J_{B} \big| A_{\mathrm{e}} + G_{\text{R}}J_A
 \big) 
\\ \times 
\exp\big [ 
i\zeta\big(
J_A + J_B
 \big) 
 \big ] . 
\hspace{0.4\columnwidth}\hspace{-0.4\twocolumnwidth}%
\label{eq:10KX} 
\end{multline}%
Unlike Eq.\ (\ref{eq:9KW}), here the integral does not factorise. However, we can write, 
\begin{multline} 
\hspace{0.4\columnwidth}\hspace{-0.4\twocolumnwidth}
p_{A}\big(
J_{A} \big| A_{\mathrm{e}} + G_{\text{R}}J_B
 \big) \\ 
= \exp\bigg(
\frac{\delta }{\delta A_{\mathrm{e}}} G_{\text{R}}J_B
 \bigg) p_{A}\big(
J_{A} \big| A_{\mathrm{e}}
 \big) , 
\hspace{0.4\columnwidth}\hspace{-0.4\twocolumnwidth}%
\label{eq:11KY} 
\end{multline}%
and 
\begin{multline} 
\hspace{0.4\columnwidth}\hspace{-0.4\twocolumnwidth}
\exp\big(
i\zeta J_B 
 \big)
\exp\bigg(
\frac{\delta }{\delta A_{\mathrm{e}}} G_{\text{R}}J_B
 \bigg) \\ 
= \exp\bigg(
-i\frac{\delta }{\delta A_{\mathrm{e}}} G_{\text{R}}
\frac{\delta }{\delta \zeta }
 \bigg)\exp\big(
i\zeta J_B 
 \big) . 
\hspace{0.4\columnwidth}\hspace{-0.4\twocolumnwidth}%
\label{eq:12KZ} 
\end{multline}%
Equation (\ref{eq:11KY}) is an application of the functional shift operator, while (\ref{eq:12KZ}) is just obvious. Relations similar to (\ref{eq:11KY}), (\ref{eq:12KZ}) may be written for the remaining two factors in the integrand in (\ref{eq:10KX}). Pulling the exponentiated differential operator out of the integral we find, 
\begin{multline} 
\Phi_{\mathrm{dev}}\big(
\zeta 
\big | 
A_{\mathrm{e}} 
 \big) = \exp\bigg(
-i\frac{\delta }{\delta A_{\mathrm{e}}'} G_{\text{R}}
\frac{\delta }{\delta \zeta } 
-i\frac{\delta }{\delta A_{\mathrm{e}}} G_{\text{R}}
\frac{\delta }{\delta \zeta' }
 \bigg) 
\\ \times 
\prod_{x,t}\bigg\{\int dJ_A(x,t)dJ_B(x,t)\bigg\} 
p_{A}\big(
J_{A} \big| A_{\mathrm{e}}
 \big) 
p_{B}\big(
J_{B} \big| A'_{\mathrm{e}}
 \big) 
\\ \times 
\exp\big(
i\zeta J_A + i\zeta' J_B
 \big)\settoheight{\auxlv}{$|$}%
\raisebox{-0.3\auxlv}{$|_{\zeta '=\zeta ,A'_{\mathrm{e}}=A_{\mathrm{e}}}$} . 
\label{eq:14LB} 
\end{multline}%
Introducing pairs of variables $\zeta(x,t) ,\zeta '(x,t)$ and $A_{\mathrm{e}}(x,t),A'_{\mathrm{e}}(x,t)$ allows all differentiations to hit the right targets. The integral in (\ref{eq:14LB}) is already factorised. Recalling (\ref{eq:11YW}) we arrive at the relation sought, 
\begin{multline} 
\hspace{0.4\columnwidth}\hspace{-0.4\twocolumnwidth}
\Phi_{\mathrm{dev}}\big(
\zeta 
\big | 
A_{\mathrm{e}} 
 \big) 
= \exp\bigg(
-i\frac{\delta }{\delta A_{\mathrm{e}}'} G_{\text{R}}
\frac{\delta }{\delta \zeta } 
-i\frac{\delta }{\delta A_{\mathrm{e}}} G_{\text{R}}
\frac{\delta }{\delta \zeta' }
 \bigg)
\\ \times 
\Phi_{\mathrm{dev}A}\big(
\zeta 
\big | 
A_{\mathrm{e}} 
 \big)\Phi_{\mathrm{dev}B}\big(
\zeta' 
\big | 
A_{\mathrm{e}}'
 \big)\settoheight{\auxlv}{$|$}%
\raisebox{-0.3\auxlv}{$|_{\zeta '=\zeta ,A'_{\mathrm{e}}=A_{\mathrm{e}}}$} . 
\hspace{0.4\columnwidth}\hspace{-0.4\twocolumnwidth}%
\label{eq:66BK} 
\end{multline}%
\subsection{Direct derivation of Eq.\ (\ref{eq:66BK})}%
\label{ch:MT}
Substituting (\ref{eq:63BF}) into (\ref{eq:23YA}) we obtain, 
\begin{multline} 
\hspace{0.4\columnwidth}\hspace{-0.4\twocolumnwidth}
\Phi_{\mathrm{dev}}\big(
\zeta \big| a_{\mathrm{e}}
 \big) 
= 
\exp\bigg(
-i\frac{\delta }{\delta a_{\mathrm{e}}} G_{\text{R}}
\frac{\delta }{\delta \zeta }
 \bigg) 
\\ \times 
\Phi_{\mathrm{dev}A} ^{\mathrm{I}}\big(
\zeta \big| a_{\mathrm{e}}
 \big) 
\Phi_{\mathrm{dev}B} ^{\mathrm{I}}\big(
\zeta \big| a_{\mathrm{e}}
 \big) . 
\hspace{0.4\columnwidth}\hspace{-0.4\twocolumnwidth}%
\label{eq:95FM} 
\end{multline}%
We now apply the relation \cite{Corresp}, 
\begin{multline} 
\hspace{0.4\columnwidth}\hspace{-0.4\twocolumnwidth}
\mathcal{F}_1\bigg(
\frac{\delta }{\delta f}
 \bigg) \mathcal{F}_2\big(
f
 \big)\mathcal{F}_3\big(
f
 \big) \\ 
= \mathcal{F}_1\bigg(
\frac{\delta }{\delta f}+
\frac{\delta }{\delta f'}
 \bigg) \mathcal{F}_2\big(
f
 \big)\mathcal{F}_3\big(
f'
 \big) \settoheight{\auxlv}{$\big|$}%
\raisebox{-0.3\auxlv}{$\big|_{f'=f}$} , 
\hspace{0.4\columnwidth}\hspace{-0.4\twocolumnwidth}%
\label{eq:96FN} 
\end{multline}%
where $\mathcal{F}_1(\cdot)$, $\mathcal{F}_2(\cdot)$, $\mathcal{F}_2(\cdot)$ are arbitrary functionals and $f(x,t),f'(x,t)$ are auxiliary functional variables. Equation (\ref{eq:96FN}) is a compact way of formulating general rules of product differentiation. It may be verified expanding $\mathcal{F}_1(\cdot)$, $\mathcal{F}_2(\cdot)$, $\mathcal{F}_2(\cdot)$ in functional Taylor series. Using it we rewrite (\ref{eq:95FM}) as, 
\begin{multline} 
\hspace{0.4\columnwidth}\hspace{-0.4\twocolumnwidth}
\Phi_{\mathrm{dev}}\big(
\zeta \big| a_{\mathrm{e}}
 \big) 
= 
\exp\bigg [ -i 
\bigg(
\frac{\delta }{\delta a_{\mathrm{e}}} + 
\frac{\delta }{\delta a_{\mathrm{e}}'} 
 \bigg) 
 G_{\text{R}}
\bigg(
\frac{\delta }{\delta \zeta } + 
\frac{\delta }{\delta \zeta '}
 \bigg) 
 \bigg ] 
\\ \times 
\Phi_{\mathrm{dev}A} ^{\mathrm{I}}\big(
\zeta \big| a_{\mathrm{e}}
 \big) 
\Phi_{\mathrm{dev}B} ^{\mathrm{I}}\big(
\zeta' \big| a_{\mathrm{e}}'
 \big) 
\settoheight{\auxlv}{$\big|$}%
\raisebox{-0.3\auxlv}{$\big|_{\zeta '=\zeta , 
a_{\mathrm{e}}'= a_{\mathrm{e}}}$}. 
\hspace{0.4\columnwidth}\hspace{-0.4\twocolumnwidth}%
\label{eq:97FP} 
\end{multline}%
Expanding the bilinear form in the exponent we have, 
\begin{multline} 
\hspace{0.4\columnwidth}\hspace{-0.4\twocolumnwidth}
\exp\bigg [ -i 
\bigg(
\frac{\delta }{\delta a_{\mathrm{e}}} + 
\frac{\delta }{\delta a_{\mathrm{e}}'} 
 \bigg) 
 G_{\text{R}}
\bigg(
\frac{\delta }{\delta \zeta } + 
\frac{\delta }{\delta \zeta '}
 \bigg) 
 \bigg ] \\ 
= 
\exp\bigg(
-i\frac{\delta }{\delta a_{\mathrm{e}}'} G_{\text{R}}
\frac{\delta }{\delta \zeta }
 \bigg)\exp\bigg(
-i\frac{\delta }{\delta a_{\mathrm{e}}} G_{\text{R}}
\frac{\delta }{\delta \zeta '}
 \bigg) 
\\ \times 
\exp\bigg(
-i\frac{\delta }{\delta a_{\mathrm{e}}} G_{\text{R}}
\frac{\delta }{\delta \zeta }
 \bigg)\exp\bigg(
-i\frac{\delta }{\delta a_{\mathrm{e}}'} G_{\text{R}}
\frac{\delta }{\delta \zeta '}
 \bigg) . 
\hspace{0.4\columnwidth}\hspace{-0.4\twocolumnwidth}%
\label{eq:98FQ} 
\end{multline}%
The last two factors here ``dress'' the devices, 
\begin{align} 
\begin{aligned} 
\exp\bigg(
-i\frac{\delta }{\delta a_{\mathrm{e}}} G_{\text{R}}
\frac{\delta }{\delta \zeta }
 \bigg) 
\Phi_{\mathrm{dev}A}^{\mathrm{I}}\big(
\zeta \big | a_{\mathrm{e}}
 \big) 
 &= 
\Phi_{\mathrm{dev}A}\big(
\zeta \big | a_{\mathrm{e}}
 \big) 
, \\ 
\exp\bigg(
-i\frac{\delta }{\delta a_{\mathrm{e}}'} G_{\text{R}}
\frac{\delta }{\delta \zeta' }
 \bigg) 
\Phi_{\mathrm{dev}B}^{\mathrm{I}}\big(
\zeta' \big | a_{\mathrm{e}}'
 \big) 
 &= 
\Phi_{\mathrm{dev}B}\big(
\zeta' \big | a_{\mathrm{e}}'
 \big) 
, 
\end{aligned} 
\label{eq:65BJ} 
\end{align}%
cf.\ Eq.\ (\ref{eq:23YA}), and we arrive at Eq.\ (\ref{eq:66BK}) with \mbox{$
A_{\mathrm{e}}\to a_{\mathrm{e}}
$}. This proves that Eq.\ (\ref{eq:8KV}) equally holds for ``dressed'' {quasiprobability distribution}s. 
\section{Discussion: cascaded systems and Sudarshan's optical equivalence theorem}\label{ch:T}
\subsection{Generalised optical equivalence theorem}%
\label{ch:CG}
\begin{figure*}
\begin{center}
\includegraphics[width=2\twocolumnwidth]{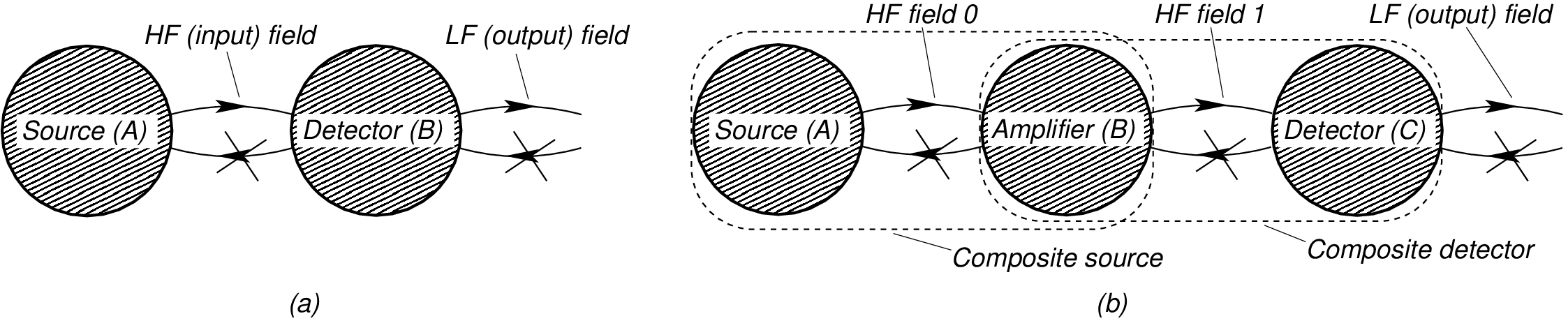}
\end{center}
\caption{Typical cascaded arrangements: (a) photodetection; (b) photodetection with coherent amplification. Shading signifies self-action problems included into models of the devices.%
\label{fig:CascadeF}}
\end{figure*}%
Analyses in Sec.\ \ref{ch:E} conclude the principal part of our investigation. They show that the {\em formal\/} response picture, postulated in \mbox{Refs.\ \cite{API,APII,APIII}} by mere analogy with the harmonic oscillator, leads to a {\em physical\/} response formulation of electromagnetic interactions of distinguishable devices. 

Moreover, quantum electrodynamics in response representation looks wholly classical. The only difference from classical statitistical electrodynamics is that the ``quantum probabilities'' $p^{\mathrm{I}}$ and $p$ are not bound to be nonnegative. {\em Any relation for P-functionals in quantum electrodynamics holds as a relation for probability distributions in classical statististical electrodynamics.\/} This statement constitutes the {\em optical equivalence theorem\/} in its most general form. 

The inverse theorem should be formulated with caution. Namely, any relation for probability distributions in classical statististical electrodynamics, which {\em (i) does not rely on their nonnegativity and (ii) is consistent with quantum electrodynamics\/}, holds as a relation for P-functionals in quantum electrodynamics. For an illustration of this statement see Sec.\ \ref{ch:TC} below. 
\subsection{Cascaded systems}\label{ch:TA}
Taken that far, the optical equivalence theorem\ is of fundamental importance but of little practicality. For an arbitrary pair of devices, solving Eq.\ (\ref{eq:66BK}) is hardly easier than solving the dressing relation (\ref{eq:23YA}) for the composite device. Things change if the macroscopic electromagnetic interaction becomes directional. Consider, for instance, the typical photodetection arrangement (Fig.\ \ref{fig:CascadeF}a). Compared to the model of Sec.\ \ref{ch:E}, this arrangement is subject to two additional approximations. Firstly, that one may distinguish the high-frequency (HF) optical field from the low-frequency (LF) photovoltage, and, secondly, that radiation of the source does not depend on the presence of the detector. Strictly speaking, these conditions introduce the concepts of {\em source\/} and {\em detector\/}. 

Note that these conditions concern not only physics but also engineering. Physics implies that the situation is macroscopic (mesoscopic): devices are distinguishable and separated by macroscopic distances, and all light beams may be controlled. Proper engineering takes care of such problems as the influence of light reflected off the detector input window on the laser source. 

The photodetection arrangement in Fig.\ \ref{fig:CascadeF}a is a particular example of a {\em cascaded quantum system\/} \cite{CascadeC,CascadeH}. In photodetection, there are two distinguishable devices, source and detector, interacting with two distinguishable electromagnetic\ fields, the input and output ones. Interactions of the devices with the fields are organised in a chain, which is made directional by neglecting the macroscopic back-action of the detector on the source. All microscopic electromagnetic\ self-actions 
are presumed to be accounted for exactly. 

For general cascaded systems, the chain of devices and fields may contain arbitrary number of links. In Fig.\ \ref{fig:CascadeF}b, we depict schematically a typical cascaded system where a field emitted by a source interacts with a coherent quantum amplifier and is then detected. 
Quantum theory of the systems in Fig.\ \ref{fig:CascadeF} is the subject of this section. 
\subsection{Generalised photodetection theory without the rotating wave approximation}%
\label{ch:TF}
\subsubsection{The model}%
\label{ch:TFM}
What makes cascaded systems simple is that radiation incident on each link may be regarded given. Nontrivial self-action (dressing) problems are hidden inside the links, and may be approached separately. Assuming that the self-action problems for devices are solved, solution to a cascaded system reduces to a large extent to its accurate formulation. 

Consider, to start with, the generalised photodetection problem (Fig.\ \ref{fig:CascadeF}a) without the RWA. 
For simplicity, we treat the HF (input, i) and LF (output, o) fields as single modes, 
\begin{align} 
\begin{aligned} 
 &\hat A_{{\mathrm{i}}}(t) = \sqrt{\frac{2\pi \hbar}{\omega_0 V_0}}
\,\hat a_0\mathrm{e}^{-i\omega _0 t} + {\mathrm{H.c.}}
\, , \\ 
 &\hat A_{{\mathrm{o}}}(t) = \sqrt{\frac{2\pi \hbar}{\omega_1 V_1}}
\,\hat a_1\mathrm{e}^{-i\omega _1 t} + {\mathrm{H.c.}}
\, , 
\end{aligned} 
\label{eq:82HF} 
\end{align}%
where $\omega _{0,1}$ are the mode frequencies and $V_{0,1}$ are the mode volumes. The field Hamiltonian reads, 
\begin{align} 
\begin{aligned} 
H_{\mathrm{f}} = \hbar \big(
\omega _0 \hat a_0^{\dag}\hat a_0 
+ 
\omega _1 \hat a_1^{\dag}\hat a_1
 \big) . 
\end{aligned} 
\label{eq:83HH} 
\end{align}%
Generalisation to a more realistic case is straightforward. 

The model of Sec.\ \ref{ch:E} applies with minor amendments, due to the presence of two field operators. Hamiltonians \mbox{$
\hat H_{\mathrm{dev}A,B}
$} and the $\rho $-matrices $\hat\rho_{\mathrm{dev}A,B} $ have the same meaning as in Sec.\ \ref{ch:E}. One may regard i,o as two values of the variable $x$, so that 
\begin{align} 
\begin{aligned} 
\hat J_A(x,t) &\to\hat J_{A{\mathrm{i}}}(t),\hat J_{A{\mathrm{o}}}(t), &
\hat J_B(x,t) &\to\hat J_{B{\mathrm{i}}}(t),\hat J_{B{\mathrm{o}}}(t). 
\end{aligned} 
\label{eq:93HT} 
\end{align}%
Device $A$ interacts with the input field by means of the current $\hat J_{A{\mathrm{i}}}(t)$, while \mbox{$
\hat J_{A{\mathrm{o}}}(t)=0
$}. Device $B$ interacts with the input field by means of the current $\hat J_{B{\mathrm{i}}}(t)$, and with the output field by means of \mbox{$\hat J_{B{\mathrm{o}}}(t)\equiv\hat J_{{\mathrm{o}}}(t)$}. 

To unify the bookkeeping we introduce a Hamiltonian with ``jumpers,'' 
\begin{multline} 
\hspace{0.4\columnwidth}\hspace{-0.4\twocolumnwidth}
\hat H(t) = \hat H_{\mathrm{f}} + \hat H_{\mathrm{dev}}(t) 
- \big [ 
\hat A_{{\mathrm{i}}}(t)+A_{{\mathrm{i}}}(t)
 \big ] \hat J_{{\mathrm{i}}}(t) 
\\ 
- \big [ 
\hat A_{{\mathrm{o}}}(t)+A_{{\mathrm{o}}}(t)
 \big ] \hat J_{{\mathrm{o}}}(t) , 
\hspace{0.4\columnwidth}\hspace{-0.4\twocolumnwidth}%
\label{eq:27KD} 
\end{multline}%
where
\begin{align} 
\begin{aligned} 
 &H_{\mathrm{dev}}(t) = {s_A}H_{\mathrm{dev}A}(t) 
+ {s_B}H_{\mathrm{dev}B}(t) , 
\\ 
 &\hat J_{\mathrm{i}}(t) = {s_A}\hat J_{\mathrm{A{\mathrm{i}}}}(t) + {s_B}\hat J_{\mathrm{B{\mathrm{i}}}}(t) . 
\\ 
 &\hat J_{\mathrm{o}}(t) = {s_B}\hat J_{\mathrm{B{\mathrm{o}}}}(t) . 
\end{aligned} 
\label{eq:28KE} 
\end{align}%
The ``jumpers'' $s_{A,B}=0,1$ ``commute'' the problems. For example, with ${s_A}=1,{s_B}=0$ we recover the problem of a solitary source. In fact we have to distinguish the problem of device $A$ and that of the light source, which differ in whether the c-number source \mbox{$
A_{{\mathrm{i}}}(t)
$} is nonzero or zero. Correspondingly we have to define two types of quantities (averages): with nonzero \mbox{$
A_{{\mathrm{i}}}(t)
$}, denoted \mbox{$
 \langle 
\cdots
 \rangle_A
$}, and with zero \mbox{$
A_{{\mathrm{i}}}(t)
$}, denoted \mbox{$
 \langle 
\cdots
 \rangle_{\mathrm{s}}
$}. 
Specifications at the averages apply in fact to averaged operators, while quantum averaging as such is always over the $\rho $-matrix, 
\begin{align} 
\begin{aligned} 
\hat \rho =  | 0 \rangle  \langle 0 | 
\otimes \hat \rho _{\mathrm{dev}} 
=  | 0 \rangle  \langle 0 | 
\otimes \hat \rho _{\mathrm{dev}A} 
\otimes \hat \rho _{\mathrm{dev}B} 
. 
\end{aligned} 
\label{eq:48AM} 
\end{align}%
Redundant degrees of freedom are traced out automatically. 
For a summary of all definitions see table \ref{T3}. 

\begin{table*}
\begin{tabular}{l@{\hspace{12pt}}l@{\hspace{12pt}}l@{\hspace{12pt}}l@{\hspace{12pt}}l}
\hline
Problem & ``Jumper'' settings & Relevant & \multicolumn{2}{l}{Notation for averages} \\ 
&&ext.\ souces&``Raw''&``Physical''\\
\hline
Light source & ${s_A}=1,{s_B}=0$ & $(A_{{\mathrm{i}}})$ & $ \langle \cdots \rangle_A$ & $ \langle \cdots \rangle_{\mathrm{s}}=(
 \langle \cdots \rangle_A
 )\settoheight{\auxlv}{$|$}%
\raisebox{-0.3\auxlv}{$|_{A_{{\mathrm{i}}}=0}$}$ \\ 
Detector & ${s_A}=0,{s_B}=1$ & $A_{{\mathrm{i}}},(A_{{\mathrm{o}}})$ & $ \langle \cdots \rangle_B$ & $ \langle \cdots \rangle_{\mathrm{d}}=(
 \langle \cdots \rangle_B
 )\settoheight{\auxlv}{$|$}%
\raisebox{-0.3\auxlv}{$|_{A_{{\mathrm{o}}}=0}$}$ \\ 
Photodet.\ arrangement & ${s_A}=1,{s_B}=1$ & $(A_{{\mathrm{i}}},A_{{\mathrm{o}}})$ & $ \langle \cdots \rangle$ & $ \langle \cdots \rangle_{\mathrm{o}}=
{
 \langle \cdots \rangle
}\settoheight{\auxlv}{$|$}%
\raisebox{-0.3\auxlv}{$|_{A_{{\mathrm{i}}}=A_{{\mathrm{o}}}=0}$}$ \\
\hline
\end{tabular}
\caption{Three problems relevant to the arrangement in Fig.\ \ref{fig:CascadeF}a. ``Raw'' averages imply the density matrix (\ref{eq:48AM}) and Hamiltonian (\ref{eq:27KD}), the latter with ``jumpers'' set to listed values. ``Physical'' averages follow by setting some or all c-number sources to zero. The table also lists the c-number sources on which the ``raw'' averages depend; those shown in brackets are set to zero in ``physical'' averages.}
\label{T3}
\end{table*}

Full formal analysis of this system is the subject of appendix \ref{ch:CB}. For purposes of this discussion, the source is described by the time-normal averages of the Heisenberg\ field operator \mbox{$
{\hat{\mathcal A}}_{{\mathrm{i}}}(t)
$}, 
\begin{multline} 
\hspace{0.4\columnwidth}\hspace{-0.4\twocolumnwidth}
\big \langle {\mathcal T}{\mbox{\rm\boldmath$:$}} 
{\hat{\mathcal A}}_{{\mathrm{i}}}(t_1)\cdots
{\hat{\mathcal A}}_{{\mathrm{i}}}(t_m)
{\mbox{\rm\boldmath$:$}} \big \rangle_{\mathrm{s}} \\ 
= \prod_{t}\bigg\{\int d A_{{\mathrm{i}}}(t)\bigg\} p_{\mathrm{s}}\big( 
A_{{\mathrm{i}}}
 \big) A_{{\mathrm{i}}}(t_1)\cdots A_{{\mathrm{i}}}(t_m) , 
\hspace{0.4\columnwidth}\hspace{-0.4\twocolumnwidth}%
\label{eq:86HL} 
\end{multline}%
where we have introduced the corresponding quasiprobability distribution. The detector and the full arrangement are both described by the time-normal averages of the Heisenberg\ current operator \mbox{$
{\hat{\mathcal J}}_{{\mathrm{o}}}(t)
$}, but defined under different conditions. For the detector, 
\begin{multline} 
\hspace{0.4\columnwidth}\hspace{-0.4\twocolumnwidth}
\big \langle {\mathcal T}{\mbox{\rm\boldmath$:$}} 
{\hat{\mathcal J}}_{{\mathrm{o}}}(t_1)\cdots
{\hat{\mathcal J}}_{{\mathrm{o}}}(t_m)
{\mbox{\rm\boldmath$:$}} \big \rangle_{\mathrm{d}} \\ 
= \prod_{t}\bigg\{\int d J_{{\mathrm{o}}}(t)\bigg\} p_{\mathrm{d}}{\big( 
J_{{\mathrm{o}}}\big| A_{{\mathrm{i}}}
 \big)} J_{{\mathrm{o}}}(t_1)\cdots J_{{\mathrm{o}}}(t_m) , 
\hspace{0.4\columnwidth}\hspace{-0.4\twocolumnwidth}%
\label{eq:89HP} 
\end{multline}%
while for the photodetection arrangement, 
\begin{multline} 
\hspace{0.4\columnwidth}\hspace{-0.4\twocolumnwidth}
\big \langle {\mathcal T}{\mbox{\rm\boldmath$:$}} 
{\hat{\mathcal J}}_{{\mathrm{o}}}(t_1)\cdots
{\hat{\mathcal J}}_{{\mathrm{o}}}(t_m)
{\mbox{\rm\boldmath$:$}} \big \rangle_{\mathrm{o}} \\ 
= \prod_{t}\bigg\{\int d J_{{\mathrm{o}}}(t)\bigg\} p_{{\mathrm{o}}}\big( 
J_{{\mathrm{o}}}
 \big) J_{{\mathrm{o}}}(t_1)\cdots J_{{\mathrm{o}}}(t_m) . 
\hspace{0.4\columnwidth}\hspace{-0.4\twocolumnwidth}%
\label{eq:92HS} 
\end{multline}%
The symbols \mbox{$
 \langle 
\cdots
 \rangle _{\mathrm{d}}
$} and \mbox{$
 \langle 
\cdots
 \rangle _{\mathrm{o}}
$} are defined in table \ref{T3}. Averages (\ref{eq:89HP}) by construction depend (are conditional) on the source \mbox{$
A_{{\mathrm{i}}}(t)
$}; this dependence is made explicit in the conditional quasiprobability distribution\ \mbox{$
p_{\mathrm{d}}{( 
J_{{\mathrm{o}}}| A_{{\mathrm{i}}}
 )}
$} characterising the detector. Averages (\ref{eq:86HL}) and (\ref{eq:92HS}) are unconditional. 
\subsubsection{Semiclassical {\em versus\/} quantum photodetection theory}\label{ch:CBS}
Following the pattern of ``doing quantum electrodynamics\ while thinking classically,'' the relation connecting observable photodetection statistics to properties of the source and detector should be, 
\begin{align} 
\begin{aligned} 
p_{{\mathrm{o}}}\big( 
J_{{\mathrm{o}}}
 \big) = \prod_{t}\bigg\{\int d A_{{\mathrm{i}}}(t)\bigg\}p_{\mathrm{s}}\big( 
A_{{\mathrm{i}}}
 \big)p_{\mathrm{d}}{\big( 
J_{{\mathrm{o}}}\big| A_{{\mathrm{i}}}
 \big)} . 
\end{aligned} 
\label{eq:91HR} 
\end{align}%
Taking notice of the definitions (\ref{eq:86HL}), (\ref{eq:89HP}) and (\ref{eq:92HS}), Eq.\ (\ref{eq:91HR}) is equivalent to the following relation for the time-normal operator averages, 
\begin{multline} 
\hspace{0.4\columnwidth}\hspace{-0.4\twocolumnwidth}
\big \langle 
{\mathcal T}{\mbox{\rm\boldmath$:$}} 
{\hat{\mathcal J}}_{\mathrm{o}}(t_1)\cdots{\hat{\mathcal J}}_{\mathrm{o}}(t_m)
{\mbox{\rm\boldmath$:$}}
 \big \rangle_{\mathrm{o}} \\ 
= \big \langle 
{\mathcal T}{\mbox{\rm\boldmath$:$}}\big [ 
\big \langle 
{\mathcal T}{\mbox{\rm\boldmath$:$}} 
{\hat{\mathcal J}}_{\mathrm{o}}(t_1)\cdots{\hat{\mathcal J}}_{\mathrm{o}}(t_m)
{\mbox{\rm\boldmath$:$}}
 \big \rangle\settoheight{\auxlv}{$|$}%
\raisebox{-0.3\auxlv}{$|_{{\mathrm{d}}}$} 
 \big ]\settoheight{\auxlv}{$|$}%
\raisebox{-0.3\auxlv}{$|_{A_{{\mathrm{i}}}\to{\hat{\mathcal A}}_{{\mathrm{i}}}}$} {\mbox{\rm\boldmath$:$}}
 \big \rangle_{\mathrm{s}} . 
\hspace{0.4\columnwidth}\hspace{-0.4\twocolumnwidth}%
\label{eq:21DQ} 
\end{multline}%
For justification of these formulae see appendix \ref{ch:CB}. 

It is instructive to compare Eq.\ (\ref{eq:21DQ}) to the corresponding classical formula. In classical stochastic electrodynamics, a photodetector may be characterised by averages of the photocurrent \mbox{$
J(t)
$} conditional on the detected field \mbox{$
A_{{\mathrm{i}}}(t)
$}, 
\begin{align} 
\begin{aligned} 
\overline{\hspace{0.1ex}J(t_1)\cdots J(t_m)\hspace{0.1ex}}\settoheight{\auxlv}{$|$}%
\raisebox{-0.3\auxlv}{$|_{[A_{{\mathrm{i}}}]}$}. 
\end{aligned} 
\label{eq:8DA} 
\end{align}%
In turn, the detected field is characterised 
by the unconditional averages, 
\begin{align} 
\begin{aligned} 
\overline{\hspace{0.1ex}A_{{\mathrm{i}}}(t_1)\cdots A_{{\mathrm{i}}}(t_m)\hspace{0.1ex}}, 
\end{aligned} 
\label{eq:72BN} 
\end{align}%
The unconditional photodetection statistics observed in the experiment is found imposing the second layer of averaging over the conditional averages (\ref{eq:8DA})
\begin{align} 
\begin{aligned} 
\overline{\hspace{0.1ex}J(t_1)\cdots J(t_m)\hspace{0.1ex}} 
= \overline{\hspace{0.1ex}\overline{\hspace{0.1ex}J(t_1)\cdots J(t_m)\hspace{0.1ex}}\settoheight{\auxlv}{$|$}%
\raisebox{-0.3\auxlv}{$|_{[A_{{\mathrm{i}}}]}$}\hspace{0.1ex}}^{\,A_{{\mathrm{i}}}} \, .
\end{aligned} 
\label{eq:75BR} 
\end{align}%
Variable at the end of the bar is the one over which the averaging is performed; such specifications help to avoid confusion. Parallelism between Eqs.\ (\ref{eq:21DQ}) and (\ref{eq:75BR}) is indisputable: they coincide up to the replacement of time-normal averages by classical statistical averages. 

The said parallelism gets even more pronounced if we note that photocurrent must be classical also in quantum theory. That is, there must exist representations of the time-normal current averages by classical statistical averages, 
\begin{align} 
\begin{aligned} 
\big \langle {\mathcal T}{\mbox{\rm\boldmath$:$}} 
{\hat{\mathcal J}}_{\mathrm{o}}(t_1)
{\hat{\mathcal J}}_{\mathrm{o}}(t_m)
{\mbox{\rm\boldmath$:$}} \big \rangle_{\mathrm{d}} 
= 
\overline{\hspace{0.1ex}J(t_1)\cdots J(t_m)\hspace{0.1ex}}\settoheight{\auxlv}{$|$}%
\raisebox{-0.3\auxlv}{$|_{[A_{{\mathrm{i}}}]}$} 
, 
\end{aligned} 
\label{eq:11DD} 
\end{align}%
and 
\begin{align} 
\begin{aligned} 
\big \langle {\mathcal T}{\mbox{\rm\boldmath$:$}} 
{\hat{\mathcal J}}_{\mathrm{o}}(t_1)
{\hat{\mathcal J}}_{\mathrm{o}}(t_m)
{\mbox{\rm\boldmath$:$}} \big \rangle_{\mathrm{o}} 
= 
\overline{\hspace{0.1ex}J(t_1)\cdots J(t_m)\hspace{0.1ex}} 
. 
\end{aligned} 
\label{eq:14DH} 
\end{align}%
The same may be expressed as nonnegativity of the corresponding P-functionals. Classical interpretation of the quantum averages (\ref{eq:11DD}) and (\ref{eq:14DH}) is possible if 
\begin{align} 
\begin{aligned} 
p_{\mathrm{o}}\big( 
J
 \big)\geq 0,
\end{aligned} 
\label{eq:16DK} 
\end{align}%
and 
\begin{align} 
\begin{aligned} 
p_{\mathrm{d}}{\big( 
J \big| A_{{\mathrm{i}}}
 \big)}\geq 0. 
\end{aligned} 
\label{eq:17DL} 
\end{align}%
Conditions (\ref{eq:16DK}) and (\ref{eq:17DL}) are a heuristic principle to be imposed on quantum models of photodetectors. 
Nothing in the above and in appendix \ref{ch:CB} depends on them. 

The only source of quantum behaviour in a photodetection experiment is possible nonpositivity of the P-functional related to averages of the detected field, 
\begin{multline} 
\hspace{0.4\columnwidth}\hspace{-0.4\twocolumnwidth}
\big \langle {\mathcal T}{\mbox{\rm\boldmath$:$}} 
{\hat{\mathcal A}}_{{\mathrm{i}}}(t_1)\cdots
{\hat{\mathcal A}}_{{\mathrm{i}}}(t_m)
{\mbox{\rm\boldmath$:$}} \big \rangle_{\mathrm{s}} \\ 
= \prod_{t}\bigg\{\int d A_{{\mathrm{i}}}(t)\bigg\} p_{\mathrm{s}}\big( 
A_{{\mathrm{i}}}
 \big) A_{{\mathrm{i}}}(t_1)\cdots A_{{\mathrm{i}}}(t_m) . 
\hspace{0.4\columnwidth}\hspace{-0.4\twocolumnwidth}%
\label{eq:22DR} 
\end{multline}%
Accounting for Eqs.\ (\ref{eq:11DD}), (\ref{eq:14DH}), Eq.\ (\ref{eq:21DQ}) may be written as, 
\begin{multline} 
\hspace{0.4\columnwidth}\hspace{-0.4\twocolumnwidth}
\overline{\hspace{0.1ex}J(t_1)\cdots J(t_m)\hspace{0.1ex}} = \prod_{t}\bigg\{\int d A_{{\mathrm{i}}}(t)\bigg\} p_{\mathrm{s}}\big( 
A_{{\mathrm{i}}}
 \big)
\\ \times 
\overline{\hspace{0.1ex}J(t_1)\cdots J(t_m)\hspace{0.1ex}}\settoheight{\auxlv}{$|$}%
\raisebox{-0.3\auxlv}{$|_{[A_{{\mathrm{i}}}]}$} . 
\hspace{0.4\columnwidth}\hspace{-0.4\twocolumnwidth}%
\label{eq:24DT} 
\end{multline}%
If \mbox{$
p_{\mathrm{s}}\big( 
A_{{\mathrm{i}}}
 \big)\geq 0
$}, functional integration over \mbox{$
p_{\mathrm{s}}\big( 
A_{{\mathrm{i}}}
 \big)
$} may be interpreted as a classical statistical averaging, 
\begin{align} 
\begin{aligned} 
\big \langle {\mathcal T}{\mbox{\rm\boldmath$:$}} 
{\hat{\mathcal A}}_{{\mathrm{i}}}(t_1)\cdots
{\hat{\mathcal A}}_{{\mathrm{i}}}(t_m)
{\mbox{\rm\boldmath$:$}} \big \rangle_{\mathrm{s}} 
= 
\overline{\hspace{0.1ex}A_{{\mathrm{i}}}(t_1)\cdots A_{{\mathrm{i}}}(t_m)\hspace{0.1ex}}^{\,A_{{\mathrm{i}}}} 
. 
\end{aligned} 
\label{eq:23DS} 
\end{align}%
In this case Eq.\ (\ref{eq:24DT}) reverts to the classical formula (\ref{eq:75BR}), and quantum mechanics becomes fully hidden from view. 

Noteworthy is that condition (\ref{eq:17DL}) does not warrant condition (\ref{eq:16DK}). Photodetection must add enough noise to counter possible nonpositiveness of \mbox{$
p_{\mathrm{s}}(A_{{\mathrm{i}}})
$}. This is the reason why a photodetector free of shot noise cannot exist. 
\subsection{Quantum and classical devices}%
\label{ch:QCD}
It is convenient to introduce terminology generalising the concepts of classical and quantum states of the harmonic oscillator to arbitrary electromagnetic devices. A device is {\em macroscopically classical\/} if the conditional P-functional characterising it is nonnegative, otherwise it is {\em macroscopically quantum\/}. Unlike the harmonic oscillator, macroscopic classicality or quantumness of a device concerns not only its quantum state but also its dynamics. We have to know both the state of the device $\hat\rho _{\mathrm{dev}}$ and the bare current operator $\hat J(x,t)$ in order to decide to which variety the device belongs. This knowledge suffices: the dressing formula (\ref{eq:87ZJ}) and Eqs.\ (\ref{eq:5KS}), (\ref{eq:8KV}) describing interactions of devices preserve nonnegativity of P-functionals (i.e., if $p^{\mathrm{I}}\geq 0$, then $p\geq 0$; if $p^{\mathrm{I}}_{A,B}\geq 0$, then $p^{\mathrm{I}}\geq 0$ and $p\geq 0$). 

The concept of macroscopically classical device calls for a word of caution. Firstly, any such device is quantum at a deeper level of insight. {\em Classicality is always an approximation.\/} Secondly, it may take quantum mechanics to explain properties of a macroscopically classical device. The best known example is the black-body radiation: by itself, it is in a classical state. Thirdly, there exist devices in classical statitistical electrodynamics which cannot be implemented as macroscopically classical devices in quantum electrodynamics. Again, examples of such devices are well known: a coherent quantum amplifier, a photodetector without shot noise, and a light-beam cloner, to name just a few. 

One more reservation one has to make is that a device may appear macroscopically classical due to limitations of an experiment. Indeed, let us have a closer look at functional $p_{\mathrm{d}}(J_{{\mathrm{o}}}|A_{{\mathrm{i}}})$. It is natural to say that the detected field is in a coherent state if 
\begin{align} 
\begin{aligned} 
p_{\mathrm{s}}(A_{{\mathrm{i}}}) = \prod_{t}\delta\big(
A_{{\mathrm{i}}}(t)-A_{\mathrm{e}}(t)
 \big) , 
\end{aligned} 
\label{eq:60SK} 
\end{align}%
where $A_{\mathrm{e}}(t)$ is a given c-number field. For such field, 
\begin{align} 
\begin{aligned} 
p_{\mathrm{o}}\big(
J_{{\mathrm{o}}}
 \big) = p_{\mathrm{d}}{\big( 
J_{{\mathrm{o}}}\big| A_{\mathrm{e}}
 \big)} . 
\end{aligned} 
\label{eq:61SL} 
\end{align}%
That is, \mbox{$p_{\mathrm{d}}(J_{{\mathrm{o}}}|A_{\mathrm{e}})$} describes results of a photodetection experiment with light in a coherent state \cite{endCoh}. The positivity condition (\ref{eq:17DL}) warrants that the photocurrent in such experiment stays classical. However, \mbox{$p_{\mathrm{d}}(J_{{\mathrm{o}}}|A_{\mathrm{e}})$} does not give a full quantum description of the detector as a quantum device. 
Positivity of $p_{\mathrm{d}}(J|E,E^*)$ may be a result of forfeiting all information about the optical mode, and/or ignoring response properties of the photocurrent mode (cf.\ appendix \ref{ch:CBA}). {\em The same device may behave classically in one experiment and in a quantum manner in another\/}, depending on which information a particular experiment allows access to. 
\subsection{Sudarshan's optical equivalence theorem}%
\label{ch:SGO}
That Eq.\ (\ref{eq:24DT}) equally holds in classical mechanics with nonnegative \mbox{$
p_{\mathrm{s}}\big( 
A_{{\mathrm{i}}}
 \big)
$} and in quantum mechanics with (possibly) nonpositive \mbox{$
p_{\mathrm{s}}\big( 
A_{{\mathrm{i}}}
 \big)
$} is {\em Sudarshan's optical equivalence theorem\/} generalised to interacting electromagnetic\ field.
In Sudarshan's seminal paper \cite{Sudarshan}, the optical equivalence theorem is formulated as follows. Let $X(\cdot,\cdot)$ be a normal representation of an operator $\hat X$, 
\begin{align} 
\begin{aligned} 
\hat X= {\mbox{\rm\boldmath$:$}}
X(\hat a,\hat a^{\dag})
{\mbox{\rm\boldmath$:$}} , 
\end{aligned} 
\label{eq:63SN} 
\end{align}%
and $P(\cdot,\cdot)$ --- the diagonal representation (P-function) of the rho-matrix $\hat \rho $, 
\begin{align} 
\begin{aligned} 
\hat\rho = \int d^2 z P(z,z^*) | 
z
 \rangle  \langle 
z
 |, 
\end{aligned} 
\label{eq:64SP} 
\end{align}%
where $ | 
z
 \rangle $ is a coherent state, 
\begin{align} 
\begin{aligned} 
 | 
z
 \rangle = \exp
(
z\hat a^{\dag}- z^*\hat a
 )  | 
0
 \rangle , 
\end{aligned} 
\label{eq:65SQ} 
\end{align}%
and $ | 
0
 \rangle $ is the vacuum state, 
\begin{align} 
\begin{aligned} 
\hat a | 
0
 \rangle = 0 . 
\end{aligned} 
\label{eq:67SS} 
\end{align}%
The quantum average of the operator $\hat X$ may then be written as a classically looking phase-space average, 
\begin{align} 
\begin{aligned} 
\big \langle 
\hat X
 \big \rangle = \text{Tr}\hat\rho \hat X= \int d^2 z P(z,z^*)X(z,z^*) . 
\end{aligned} 
\label{eq:66SR} 
\end{align}%
Equation (\ref{eq:66SR}) is a somewhat modernised version of eq.\ (5) in \mbox{Ref.\ \cite{Sudarshan}}, which was written for a special case $\hat X= \hat a^{\dag \lambda}\hat a^{\mu }$. Parallelism between Eqs.\ (\ref{eq:24DT}) and (\ref{eq:66SR}) is evident. 

Sudarshan's optical equivalence theorem\ for interacting fields equally applies to classical and quantum devices. Indeed, Eq.\ (\ref{eq:91HR}) does not rely on heusistic conditions (\ref{eq:16DK}), (\ref{eq:17DL}). Equation (\ref{eq:21DQ}) may be written in yet another form, 
\begin{multline} 
\hspace{0.4\columnwidth}\hspace{-0.4\twocolumnwidth}
\big \langle 
{\mathcal T}{\mbox{\rm\boldmath$:$}} 
{\hat{\mathcal J}}_{\mathrm{o}}(t_1)\cdots{\hat{\mathcal J}}_{\mathrm{o}}(t_m)
{\mbox{\rm\boldmath$:$}}
 \big \rangle_{\mathrm{o}} \\ 
= \prod_{t}\bigg\{\int d A_{{\mathrm{i}}}(t)\bigg\} p_{\mathrm{s}}\big( 
A_{{\mathrm{i}}}
 \big) 
\big \langle 
{\mathcal T}{\mbox{\rm\boldmath$:$}} 
{\hat{\mathcal J}}_{\mathrm{o}}(t_1)\cdots{\hat{\mathcal J}}_{\mathrm{o}}(t_m)
{\mbox{\rm\boldmath$:$}}
 \big \rangle\settoheight{\auxlv}{$|$}%
\raisebox{-0.3\auxlv}{$|_{{\mathrm{d}}}$} 
. 
\hspace{0.4\columnwidth}\hspace{-0.4\twocolumnwidth}%
\label{eq:97HX} 
\end{multline}%
This relation holds irrespective of whether the time-normal averages here may be interpreted classically according to Eqs.\ (\ref{eq:11DD}), (\ref{eq:14DH}). {\em Any macroscopic device, no matter whether quantum or classical, is fully specified by its response to input field in classical states.\/}
\subsection{Semiclassical theory of photodetection with coherent quantum amplification}\label{ch:TC}
\subsubsection{Preliminary remarks}\label{ch:TCP}
Utility of classical devices is in that they afford a phenomenological classical description {\em within quantum electrodynamics\/}. For a photodetector, this description is given by conditional statistical averages (\ref{eq:11DD}), or, which is the same, by the functional \mbox{$p_{\mathrm{d}}(J_{{\mathrm{o}}}|A_{{\mathrm{i}}})$}. It is often possible to construct a simple phenomenological model which correctly reproduces averages (\ref{eq:11DD}), or analogous quantities in other situations. If two or more classical devices are combined in a set-up with some quantum devices, a classical approach to the classical part of the set-up may preceed a fully quantum treatment of the set-up as a whole {\em without any loss in rigour\/}. 

As an example we consider the cascaded system in Fig.\ \ref{fig:CascadeF}b. Subject to self-noise of the amplifier being in a classical state, both the amplifier and detector are classical devices: if their input signal is in a classical state, their output signal is also in a classical state. Since, according to the optical equivalence theorem, full quantum description of a device is given by its reaction to classical fields, theory of the composite detector (cf.\ Fig.\ \ref{fig:CascadeF}b) may be formulated in purely classical terms, which is next to trivial. 

We also take this opportunity to generalise our analyses to the rotating wave approximation\ for optical fields. Consequently the HF fields 0 and 1 (cf.\ Fig.\ \ref{fig:CascadeF}b) are described by slow amplitudes \mbox{$
E_0(t)
$} and \mbox{$
E_1(t)
$}, which in quantum treatment become Heisenberg\ operators \mbox{$
{\hat{\mathcal E}}_0(t)
$} and \mbox{$
{\hat{\mathcal E}}_1(t)
$}. The photocurrent \mbox{$
J(t)
$}, which in quantum treatment is represented by the operator \mbox{$
{\hat{\mathcal J}}_{{\mathrm{o}}}(t)
$}, is a broad-band\ process and thus in no way is a subject to rotating wave approximation. 

\subsubsection{Semiclassical {\em versus\/} quantum theory}\label{ch:TCV}
Full quantum treatment of the system in Fig.\ \ref{fig:CascadeF}b may be found in appendix \ref{ch:A}. The result is very much as expected: ``doing quantum electrodynamics\ while thinking classically'' is equally applicable to this system. In classical stochastic electrodynamics, the detector is described by the probability distribution of the photocurrent, conditional on the detected field \mbox{$
E_1(t)
$}, \mbox{$p_{\mathrm{d}}{\big( 
J_{\mathrm{o}}\big| E_1,E_1^*
 \big)} $}. The amplifier is described by the probability distribution of the output field, conditional on the input field \mbox{$
E_0(t)
$}, \mbox{$p_{\mathrm{a}}{\big( 
E_1,E_1^* \big| E_0,E_0^*
 \big)} $}. As to the source, it is described by the unconditional probability distribution of the radiated field, \mbox{$p_{\mathrm{s}}\big( 
E_0,E_0^*
 \big) $}. It is convenient to start from solving for the composite detector (cf.\ Fig.\ \ref{fig:CascadeF}b), characterised by the probability distribution of the photocurrent, conditional on the detected field \mbox{$
E_0(t)
$}, \mbox{$p_{\mathrm{cd}}{\big( 
J_{\mathrm{o}}\big| E_0,E_0^*
 \big)} $}. Obviously, 
\begin{multline} 
\hspace{0.4\columnwidth}\hspace{-0.4\twocolumnwidth}
p_{\mathrm{cd}}{\big( 
J_{\mathrm{o}}\big| E_0,E_0^*
 \big)} = \prod_{t}\bigg\{\int d^2E_1(t)\bigg\}
\\ \times 
p_{\mathrm{a}}{\big( 
E_1,E_1^* \big| E_0,E_0^*
 \big)} 
p_{\mathrm{d}}{\big( 
J_{\mathrm{o}}\big| E_1,E_1^*
 \big)}
. 
\hspace{0.4\columnwidth}\hspace{-0.4\twocolumnwidth}%
\label{eq:29KF} 
\end{multline}%
Statistical properties of the photocurrent, characterised by the unconditional probability distribution \mbox{$p_{\mathrm{o}}\big( 
J_{\mathrm{o}}\big)$}, then follow with ease, 
\begin{multline} 
\hspace{0.4\columnwidth}\hspace{-0.4\twocolumnwidth}
p_{\mathrm{o}}\big( 
J_{\mathrm{o}}\big) = \prod_{t}\bigg\{\int d^2E_0(t)\bigg\} 
p_{\mathrm{s}}\big( 
E_0,E_0^*
 \big) p_{\mathrm{cd}}{\big( 
J_{\mathrm{o}}\big| E_0,E_0^*
 \big)} . 
\hspace{0.4\columnwidth}\hspace{-0.4\twocolumnwidth}%
\label{eq:30KH} 
\end{multline}%
Alternatively, one may solve for the composite source (cf.\ Fig.\ \ref{fig:CascadeF}b), characterised by the unconditional probability distribution \mbox{$p_{\mathrm{cs}}\big( 
E_1,E_1^*
 \big) $}, 
\begin{multline} 
\hspace{0.4\columnwidth}\hspace{-0.4\twocolumnwidth}
p_{\mathrm{cs}}\big( 
E_1,E_1^*
 \big) = \prod_{t}\bigg\{\int d^2E_0(t)\bigg\}
\\ \times 
p_{\mathrm{s}}\big( 
E_0,E_0^*
 \big) 
p_{\mathrm{a}}{\big( 
E_1,E_1^* \big| E_0,E_0^*
 \big)} 
, 
\hspace{0.4\columnwidth}\hspace{-0.4\twocolumnwidth}%
\label{eq:31KJ} 
\end{multline}%
and, 
\begin{multline} 
\hspace{0.4\columnwidth}\hspace{-0.4\twocolumnwidth}
p_{\mathrm{o}}\big( 
J_{\mathrm{o}}
 \big) = \prod_{t}\bigg\{\int d^2E_1(t)\bigg\}
\\ \times 
p_{\mathrm{cs}}\big( 
E_1,E_1^*
 \big) 
p_{\mathrm{d}}{\big( 
J_{\mathrm{o}}\big| E_1,E_1^*
 \big)}
. 
\hspace{0.4\columnwidth}\hspace{-0.4\twocolumnwidth}%
\label{eq:32KK} 
\end{multline}%
Either way, we find, 
\begin{multline} 
\hspace{0.4\columnwidth}\hspace{-0.4\twocolumnwidth}
p_{\mathrm{o}}\big( 
J_{\mathrm{o}}\big) 
= \prod_{t}\bigg\{\int d^2E_0(t)d^2E_1(t)\bigg\}
\\ \times 
p_{\mathrm{s}}\big( 
E_0,E_0^*
 \big) 
p_{\mathrm{a}}{\big( 
E_1,E_1^* \big| E_0,E_0^*
 \big)} 
p_{\mathrm{d}}{\big( 
J_{\mathrm{o}}\big| E_1,E_1^*
 \big)} . 
\hspace{0.4\columnwidth}\hspace{-0.4\twocolumnwidth}%
\label{eq:33KL} 
\end{multline}%

In appendix \ref{ch:A} we show that Eqs.\ (\ref{eq:29KF})--(\ref{eq:33KL}) equally hold as relations for P-functionals. A major fraction of the effort is spent on proper quantum definitions. However, as soon as we become aware of this result, 90\% of the quantum treatment turns redundant. Subject to self-noise of the amplifier being in a classical state, both amplifier and detector are classical devices. Their properties may be described within simple semiclassical models. The only possible source of quantum behaviour is the light source. To include it in our analyses, it suffices to assume that there exists a quantum framework, in which one can define time-normal averages of \mbox{$
{\hat{\mathcal E}}_0(t)
$}, 
\begin{align} 
\begin{aligned} 
\big \langle {\mathcal T}{\mbox{\rm\boldmath$:$}} 
{\hat{\mathcal E}}_0(t_1)\cdots{\hat{\mathcal E}}_0(t_m)
{\hat{\mathcal E}}_0^{\dag}(t'_1)\cdots{\hat{\mathcal E}}_0^{\dag}(t'_n)
{\mbox{\rm\boldmath$:$}} \big \rangle_{\mathrm{s}} 
.
\end{aligned} 
\label{eq:96HW} 
\end{align}%
Under the rotating wave approximation, one has to apply apply the Kelley-Kleiner definition \cite{KelleyKleiner,GlauberTN}, cf.\ appendix \ref{ch:DA}. Equation (\ref{eq:96HW}) is the only quantum formula we need in this section.
\subsubsection{Classical coherent amplifier with spontaneous noise}\label{ch:TCA}
The system in Fig.\ \ref{fig:CascadeF}b is rich enough in the sense that it allows one to illustrate restrictions to ``doing quantum electrodynamics\ while thinking classically.'' These restrictions are two: quantum mechanics does not allow for noiseless amplification nor for noiseless detection. So, the simplest semiclassical model of the coherent amplifier consistent with quantum mechanics reads, 
\begin{align} 
\begin{aligned} 
E_1(t) = \sqrt{T_{\mathrm{a}}}E_0(t) + E_{\mathrm{a}}(t) , 
\end{aligned} 
\label{eq:25DU} 
\end{align}%
where $T_{\mathrm{a}}$ is the transfer coefficient and \mbox{$
E_{\mathrm{a}}(t)
$} is the random radiation (noise) added by the amplifier. This noise does not depend on \mbox{$
E_0(t)
$}. If \mbox{$
T_{\mathrm{a}}\leq 1
$} (attenuator), the noise may be neglected without getting into contradiction with quantum mechanics. (This does not mean that for \mbox{$
T_{\mathrm{a}}\leq 1
$} the noise is always negligible: e.g., active medium without inversion attenuates the signal {\em and\/} adds spontaneous noise.) If \mbox{$
T_{\mathrm{a}}>1
$} (amplifier proper), there is a well known limit \cite{Caves} to how small the noise may be (discussed in Sec.\ \ref{ch:TL} below). This limit cannot be found classically, but there is no problem with including it into the classical Eq.\ (\ref{eq:25DU}). This is the reason we call the model (\ref{eq:25DU}) not classical but semiclassical. It operates with classical notions while making provisions for later ``promotion'' to quantum mechanics. By itself, Eq.\ (\ref{eq:25DU}) is fully classical, and may be handled without any reference to quantum mechanics. 

For $E_{\mathrm{a}}(t)$, we assume Gaussian statistics specified by the averages, 
\begin{align} 
\begin{aligned} 
 &%
\begin{aligned} 
 &\overline{\hspace{0.1ex}E_{\mathrm{a}}(t)\hspace{0.1ex}} = 0, &
 &\overline{\hspace{0.1ex}E_{\mathrm{a}}(t)E_{\mathrm{a}}(t')\hspace{0.1ex}} = 0, 
\end{aligned} \\ 
 &\overline{\hspace{0.1ex}E^*_{\mathrm{a}}(t)E_{\mathrm{a}}(t')\hspace{0.1ex}} 
= 2\pi I_{\mathrm{a}}\mathrm{e}^{-(\gamma_{\mathrm{a}}/2)|t-t'|} 
. 
\end{aligned} 
\label{eq:30DZ} 
\end{align}%
We treat $I_{\mathrm{a}}$, $\gamma_{\mathrm{a}}$ and $T_{\mathrm{a}}$ as phenomenological constants, available, e.g., from experiments with the amplifier. For consistency we have to assume that \mbox{$
\gamma_{\mathrm{a}}
$} is large compared to the characteristic spectral width of the amplified signal (otherwise one cannot regard \mbox{$
T_{\mathrm{a}}
$} as frequency-independent). That is, for our purposes, 
\begin{align} 
\begin{aligned} 
 &\overline{\hspace{0.1ex}E^*_{\mathrm{a}}(t)E_{\mathrm{a}}(t')\hspace{0.1ex}} 
= 2\pi I_{\mathrm{a}\omega }\delta(t-t') . 
\end{aligned} 
\label{eq:32EB} 
\end{align}%
where 
\begin{align} 
\begin{aligned} 
 &I_{\mathrm{a}\omega } = \frac{4I_{\mathrm{a}}}{\gamma_{\mathrm{a}}}, &
 &\omega \ll \gamma_{\mathrm{a}} . 
\end{aligned} 
\label{eq:33EC} 
\end{align}%
is the spectral density of the noise. Of use will also be the average, (recall that the noise is assumed Gaussian)
\begin{align} 
\begin{aligned} 
\overline{\hspace{0.1ex}|E_{\mathrm{a}}(t)|^2 |E_{\mathrm{a}}(t')|^2\hspace{0.1ex}} = 
\overline{\hspace{0.1ex}|E_{\mathrm{a}}(t)|^2\hspace{0.1ex}}\ \overline{\hspace{0.1ex}|E_{\mathrm{a}}(t')|^2\hspace{0.1ex}}
+ \big|\overline{\hspace{0.1ex}E^*_{\mathrm{a}}(t)E_{\mathrm{a}}(t')\hspace{0.1ex}}\big|^2 
\\ 
= 4\pi ^2 I_{\mathrm{a}}^2\big(
1+\mathrm{e}^{-\gamma_{\mathrm{a}}|t-t'|}
 \big) \sim 4\pi ^2 I_{\mathrm{a}}^2\big [ 
1+2\gamma_{\mathrm{a}}^{-1}\delta(t-t')
 \big ] . 
\end{aligned} 
\label{eq:53EZ} 
\end{align}%

\subsubsection{Classical photodetector with shot noise}\label{ch:TCD}
In the conventional semiclassical theory of ideal photodetection \cite{MandelWolf}, the inherently quantum nature of detection process is accounted for phenomenologically, postulating that photocurrent is a sequence of infinitesimally short photodetection pulses. The probability for a photodetection pulse to arrive within a sufficiently small time interval 
$t, t+\Delta t$ equals 
\begin{align} 
\begin{aligned} 
\bar p(t,t+\Delta t) = \chi | E_1(t) | ^2\,\Delta t .
\end{aligned} 
\label{eq:DetP1} 
\end{align}%
Here, 
\begin{align} 
\begin{aligned} 
\chi = \frac{\eta S}{2\pi \hbar \omega_0}, 
\end{aligned} 
\label{eq:53QT} 
\end{align}%
where $\eta $ is the detector efficiency and $S$ is the beam area (assumed smaller than the detection surface). Planck's constant in (\ref{eq:53QT}) is all that survives of the quantum nature of the detector. If $\bar p(t,t+\Delta t)$ is not small but $E_1(t)$ may still be regarded constant on the scale of $\Delta t$, photodetection pulses within $\Delta t$ obey Poissonian statistics. The bottom line is, all photodetection events (pulses) are independent of each other, and all correlations in the pulse sequence may only be due to correlations in the detected beam. 
\subsubsection{Photocurrent statistics}\label{ch:TCS}
Based on these assumptions we can calculate all statistical averages of the photocurrent. For averages conditional on \mbox{$
E_1(t)
$} we have, 
\begin{align} 
\begin{aligned} 
q^{-1}\,\overline{\hspace{0.1ex} J(t)\hspace{0.1ex}}\settoheight{\auxlv}{$|$}%
\raisebox{-0.3\auxlv}{$|_{[E_1,E_1^*]}$} &= \chi |E_1(t)|^2 , \\ 
q^{-2}\,\overline{\hspace{0.1ex} J(t) J(t')\hspace{0.1ex}}\settoheight{\auxlv}{$|$}%
\raisebox{-0.3\auxlv}{$|_{[E_1,E_1^*]}$} &
= \chi |E_1(t)|^2\delta(t-t') 
\\ &\quad\qquad + 
\chi^2 |E_1(t)|^2|E_1(t')|^2 , 
\end{aligned} 
\label{eq:27DW} 
\end{align}%
etc., where $q$ is the charge in a pulse. Using Eq.\ (\ref{eq:25DU}) and averaging over \mbox{$
E_0(t)
$} and \mbox{$
E_{\mathrm{a}}(t)
$} we find the unconditional (observed) photocurrent averages, 
\begin{align} 
\begin{aligned} 
 &q^{-1}\,\overline{\hspace{0.1ex} J(t)\hspace{0.1ex}} = \eta \big [ 
T_{\mathrm{a}}n_0 + n_{\mathrm{a}}
 \big ] , \\ 
 &q^{-2}\,\overline{\hspace{0.1ex} J(t) J(t')\hspace{0.1ex}} 
= \eta \big(
T_{\mathrm{a}}n_0 + n_{\mathrm{a}}
 \big)\delta(t-t') 
\\ &\quad
+ 
\eta^2 \Big \{ 
T_{\mathrm{a}}^2 n^2_0(t-t') 
+ 2T_{\mathrm{a}}n_0 n_{\mathrm{a}} 
\big [ 
1+4\gamma_{\mathrm{a}}^{-1}\delta(t-t')
 \big ] 
\\ &\quad
+ n_{\mathrm{a}}^2 
\big [ 
1+2\gamma_{\mathrm{a}}^{-1}\delta(t-t')
 \big ] 
 \Big \} 
, 
\end{aligned} 
\label{eq:31EA} 
\end{align}%
where we have introduced the notation, 
\begin{align} 
\begin{aligned} 
 &%
\begin{aligned} 
 &n_0 = \frac{S\overline{\hspace{0.1ex}|E_0(t)|^2\hspace{0.1ex}}}{2\pi \hbar \omega _0}, &
 &n_{\mathrm{a}} = \frac{S\overline{\hspace{0.1ex}|E_a(t)|^2\hspace{0.1ex}}}{2\pi \hbar \omega _0}, 
\end{aligned} \\ 
 &n^2_0(t-t') = \frac{S^2\overline{\hspace{0.1ex}|E_0(t)|^2|E_0(t')|^2\hspace{0.1ex}}}{4\pi^2 \hbar^2 \omega _0^2} . 
\end{aligned} 
\label{eq:34ED} 
\end{align}%
For simplicity we assumed that \mbox{$
E_0(t)
$} is a stationary random process. 

From a quantum perspective, $n_0$ and $n_{\mathrm{a}}$ are the initial and added-noise photon fluxes and \mbox{$
n^2_0(t-t')
$} are autocorrelations in the former. Strictly speaking, at this stage we are not allowed to use this terminology, but we forfeit full rigour in favour of clarity. Another incidence of ``premature quantum awareness'' is separation of $\eta$ from $\hbar $ in Eqs.\ (\ref{eq:31EA}), (\ref{eq:34ED}). The parameter characterising the detector semiclassically is their ratio $\hbar /\eta$. 

\subsubsection{Full classical solution}\label{ch:TCF}
General formulae for photocurrent averages are rather tangled, but a solution in terms of characteristic functionals is constructed with ease. Using the well-known characteristic function of the Poissonian process \cite{ProbText}, for the functional \mbox{$
\Phi_{\mathrm{d}}{\big( 
\zeta \big| E_1,E_1^* \big)}
$}, replacing (\ref{eq:99CS}) under the RWA for the detected field, we obtain, 
\begin{align} 
\begin{aligned} 
\Phi_{\mathrm{d}}{\big( 
\zeta \big| E_1,E_1^* \big)} = \exp
\int dt\, \chi |E_1(t)|^2 \big \{ 
\exp [ 
q\zeta(t)
 ] - 1
 \big \} , 
\end{aligned} 
\label{eq:54QU} 
\end{align}%
It is equally straightforward to obtain photocurrent averages conditional on \mbox{$
E_0(t)
$}. Formally, this corresponds to regarding the amplifier and detector as a composite detector (cf.\ Fig.\ \ref{fig:CascadeF}b). A generating functional of such averages follows by making substution (\ref{eq:25DU}) in (\ref{eq:54QU}) and averaging over \mbox{$
E_{\mathrm{a}}(t)
$}, namely, 
\begin{multline} 
\hspace{0.4\columnwidth}\hspace{-0.4\twocolumnwidth}
\Phi_{\mathrm{cd}}{\big( 
\zeta \big| E_0,E_0^* \big)} \\ 
= \overline{\hspace{0.1ex}\exp
\int dt\, \chi \big|\sqrt{T_{\mathrm{a}}}E_0(t)+E_{\mathrm{a}}(t)\big|^2 \big \{ 
\exp [ 
q\zeta(t)
 ] - 1
 \big \} 
\hspace{0.1ex}}^{\,E_{\mathrm{a}}} . 
\hspace{0.4\columnwidth}\hspace{-0.4\twocolumnwidth}%
\label{eq:28DX} 
\end{multline}%
Observed photocurent statistics follows by averaging over the statistics of radiation of the source, 
\begin{align} 
\begin{aligned} 
\Phi_{\mathrm{o}}\big( 
\zeta 
 \big) 
= \overline{\hspace{0.1ex}\Phi_{\mathrm{cd}}{\big( 
\zeta \big| E_0,E_0^* \big)}\hspace{0.1ex}}^{\,E_0} . 
\end{aligned} 
\label{eq:29DY} 
\end{align}%
\subsubsection{``Quantisation'' of the problem}\label{ch:TQS}
``Upgrade'' of the semiclassical theory to quantum fields is trivial. The composite detector (cf.\ Fig.\ \ref{fig:CascadeF}b) is a classical device, and the functional (\ref{eq:28DX}) may be ``imported'' to quantum electrodynamics\ as is. Proof of this statement, which takes quantum formulation of the arrangement in Fig.\ \ref{fig:CascadeF}b, may be found in appendix \ref{ch:A} (see also Sec.\ \ref{ch:TCV}). 
All we need is to redefine averages (\ref{eq:29DY}), namely, 
\begin{align} 
\begin{aligned} 
\Phi_{\mathrm{o}}\big( 
\zeta 
 \big) 
= \big \langle {\mathcal T}{\mbox{\rm\boldmath$:$}}\Phi_{\mathrm{cd}}{\big( 
\zeta \big| {\hat{\mathcal E}}_0,{\hat{\mathcal E}}_0^{\dag}\big)}{\mbox{\rm\boldmath$:$}} \big \rangle_{\mathrm{s}} . 
\end{aligned} 
\label{eq:35EE} 
\end{align}%
The Heisenberg field operator \mbox{$
{\hat{\mathcal E}}_0(t)
$} is defined for the solitary source interacting with a narrow-band\ field. Hence, for purposes of this discussion, it suffices to assume that {\em there exists\/} a quantum formulation of the source, assigning meaning to averages (\ref{eq:96HW}) and thus to Eq.\ (\ref{eq:35EE}). For formal particulars see appendix \ref{ch:A}. 

For Eqs.\ (\ref{eq:31EA}), ``quantisation of the source'' reduces to redefinition of quantitites (\ref{eq:34ED}): 
\begin{align} 
\begin{aligned} 
 &n_0 = \frac{S\big \langle {\hat{\mathcal E}}_0^{\dag}(t){\hat{\mathcal E}}_0(t) \big \rangle_{\mathrm{s}}}{2\pi \hbar \omega _0}, \\ 
 &n^2_0(t-t') = \frac{S^2\big \langle {\mathcal T}{\mbox{\rm\boldmath$:$}}{\hat{\mathcal E}}_0^{\dag}(t){\hat{\mathcal E}}_0^{\dag}(t'){\hat{\mathcal E}}_0(t'){\hat{\mathcal E}}_0(t){\mbox{\rm\boldmath$:$}} \big \rangle_{\mathrm{s}}}{4\pi^2 \hbar^2 \omega _0^2} . 
\end{aligned} 
\label{eq:37EH} 
\end{align}%
We omit the symbol of time-normal ordering where it is redundant. If necessary, averages of \mbox{$
E_{\mathrm{a}}(t)
$} may also be redefined, e.g., 
\begin{align} 
\begin{aligned} 
 &n_{\mathrm{a}} = \frac{S\big [ 
\big \langle {\hat{\mathcal E}}_1^{\dag}(t){\hat{\mathcal E}}_1(t) \big \rangle_{\mathrm{a}}
 \big ]\settoheight{\auxlv}{$|$}%
\raisebox{-0.3\auxlv}{$|_{E_0=0}$} 
}{2\pi \hbar \omega _0}. 
\end{aligned} 
\label{eq:38EJ} 
\end{align}%
The average here is for a solitary amplifier without the input signal; as a formal quantity it is defined in appendix \ref{ch:A}. While aesthetically satisfying, such redefinition is utterly unnecessary (so far as the self-noise of the amplifier is in a classical state). 

Alternatively, we can find photocurrent as dependent on radiation of the composite source (cf.\ Fig.\ \ref{fig:CascadeF}b), 
\begin{multline} 
\hspace{0.4\columnwidth}\hspace{-0.4\twocolumnwidth}
\Phi_{\mathrm{o}}\big( 
\zeta 
 \big) \\ 
= \bigg \langle 
\mathcal{T}{\mbox{\rm\boldmath$:$}}
\exp
\int dt\, \chi{\hat{\mathcal E}}_1^{\dag}(t){\hat{\mathcal E}}_1(t) \big \{ 
\exp [ 
q\zeta(t)
 ] - 1
 \big \}
{\mbox{\rm\boldmath$:$}} 
 \bigg \rangle_{\mathrm{cs}} . 
\hspace{0.4\columnwidth}\hspace{-0.4\twocolumnwidth}%
\label{eq:55QV} 
\end{multline}%
The symbol \mbox{$
 \langle \cdots \rangle_{\mathrm{cs}}
$} is defined regarding the source and detector and the electromagnetic\ interaction between them as a composite source (cf.\ Fig.\ \ref{fig:CascadeF}b). For formal particulars see appendix \ref{ch:A}. 
From Eq.\ (\ref{eq:55QV}) relation we find photocurrent statistics in terms of radiation of the composite source, 
\begin{align} 
\begin{aligned} 
q^{-1}\,\overline{\hspace{0.1ex} J(t)\hspace{0.1ex}} &= \chi \big \langle 
{\hat{\mathcal E}}_1^{\dag}(t){\hat{\mathcal E}}_1(t)
 \big \rangle_{\mathrm{cs}} , \\ 
q^{-2}\,\overline{\hspace{0.1ex} J(t) J(t')\hspace{0.1ex}} &
= \chi \big \langle 
{\hat{\mathcal E}}_1^{\dag}(t){\hat{\mathcal E}}_1(t)
 \big \rangle_{\mathrm{cs}}\delta(t-t') 
\\ &\quad
+ 
\chi^2 \big \langle 
\mathcal{T}{\mbox{\rm\boldmath$:$}}
{\hat{\mathcal E}}_1^{\dag}(t){\hat{\mathcal E}}_1^{\dag}(t'){\hat{\mathcal E}}_1(t'){\hat{\mathcal E}}_1(t)
{\mbox{\rm\boldmath$:$}} 
 \big \rangle_{\mathrm{cs}} . 
\end{aligned} 
\label{eq:1D} 
\end{align}%
As expected, Eqs.\ (\ref{eq:1D}) are standard Glauber-Kelley-Kleiner's photodetection formulae \cite{MandelWolf}. 
\subsubsection{Limits on spontaneous noise}\label{ch:TL}
Quantum mechanics imposes a lower limit on the spectral density of spontaneous noise, 
\begin{align} 
\begin{aligned} 
\frac{4n_{\mathrm{a}}}{\gamma_{\mathrm{a}}}\geq \frac{1}{2} 
\big [ \big(
T_{\mathrm{a}}-1
 \big) + 
\big|
T_{\mathrm{a}}-1
\big|
 \big ] = 
\begin{cases}
T_{\mathrm{a}}-1, & T_{\mathrm{a}}>1, \\
0, & T_{\mathrm{a}}\leq 1. 
\end{cases} . 
\end{aligned} 
\label{eq:48EU} 
\end{align}%
For amplifiers based on stimulated emission, this is a consequence of the connection between the Einstein coeffcients, cf.\ eq.\ (44.9) in Landau and Lifshitz's textbook \cite{LLIV}. As a general quantum relation, this formula was derived, e.g., by Caves (specifically, this is eq.\ (4.21) of \mbox{Ref.\ \cite{Caves}} rewritten in our notation, cf.\ endnote \cite{endCaves}). 

To see how condition (\ref{eq:48EU}) serves to keep the semiclassical theory consistent, consider the spectrum of fluctuations of the photocurrent, 
\begin{align} 
\begin{aligned} 
\overline{\hspace{0.1ex}J^2_{\omega }\hspace{0.1ex}} 
= \int d\tau\, \mathrm{e}^{i\omega \tau }\, \overline{\hspace{0.1ex} J(t) J(t+\tau )\hspace{0.1ex}} . 
\end{aligned} 
\label{eq:39EK} 
\end{align}%
Using Eq.\ (\ref{eq:31EA}) and ignoring the zero-frequency (constant) contribution we obtain, 
\begin{multline} 
\hspace{0.4\columnwidth}\hspace{-0.4\twocolumnwidth}
q^{-2}\overline{\hspace{0.1ex}J^2_{\omega }\hspace{0.1ex}} = \eta \big(
T_{\mathrm{a}}n_0 + n_{\mathrm{a}}
 \big) \\ 
+ 
\eta^2 \big(
T_{\mathrm{a}}^2 n^2_{0\omega } 
+ 8T_{\mathrm{a}}n_0 n_{\mathrm{a}} 
\gamma_{\mathrm{a}}^{-1}
+ 2n_{\mathrm{a}}^2 
\gamma_{\mathrm{a}}^{-1}
 \big), \ \ \omega >0, 
\hspace{0.4\columnwidth}\hspace{-0.4\twocolumnwidth}%
\label{eq:40EL} 
\end{multline}%
where 
\begin{multline} 
\hspace{0.4\columnwidth}\hspace{-0.4\twocolumnwidth}
n^2_{0\omega } = \int d\tau\, \mathrm{e}^{i\omega \tau }n^2_0(\tau ) \\ 
= \frac{S^2}{4\pi^2 \hbar^2 \omega _0^2}\int d\tau\, \mathrm{e}^{i\omega \tau }
\big \langle {\mathcal T}{\mbox{\rm\boldmath$:$}}{\hat{\mathcal E}}_0^{\dag}(t){\hat{\mathcal E}}_0^{\dag}(t'){\hat{\mathcal E}}_0(t'){\hat{\mathcal E}}_0(t){\mbox{\rm\boldmath$:$}} \big \rangle_{\mathrm{s}} . 
\hspace{0.4\columnwidth}\hspace{-0.4\twocolumnwidth}%
\label{eq:41EM} 
\end{multline}%
Unlike in classical mechanics, in quantum mechanics \mbox{$
n^2_{0\omega }
$} is not bound to be positive. There is an obvious limit to how negative this quantity may be. Without the amplifier, photocurrent spectrum reads, 
\begin{align} 
\begin{aligned} 
 &q^{-2}\overline{\hspace{0.1ex}J^2_{\omega }\hspace{0.1ex}} = \eta n_0 + 
\eta^2 n^2_{0\omega } . 
\end{aligned} 
\label{eq:43EP} 
\end{align}%
Photocurrent is classical, so that its spectrum should be positive. Assuming an ideal photodetector, \mbox{$
\eta=1
$}, we find the lower limit on \mbox{$
n^2_{0\omega }
$}, 
\begin{align} 
\begin{aligned} 
n^2_{0\omega }\geq -n_0 . 
\end{aligned} 
\label{eq:44EQ} 
\end{align}%
Furthermore, assume that at $\omega =\tilde\omega $ this limit is actually reached, 
\begin{align} 
\begin{aligned} 
n^2_{0\tilde\omega }= -n_0 . 
\end{aligned} 
\label{eq:45ER} 
\end{align}%
At this frequency, Eq.\ (\ref{eq:40EL}) yields,
\begin{multline} 
\hspace{0.4\columnwidth}\hspace{-0.4\twocolumnwidth}
q^{-2}\overline{\hspace{0.1ex}J^2_{\tilde\omega }\hspace{0.1ex}} = \eta T_{\mathrm{a}}n_0 
\big(
1 - \eta T_{\mathrm{a}} + 8\eta n_{\mathrm{a}} 
\gamma_{\mathrm{a}}^{-1} 
 \big) \\ 
+ 
\eta n_{\mathrm{a}}\big(
1+
2\eta n_{\mathrm{a}} 
\gamma_{\mathrm{a}}^{-1} 
 \big) 
. 
\hspace{0.4\columnwidth}\hspace{-0.4\twocolumnwidth}%
\label{eq:46ES} 
\end{multline}%
Having this quantity positive for arbitrary $n_0$ \cite{endLin} and \mbox{$
\eta\leq 1
$} imposes a lower limit on the spectral density of spontaneous noise, 
\begin{align} 
\begin{aligned} 
\frac{4n_{\mathrm{a}}}{\gamma_{\mathrm{a}}}\geq \frac{1}{2} 
\big(
T_{\mathrm{a}}-1
 \big) . 
\end{aligned} 
\label{eq:47ET} 
\end{align}%
This condition is weaker than (\ref{eq:48EU}), so that amplification is bound to degrade quantum properties of the signal. 

\section{Conclusion}
So what is it that we wish to say and that, we believe, has not been fully appreciated for more than 40 years? Conventional wisdom is that relation like Eq.\ (\ref{eq:54QU}) are limited to classical input fields. This is indeed the case if we regard it as characterisation of the {\em photocurrent\/}. However, regarded as characterisation of the {\em photodetector\/}, Eq.\ (\ref{eq:54QU}) is a wholesome quantum formula. Moreover, it is a {\em complete\/} quantum formula: owing to Sudarshan's optical equivalence theorem, there is no freedom in generalising semiclassical photodetection theory to quantum fields. It is this {\em absence of freedom\/} that we wish to stress. It applies to any device: if we know how it reacts to classical fields, we know how it will react to quantum fields. In principle, it applies also to quantum devices, but its most dramatic consequences are for classical devices. Sudarshan's optical equivalence theorem\ turns their semiclassical models into quantum approaches. 

In conclusion, quantum theory of the electromagnetic\ interaction under macroscopic conditions of distinguishability of devices and of controlled actions and back-actions between them is constructed. This theory is subject to ``doing quantum electrodynamics\ while thinking classically,'' which allows one to substitute essentally classical considerations for quantum ones without any loss in generality. 

\appendix
\section{Summary of formal definitions}%
\label{ch:SS}
\subsection{Time orderings of operators}%
\label{ch:DO}
By definition, bosonic operators commute under all kinds of ordering. The standard time-ordering, denoted $T_+$, puts operators in the order of decreasing time arguments, 
\begin{multline} 
\hspace{0.4\columnwidth}\hspace{-0.4\twocolumnwidth}
T_+{\hat{\mathcal X}}_1(t_1){\hat{\mathcal X}}_2(t_2) = \theta(t_1-t_2){\hat{\mathcal X}}_1(t_1){\hat{\mathcal X}}_2(t_2) 
+\big \{ 
1\leftrightarrow 2
 \big \} , 
\hspace{0.4\columnwidth}\hspace{-0.4\twocolumnwidth}%
\label{eq:10TD} 
\end{multline}%
and similarly for more operators. The ``reverse'' time ordering, denoted $T_-$, may be introduced by the relation, 
\begin{multline} 
\hspace{0.4\columnwidth}\hspace{-0.4\twocolumnwidth}
{[T_{+}{\hat{\mathcal X}}_1(t_1){\hat{\mathcal X}}_2(t_2)\cdots{\hat{\mathcal X}}_m(t_m)]}^{\dag}\\ 
= T_{-}{\hat{\mathcal X}}_1^{\dag}(t_1){\hat{\mathcal X}}_2^{\dag}(t_2)\cdots{\hat{\mathcal X}}_m^{\dag}(t_m). 
\hspace{0.4\columnwidth}\hspace{-0.4\twocolumnwidth}%
\label{eq:94HX} 
\end{multline}%
The closed-time-loop, or C-contour, ordering \cite{SchwingerC,Perel,Keldysh} is a way of writing the {\em double time ordered\/} operator structures, 
\begin{multline} 
\hspace{0.4\columnwidth}\hspace{-0.4\twocolumnwidth}
T_C{\hat{\mathcal X}}_{1-}(t_1)\cdots{\hat{\mathcal X}}_{m-}(t_m) 
{\hat{\mathcal Y}}_{1+}(t'_1)\cdots{\hat{\mathcal Y}}_{n+}(t'_n) 
\\ 
= 
T_-{\hat{\mathcal X}}_1(t_1)\cdots{\hat{\mathcal X}}_m(t_m) \, 
T_+{\hat{\mathcal Y}}_1(t'_1)\cdots{\hat{\mathcal Y}}_n(t'_n) 
. 
\hspace{0.4\columnwidth}\hspace{-0.4\twocolumnwidth}%
\label{eq:10BU} 
\end{multline}%
The ${}_{\pm}$ indices in (\ref{eq:10BU}) serve only for ordering purposes and otherwise should be disregarded. We note in passing that definitions like (\ref{eq:10TD}), (\ref{eq:10BU}) may cause mathematical problems, see the concluding remark in 
appendix A1 in \cite{WickCaus}. 

\subsection{The frequency-positive and negative\ parts}%
\label{ch:DF}
The symbols ${}^{(\pm)}$ denote separation of the frequency-positive and negative\ parts of functions, 
\begin{align} 
\begin{aligned} 
 &f(t) = f^{(+)}(t) + f^{(-)}(t) , 
\\ 
 &f^{(\pm)}(t) = \int_{-\infty}^{+\infty}\frac{d\omega }{2\pi }\mathrm{e}^{-i\omega t}\theta(\pm\omega )
f_{\omega }, &
 &f_{\omega } = \int_{-\infty}^{+\infty}dt \mathrm{e}^{i\omega t}f(t) . 
\end{aligned} 
\label{eq:4JH} 
\end{align}%
This operation is alternatively expressed as 
an integral transformation, 
\begin{align} 
\begin{aligned} 
f ^{(\pm)}(t) = \int dt' \delta ^{(\pm)}(t-t') f (t'), 
\end{aligned} 
\label{eq:39KV} 
\end{align}%
where 
\begin{align} 
\begin{aligned} 
\delta ^{(\pm)}(t) = \delta ^{(\mp)}(-t) = \big [ 
\delta ^{(\mp)}(t)
 \big ] ^* = \pm\frac{1}{2\pi i(t\mp i0^+)} 
\end{aligned} 
\label{eq:40KW} 
\end{align}%
are the frequency-positive and negative\ parts of the delta-function. 
For more details on this operation see \mbox{Ref.\ \cite{APII}}, appendix A. 
\subsection{Time-normal ordering}%
\label{ch:DA}
Definition of the time-normal operator ordering, denoted \mbox{$
{\mathcal T}{\mbox{\rm\boldmath$:$}}\cdots{\mbox{\rm\boldmath$:$}}
$}, is different for slow amplitudes, such as ${\hat{\mathcal E}}(x,t)$ and ${\hat{\mathcal D}}(x,t)$, and for Hermitian broad-band\ fields, such as ${\hat{\mathcal A}}(x,t)$ and ${\hat{\mathcal J}}(x,t)$ (for purposes of this discussion, their physical nature is irrelevant). To the former, one applies the conventional definition of Kelley and Kleiner \cite{KelleyKleiner,GlauberTN}, while for the latter, the amended definition of \mbox{Refs.\ \cite{APII,APIII}} must be used. Most cases of interest in the paper are covered postulating the operator-valued characteristic functional, 
\begin{multline} 
\hspace{0.4\columnwidth}\hspace{-0.4\twocolumnwidth}
{\mathcal T}{\mbox{\rm\boldmath$:$}}\exp\big(
i\eta{\hat{\mathcal A}}+i\zeta{\hat{\mathcal J}}+i\mu^*{\hat{\mathcal E}}-i\mu{\hat{\mathcal E}}^{\dag}
+i\nu^* {\hat{\mathcal D}}-i\nu{\hat{\mathcal D}}^{\dag}
 \big) 
{\mbox{\rm\boldmath$:$}} \\ 
\begin{aligned} 
 &= 
T_C \exp\big [ 
i\eta ^{(-)}{\hat{\mathcal A}}_+
+i\eta ^{(+)}{\hat{\mathcal A}}_-
+i\zeta ^{(-)}{\hat{\mathcal J}}_+
+i\zeta ^{(+)}{\hat{\mathcal J}}_-
\\ &\quad\qquad\qquad\qquad 
+i\mu^* {\hat{\mathcal E}}_+
-i\mu {\hat{\mathcal E}}^{\dag}_-
+i\nu^* {\hat{\mathcal D}}_+
-i\nu {\hat{\mathcal D}}^{\dag}_-
 \big ] 
 \\ 
 &= 
T_- \exp\big [ 
i\eta ^{(+)}{\hat{\mathcal A}}
+i\zeta ^{(+)}{\hat{\mathcal J}}
-i\mu {\hat{\mathcal E}}^{\dag}
-i\nu {\hat{\mathcal D}}^{\dag}
 \big ] 
\end{aligned}
\\ \times
T_+ \exp\big [ 
i\eta ^{(-)}{\hat{\mathcal A}}
+i\zeta ^{(-)}{\hat{\mathcal J}}
+i\mu^* {\hat{\mathcal E}}
+i\nu^* {\hat{\mathcal D}}
 \big ] 
 , 
\hspace{0.4\columnwidth}\hspace{-0.4\twocolumnwidth}%
\label{eq:17ZC} 
\end{multline}%
where 
\mbox{$
\eta (x,t)
$}, 
\mbox{$
\zeta (x,t)
$}, 
\mbox{$
\mu (x,t)
$}, 
\mbox{$
\nu (x,t)
$} are auxiliary c-number functions. We use here notation (\ref{eq:3VS}); separation of the frequency-positive and negative\ parts was defined in Sec.\ \ref{ch:DF}. Similar definitions apply to free operators; for \mbox{$
E(x,t)
$} and \mbox{$
A(x,t)
$} defined by (\ref{eq:4SX}) the time-normal ordering reduces to the standard normal ordering. For an in-depth discusion see \mbox{Ref.\ \cite{PFunc}}. 
Causality properties of the time-normal ordering, which are of crucial importance for physical consistency of our analyses, were the subject of \mbox{Ref.\ \cite{RelCaus}}. 
\section{The general case of interaction of two distinguishable devices}%
\label{ch:D}
\subsection{The model}%
\label{ch:DM}
In the general case, a subset of oscillator modes (\ref{eq:87FC}) is made subject to the rotating wave approximation\ (RWA). All oscillators are organised in two quantised fields: the {\em narrow-band\/}, or {\em resonant\/}, field $\hat E(x,t)$, and the {\em broad-band\/}, or {\em nonresonant\/}, field $\hat A(x,t)$, ($1<M<N$)
\begin{align} 
\begin{aligned} 
 &\hat E(x,t) = i\sum_{\kappa =1}^{M} 
\sqrt{\frac{\hbar\omega_{\kappa}}{2}} u_{\kappa}(x) \hat a_{\kappa} \mathrm{e}^{-i(\omega_{\kappa}-\omega _0)t}, 
\\ 
 &\hat E^{\dag}(x,t) = -i\sum_{\kappa =1}^{M} 
\sqrt{\frac{\hbar\omega_{\kappa}}{2}} u^*_{\kappa}(x) \hat a_{\kappa}^{\dag}\mathrm{e}^{i(\omega_{\kappa}-\omega _0)t}, 
\\ 
 &\hat A(x,t) = \sum_{\kappa =M+1}^{N}
\sqrt{\frac{\hbar}{2\omega_{\kappa}}}
 u_{\kappa}(x) \hat a_{\kappa}\mathrm{e}^{-i\omega_{\kappa}t} + \mathrm{H.c.}\, . 
\end{aligned} 
\label{eq:4SX} 
\end{align}%
Frequencies $\omega _{\kappa }$, \mbox{$
1\leq \kappa \leq M
$}, are supposed to occupy a narrow band centered at $\omega _0$, so that \mbox{$
\hat E(x,t)
$} is by definition a slow amplitide. This assumption only matters for physics; formally, it may be disregarded. 

The general case is equally governed by the generic Hamiltonian (\ref{eq:86FB}), where the electromagnetic interaction is now split into the narrow-band\ (RWA) and the broad-band\ (no-RWA) parts, 
\begin{multline} 
\hspace{0.4\columnwidth}\hspace{-0.4\twocolumnwidth}
\hat H_{\text{I}}(t) = 
- 
\int dx \Big(
\big [ 
\hat A(x,t)+A_{\mathrm{e}}(x,t)
 \big ]\hat J(x,t) \\ 
+\Big \{ 
\big [ 
\hat E(x,t) + E_{\mathrm{e}}(x,t) 
 \big ]\hat D^{\dag}(x,t)+\mathrm{H.c.}
 \Big \} \Big) 
. 
\hspace{0.4\columnwidth}\hspace{-0.4\twocolumnwidth}%
\label{eq:83AR} 
\end{multline}%
The Hamiltonian $\hat H_{\mathrm{dev}}(t)$ in (\ref{eq:86FB}), the dipole momentum $\hat D(x,t)$ and the current operator $\hat J(x,t)$ describe the device. They commute with all \mbox{$
\hat a_{\kappa },\hat a_{\kappa }^{\dag}
$} and otherwise remain arbitrary. The c-number external sources $E_{\mathrm{e}}(x,t)$ and $ A_{\mathrm{e}}(x,t)$ are added for formal purposes. For a discussion of this model see our \mbox{Ref.\ \cite{QDynResp}}, sections 
II and III. Compared to \mbox{Refs.\ \cite{QDynResp,PFunc}}, interaction (\ref{eq:83AR}) lacks the c-number dipole and current which are of no use in this paper. 

The broad-band\ field enters the theory through the retarded Green (response) function (\ref{eq:83LF}), while the narrow-band\ one --- through another response function, 
\begin{align} 
\begin{aligned} 
 &\Delta_{\text{R}}(x,x',t-t') = \frac{i}{\hbar}\theta(t-t')
\big [ 
\hat E(x,t),\hat E^{\dag}(x',t')
 \big ] 
. 
\end{aligned} 
\label{eq:85LJ} 
\end{align}%

The operators \mbox{$\hat E(x,t)
$}, \mbox{$
\hat A(x,t)
$}, \mbox{$
\hat D(x,t)
$}, and \mbox{$
\hat J(x,t)
$} are by definition the interaction-picture (free) ones. Their {Heisenberg}\ counterparts will be dehoted as, respectively, \mbox{${\hat{\mathcal E}}(x,t)
$}, \mbox{$
{\hat{\mathcal A}}(x,t)
$}, \mbox{$
{\hat{\mathcal D}}(x,t)
$}, and \mbox{$
{\hat{\mathcal J}}(x,t)
$}. 
\subsection{Quantum electrodynamics of a solitary device revisited}%
\label{ch:G}
Extension of formulae of Sec.\ \ref{ch:S} to the general case reduces to a large extent to triplicating all variables. In place of definitions (\ref{eq:96KJ}), (\ref{eq:97KK}) we have, 
\begin{widetext} 
\begin{multline} 
\Phi^{\mathrm{I}}_{\mathrm{dev}}\big(
\zeta,\nu ,\nu ^* 
\big | 
A_{\mathrm{e}},E_{\mathrm{e}},E_{\mathrm{e}}^* 
 \big) 
= \text{Tr}\hat\rho_{\mathrm{dev}}
{\mathcal T}{\mbox{\rm\boldmath$:$}}\exp\big(
i{\zeta} \hat J' 
+i\bar\nu^* \hat D' -i\nu \hat D^{\prime\dag}
 \big) 
{\mbox{\rm\boldmath$:$}} 
\\ 
= \prod_{x,t}\bigg\{\int dJ(x,t)d^2D(x,t)\bigg\}
p^{\mathrm{I}}\big(
J,D,D^* \big| A_{\mathrm{e}},E_{\mathrm{e}},E_{\mathrm{e}}^*
 \big) 
\exp\big(
i\zeta J+i\nu ^*D-i\nu D^*
 \big)
, 
\label{eq:79NX} 
\end{multline}%
where the time-normal ordering is defined by Eq.\ (\ref{eq:17ZC}), and the primed operators are defined as {Heisenberg}\ ones with respect to the Hamiltonian, 
\begin{align} 
\begin{aligned} 
\hat H(t) = \hat H_{\textrm{dev}}(t) - \int dx \big [ 
A_{\textrm{e}}(x,t)\hat J(x,t)
+E^*_{\textrm{e}}(x,t)\hat D(x,t)
+E_{\textrm{e}}(x,t)\hat D^{\dag}(x,t)
 \big ] . 
\end{aligned} 
\label{eq:63HX} 
\end{align}%
This is Hamiltonian (\ref{eq:86FB}) with field operators set to zero. In place of (\ref{eq:53HM}) we find, 
\begin{multline} 
\Phi_{\mathrm{dev}}\big(
\zeta,\nu ,\nu ^* 
\big | 
A_{\mathrm{e}},E_{\mathrm{e}},E_{\mathrm{e}}^* 
 \big) 
= 
\Big \langle {\mathcal T}{\mbox{\rm\boldmath$:$}}\exp\big(
i\zeta{\hat{\mathcal J}}
+i\nu^*{\hat{\mathcal D}}
-i\nu{\hat{\mathcal D}}^*
 \big) 
{\mbox{\rm\boldmath$:$}}
 \Big \rangle \\ 
= 
\prod_{x,t}\bigg\{\int dJ(x,t)d^2D(x,t)\bigg\}
p\big(
J,D,D^* \big| A_{\mathrm{e}},E_{\mathrm{e}},E_{\mathrm{e}}^*
 \big) 
\exp\big(
i\zeta J+i\nu ^*D-i\nu D^*
 \big)
. 
\label{eq:95KH} 
\end{multline}%
The dressing formula reads \cite{QDynResp},
\begin{align} 
\begin{aligned} 
\Phi_{\mathrm{dev}}\big(
\zeta ,\nu,\nu ^*\big|a_{\mathrm{e}},e_{\mathrm{e}},e_{\mathrm{e}}^*
 \big) 
= 
\exp\bigg(
-i\frac{\delta }{\delta a_{\mathrm{e}}} 
G_{\text{R}}
\frac{\delta }{\delta \zeta }
-i 
\frac{\delta }{\delta e_{\mathrm{e}}} 
\Delta _{\text{R}}
\frac{\delta }{\delta \nu^* }
+i \frac{\delta }{\delta e_{\mathrm{e}}^*} 
\Delta _{\text{R}}^*
\frac{\delta }{\delta \nu }
 \bigg)
\Phi_{\mathrm{dev}}^{\mathrm{I}}\big(
\zeta ,\nu,\nu ^* \big| a_{\mathrm{e}},e_{\mathrm{e}},e_{\mathrm{e}}^*
 \big) , 
\end{aligned} 
\label{eq:91FH} 
\end{align}%
with the equivalent formula for the {quasiprobability distribution}s being \cite{PFunc}, 
\begin{align} 
\begin{aligned} 
p\big(
 J,D,D^*\big | A_{\mathrm{e}},E_{\mathrm{e}},E_{\mathrm{e}}^*
 \big) 
= p^{\mathrm{I}}\big(
 J,D,D^*\big | A_{\mathrm{e}} + G_{\text{R}}J, 
 E_{\mathrm{e}}+\Delta _{\text{R}}D,E_{\mathrm{e}}^*+\Delta _{\text{R}}^* D^*
 \big) . 
\end{aligned} 
\label{eq:57HR} 
\end{align}%
Functionals $\Phi^{\mathrm{I}}$ and $\Phi_{\mathrm{dev}}$ may equally be defined with operators without sources. For the bare device, 
\begin{align} 
\begin{aligned} 
\Phi_{\mathrm{dev}}^{\mathrm{I}}\big(
\zeta ,\nu ,\nu ^*\big| a_{\mathrm{e}},e_{\mathrm{e}},e_{\mathrm{e}}^*
 \big) 
= \text{Tr}\hat\rho_{\mathrm{dev}}
T_C\exp\big(
i{\zeta}_+ \hat J_{+} 
-i{\zeta}_- \hat J_{-}
+i\bar\nu_+ \hat D_{+} +i\nu_+ \hat D_{+}^{\dag}
-i\bar\nu_- \hat D_{-} -i\nu_- \hat D_{-}^{\dag}
 \big) 
\settoheight{\auxlv}{$\big|$}%
\raisebox{-0.3\auxlv}{$\big|_{\mathrm{c.v.}}$} , 
\end{aligned} 
\label{eq:92FJ} 
\end{align}%
cf.\ Eq.\ (\ref{eq:98KL}), where c.v.\ (short for {\em causal variables\/}) refers to the union of the nonresonant\ response substitution (\ref{eq:12YX}) and of the following resonant\ one, 
\begin{align} 
\begin{aligned} 
 &%
\begin{aligned}\nu_+
(x,t) 
 &= \frac{e_{\mathrm{e}}
(x,t) 
}{\hbar }, &
\bar\nu_+
(x,t) 
 &= \nu^*
(x,t) 
+ \frac{e_{\mathrm{e}}^*
(x,t) 
}{\hbar }, 
\end{aligned} \\ 
 &%
\begin{aligned}\bar\nu_-
(x,t) 
 &= \frac{e_{\mathrm{e}}^*
(x,t) 
}{\hbar }, &
\nu_-
(x,t) 
 &= \nu
(x,t) 
 + \frac{e_{\mathrm{e}}
(x,t) 
}{\hbar }. 
\end{aligned}
\end{aligned} 
\label{eq:21FB} 
\end{align}%
For the dressed device an extension of Eq.\ (\ref{eq:20ZF}) may be derived, 
\begin{multline} 
\Phi_{\mathrm{dev}}^{\mathrm{I}}\big(
\zeta ,\nu ,\nu ^*\big| a_{\mathrm{e}}+A_{\mathrm{e}},e_{\mathrm{e}}+E_{\mathrm{e}},e_{\mathrm{e}}^*+E_{\mathrm{e}}^*
 \big) \\ 
= \text{Tr}\hat\rho_{\mathrm{dev}}
T_C\exp\big(
i{\zeta}_+ {\hat{\mathcal J}}_{+} 
-i{\zeta}_- {\hat{\mathcal J}}_{-}
+i\bar\nu_+ {\hat{\mathcal D}}_{+} +i\nu_+ {\hat{\mathcal D}}_{+}^{\dag}
-i\bar\nu_- {\hat{\mathcal D}}_{-} -i\nu_- {\hat{\mathcal D}}_{-}^{\dag}
 \big) 
\settoheight{\auxlv}{$\big|$}%
\raisebox{-0.3\auxlv}{$\big|_{\mathrm{c.v.}}$} . 
\label{eq:24ZL} 
\end{multline}%
For a verification of Eqs.\ (\ref{eq:79NX})--(\ref{eq:24ZL}) see \mbox{Refs.\ \cite{APII,QDynResp,PFunc}}. 

\subsection{Interaction of distinguishable devices}%
\label{ch:GD}
Conditions (\ref{eq:1FS})--(\ref{eq:62BE}), (\ref{eq:5YQ})--(\ref{eq:7YS})\ and (\ref{eq:8YT})--(\ref{eq:10YV})\ are supplemented by, respectively, \mbox{$
\hat D(t) = \hat D_A(x,t) + \hat D_B(x,t)
$}, \mbox{$
\hat D(t) = \hat D_A(x,t)
$} and \mbox{$
\hat D(t) = \hat D_B(x,t)
$}. 
The bare and dressed functionals characterising the components of the device are defined inserting indices $A,B$ into Eqs.\ (\ref{eq:79NX})--(\ref{eq:57HR}). 
The connection between $p^{\mathrm{I}}$ and $p^{\mathrm{I}}_{A,B}$ is a natural generalisation of Eq.\ (\ref{eq:5KS}), 
\begin{multline} 
p^{\mathrm{I}}\big(
J ,E,E^*\big| A_{\mathrm{l}}, D_{\mathrm{l}}, D_{\mathrm{l}}^*
 \big) = \prod_{x,t}\bigg\{\int 
dJ_A(x,t)dJ_B(x,t)d^2D_A(x,t)d^2D_B(x,t)
\\ \times 
\delta\big(
J(x,t)-J_A(x,t)-J_B(x,t)
 \big) 
\delta^{(2)}\big(
D(x,t)-D_A(x,t)-D_B(x,t)
 \big) 
\bigg\} 
\\ \times 
p^{\mathrm{I}}_{A}\big(
J_A ,D_A,D_A^*\big| 
A_{\mathrm{l}}, E_{\mathrm{l}}, E_{\mathrm{l}}^*
 \big) 
p^{\mathrm{I}}_{B}\big(
J_B ,D_B,D_B^*\big| 
A_{\mathrm{l}}, E_{\mathrm{l}}, E_{\mathrm{l}}^*
 \big) . 
\label{eq:85PD} 
\end{multline}%
Using the general dressing formula (\ref{eq:57HR}) and proceeding as in Sec.\ \ref{ch:MS} we arrive at a generalisation of Eq.\ (\ref{eq:8KV}), 
\begin{multline} 
p\big(
J ,D,D^*\big| A_{\mathrm{e}}, D_{\mathrm{e}}, D_{\mathrm{e}}^*
 \big) = \prod_{x,t}\bigg\{\int 
dJ_A(x,t)dJ_B(x,t)d^2D_A(x,t)d^2D_B(x,t)
\\ \times 
\delta\big(
J(x,t)-J_A(x,t)-J_B(x,t)
 \big) 
\delta^{(2)}\big(
D(x,t)-D_A(x,t)-D_B(x,t)
 \big) 
\bigg\} 
\\ \times 
p_{A}\big(
J_A ,D_A,D_A^*\big| 
A_{\mathrm{e}A}, E_{\mathrm{e}A}, E_{\mathrm{e}A}^*
 \big) 
p_{B}\big(
J_B ,D_B,D_B^*\big| 
A_{\mathrm{e}B}, E_{\mathrm{e}B}, E_{\mathrm{e}B}^*
 \big) , 
\label{eq:87PF} 
\end{multline}%
where $A_{\mathrm{e}A,B}$ are given by Eqs.\ (\ref{eq:2ZY}), and $E_{\mathrm{e}A,B}$ --- by the analogous formulae, 
\begin{align} 
\begin{aligned} 
 E_{\mathrm{e}A}(x,t) &= E_{\mathrm{e}}(x,t) 
+ \int dx'dt'\Delta _{\text{R}}(x,x',t-t') D_B(x,t') , \\ 
 E_{\mathrm{e}B}(x,t) &= E_{\mathrm{e}}(x,t) 
+ \int dx'dt'\Delta _{\text{R}}(x,x',t-t') D_A(x,t') . 
\end{aligned} 
\label{eq:88PH} 
\end{align}%
In operator terms Eqs.\ (\ref{eq:85PD}) and (\ref{eq:87PF}) are equivalent to, correspondingly, 
\begin{align} 
\begin{aligned} 
\Phi_{\mathrm{dev}}^{\mathrm{I}}\big(
\zeta ,\nu ,\nu ^*\big| a_{\mathrm{e}},e_{\mathrm{e}},e_{\mathrm{e}}^*
 \big) = 
\Phi_{\mathrm{dev}A}^{\mathrm{I}}\big(
\zeta ,\nu ,\nu ^*\big| a_{\mathrm{e}},e_{\mathrm{e}},e_{\mathrm{e}}^*
 \big) 
\Phi_{\mathrm{dev}B}^{\mathrm{I}}\big(
\zeta ,\nu ,\nu ^*\big| a_{\mathrm{e}},e_{\mathrm{e}},e_{\mathrm{e}}^*
 \big) , 
\end{aligned} 
\label{eq:60CM} 
\end{align}%
and 
\begin{multline} 
\Phi_{\mathrm{dev}}\big(
\zeta ,\nu,\nu ^*\big| 
 a_{\mathrm{e}},e_{\mathrm{e}},e_{\mathrm{e}}^*
 \big) 
= 
\exp\bigg(
-i 
\frac{\delta }{\delta e_{\mathrm{e}}} 
\Delta _{\text{R}}
\frac{\delta }{\delta \nu^{\prime *} }
-i 
\frac{\delta }{\delta e_{\mathrm{e}}'} 
\Delta _{\text{R}}
\frac{\delta }{\delta \nu^* }
+i \frac{\delta }{\delta e_{\mathrm{e}}^*} 
\Delta _{\text{R}}^*
\frac{\delta }{\delta \nu' } 
+i \frac{\delta }{\delta e_{\mathrm{e}}^{\prime *}} 
\Delta _{\text{R}}^*
\frac{\delta }{\delta \nu }
\\ 
-i\frac{\delta }{\delta a_{\mathrm{e}}} 
G_{\text{R}}
\frac{\delta }{\delta \zeta'}
-i\frac{\delta }{\delta a_{\mathrm{e}}'} 
G_{\text{R}}
\frac{\delta }{\delta \zeta }
 \bigg) 
\Phi_{\mathrm{dev}A}\big(
\zeta ,\nu,\nu^{ *} \big| 
 a_{\mathrm{e}},e_{\mathrm{e}},e_{\mathrm{e}}^{ *} 
 \big) 
\Phi_{\mathrm{dev}B}\big(
\zeta ',\nu',\nu^{\prime *} \big| 
 a_{\mathrm{e}}',e_{\mathrm{e}}',e_{\mathrm{e}}^{\prime *} 
 \big) 
\settoheight{\auxlv}{$\big|$}%
\raisebox{-0.3\auxlv}{$\big|_{\zeta '=\zeta , 
\nu '=\nu ,
 a_{\mathrm{e}}'= a_{\mathrm{e}}, 
 e_{\mathrm{e}}'=e_{\mathrm{e}}}$}
, 
\label{eq:4FV} 
\end{multline}%
\end{widetext}%
cf.\ Eqs.\ (\ref{eq:63BF}) and (\ref{eq:66BK}). 
\section{Photodetection problem without the rotating wave approximation}\label{ch:CB}
\subsection{Formal quantum solution}\label{ch:CBF}
In this appendix we give a formal justification to Eqs.\ (\ref{eq:91HR}) and (\ref{eq:21DQ}). In physical terms, this demonstrates consistency of ``doing quantum electrodynamics\ while thinking classically'' with the approximation of avoided macroscopic back-action of the detector on the source, 
which underlies the very concept of photodetection. For all definitions see Sec.\ \ref{ch:TFM} and table \ref{T3}. 

We adapt the algebra of Sec.\ \ref{ch:MT}. The presence of two broad-band\ fields is formally accounted for assigning indices i,o to the field and current operators and all auxiliary quantities, and redefining condensed notation accordingly, e.g., 
\begin{align} 
\begin{aligned} 
\zeta \hat J
= 
\int dt \big [ 
\zeta _{\mathrm{i}}(t) \hat J_{{\mathrm{i}}}(t) + 
\zeta _{\mathrm{o}}(t) \hat J_{{\mathrm{o}}}(t)
 \big ] , 
\end{aligned} 
\label{eq:51AQ} 
\end{align}%
etc. Such redefinitions automatically extend all results of Sec.\ \ref{ch:E} for the broad-band\ case\ to the problem at hand. The emerging formal structure is anything but transparent, and, worse, takes no heed of specifics of the problem at hand. We therefore keep indices i,o explicit. 

Bare devices are characterised by the functionals, 
\begin{align} 
\begin{aligned} 
 &%
\begin{aligned} &\Phi ^{\mathrm{I}}_{\mathrm{dev}A}\big(
\zeta _{\mathrm{i}}\big|A_{{\mathrm{i}}}
 \big) = \big \langle {\mathcal T}{\mbox{\rm\boldmath$:$}} 
\exp\big(
i\zeta_{\mathrm{i}}\hat J_{{\mathrm{i}}}'
 \big) 
{\mbox{\rm\boldmath$:$}} \big \rangle_A, \\ 
 &\Phi ^{\mathrm{I}}_{\mathrm{dev}B}\big(
\zeta _{\mathrm{i}},\zeta _{\mathrm{o}}\big|A_{{\mathrm{i}}},A_{{\mathrm{o}}}
 \big) = \big \langle {\mathcal T}{\mbox{\rm\boldmath$:$}} 
\exp\big(
i\zeta_{\mathrm{i}}\hat J_{{\mathrm{i}}}' + 
i\zeta_{\mathrm{o}}\hat J_{{\mathrm{o}}}'
 \big) 
{\mbox{\rm\boldmath$:$}} \big \rangle_B, \\ 
 &\Phi ^{\mathrm{I}}_{\mathrm{dev}}\big(
\zeta _{\mathrm{i}},\zeta _{\mathrm{o}}\big|A_{{\mathrm{i}}},A_{{\mathrm{o}}}
 \big) = \big \langle {\mathcal T}{\mbox{\rm\boldmath$:$}} 
\exp\big(
i\zeta_{\mathrm{i}}\hat J_{{\mathrm{i}}}' + 
i\zeta_{\mathrm{o}}\hat J_{{\mathrm{o}}}'
 \big) 
{\mbox{\rm\boldmath$:$}} \big \rangle 
\end{aligned} 
\\ &\qquad\qquad\qquad 
= \Phi ^{\mathrm{I}}_{\mathrm{dev}A}\big(
\zeta _{\mathrm{i}}\big|A_{{\mathrm{i}}}
 \big) 
\Phi ^{\mathrm{I}}_{\mathrm{dev}B}\big(
\zeta _{\mathrm{i}},\zeta _{\mathrm{o}}\big|A_{{\mathrm{i}}},A_{{\mathrm{o}}}
 \big) , 
\end{aligned} 
\label{eq:58AX} 
\end{align}%
where the primed operators are the Heisenberg\ ones for the Hamiltonian (\ref{eq:27KD}) with $\hat A_{{\mathrm{i}}}(t)=\hat A_{{\mathrm{o}}}(t)=0$. Device $A$ does not interact with the output field; consequently $\Phi ^{\mathrm{I}}_{\mathrm{dev}A}$ does not depend on \mbox{$
\zeta _{\mathrm{o}}(t)
$} nor on \mbox{$
A_{{\mathrm{o}}}(t)
$}. 

For the dressed devices, use will be made of the functionals, 
\begin{align} 
\begin{aligned} 
 &\Phi_{\mathrm{dev}A}\big(
\zeta _{\mathrm{i}}\big|A_{{\mathrm{i}}}
 \big) = \big \langle {\mathcal T}{\mbox{\rm\boldmath$:$}} 
\exp\big(
i\zeta_{\mathrm{i}}{\hat{\mathcal J}}_{{\mathrm{i}}}
 \big) 
{\mbox{\rm\boldmath$:$}} \big \rangle_A, \\ 
 &\Phi_{\mathrm{dev}B}\big(
\zeta _{\mathrm{i}},\zeta _{\mathrm{o}}\big|A_{{\mathrm{i}}},A_{{\mathrm{o}}}
 \big) = \big \langle {\mathcal T}{\mbox{\rm\boldmath$:$}} 
\exp\big(
i\zeta_{\mathrm{i}}{\hat{\mathcal J}}_{{\mathrm{i}}}+ 
i\zeta_{\mathrm{o}}{\hat{\mathcal J}}_{{\mathrm{o}}}
 \big) 
{\mbox{\rm\boldmath$:$}} \big \rangle_B, \\ 
 &\Phi_{\mathrm{dev}}\big(
\zeta _{\mathrm{i}},\zeta _{\mathrm{o}}\big|A_{{\mathrm{i}}},A_{{\mathrm{o}}}
 \big) = \big \langle {\mathcal T}{\mbox{\rm\boldmath$:$}} 
\exp\big(
i\zeta_{\mathrm{i}}{\hat{\mathcal J}}_{{\mathrm{i}}}+ 
i\zeta_{\mathrm{o}}{\hat{\mathcal J}}_{{\mathrm{o}}}
 \big) 
{\mbox{\rm\boldmath$:$}} \big \rangle 
. 
\end{aligned} 
\label{eq:59AY} 
\end{align}%
Calligraphic letters denote Heisenberg\ operators. These are in essence placeholders, the exact meaning of which depends on ``jumper'' settings (see Sec.\ \ref{ch:TFM}). 

Response function $G_{\text{R}}$ given by (\ref{eq:83LF}) is now a $2\times 2$ matrix of kernels, however, since $\hat A_{{\mathrm{i}}}(t)$ and $\hat A_{{\mathrm{o}}}(t)$ by construction commute, this matrix is diagonal. All quadratic forms involving $G_{\text{R}}$ split in two, 
\begin{align} 
\begin{aligned} 
\eta G_{\text{R}}j = \eta_{\mathrm{i}}G_{\mathrm{R}{\mathrm{i}}}j_{\mathrm{i}}+\eta_{\mathrm{o}}G_{\mathrm{R}{\mathrm{o}}}j_{\mathrm{o}},
\end{aligned} 
\label{eq:53AS} 
\end{align}%
and the causal reordering exponent factorises, 
\begin{multline} 
\hspace{0.4\columnwidth}\hspace{-0.4\twocolumnwidth}
\exp\bigg(
-i\frac{\delta}{\delta A}G_{\text{R}}\frac{\delta}{\delta \zeta} 
 \bigg) \\ 
= 
\exp\bigg(
-i\frac{\delta}{\delta A_{{\mathrm{i}}}}G_{\mathrm{R}{\mathrm{i}}}\frac{\delta}{\delta \zeta_{\mathrm{i}}} 
 \bigg)\exp\bigg(
-i\frac{\delta}{\delta A_{{\mathrm{o}}}}G_{\mathrm{R}{\mathrm{o}}}\frac{\delta}{\delta \zeta_{\mathrm{o}}} 
 \bigg) . 
\hspace{0.4\columnwidth}\hspace{-0.4\twocolumnwidth}%
\label{eq:52AR} 
\end{multline}%
The kernels $G_{\mathrm{R}{\mathrm{i}}}$ and $G_{\mathrm{R}{\mathrm{o}}}$ are defined applying (\ref{eq:83LF}) to the input and output fields, 
\begin{align} 
\begin{aligned} 
G_{\mathrm{R}{\mathrm{i}}}(t-t') &= \frac{i}{\hbar }\theta(t-t') 
\big [ 
\hat A_{{\mathrm{i}}}(t),\hat A_{{\mathrm{i}}}(t')
 \big ] , \\ 
G_{\mathrm{R}{\mathrm{o}}}(t-t') &= \frac{i}{\hbar }\theta(t-t') 
\big [ 
\hat A_{{\mathrm{o}}}(t),\hat A_{{\mathrm{o}}}(t')
 \big ] . 
\end{aligned} 
\label{eq:54AT} 
\end{align}%
In place of Eq.\ (\ref{eq:95FM}) we thus have, 
\begin{multline} 
\hspace{0.4\columnwidth}\hspace{-0.4\twocolumnwidth}
\Phi_{\mathrm{dev}}\big(
\zeta _{\mathrm{i}},\zeta _{\mathrm{o}}\big|A_{{\mathrm{i}}},A_{{\mathrm{o}}}
 \big) = \\ 
\exp\bigg(
-i\frac{\delta}{\delta A_{{\mathrm{i}}}}G_{\mathrm{R}{\mathrm{i}}}\frac{\delta}{\delta \zeta_{\mathrm{i}}} 
 \bigg)\exp\bigg(
-i\frac{\delta}{\delta A_{{\mathrm{o}}}}G_{\mathrm{R}{\mathrm{o}}}\frac{\delta}{\delta \zeta_{\mathrm{o}}} 
 \bigg) 
\\ \times 
\Phi ^{\mathrm{I}}_{\mathrm{dev}A}\big(
\zeta _{\mathrm{i}}\big|A_{{\mathrm{i}}}
 \big) 
\Phi ^{\mathrm{I}}_{\mathrm{dev}B}\big(
\zeta _{\mathrm{i}},\zeta _{\mathrm{o}}\big|A_{{\mathrm{i}}},A_{{\mathrm{o}}}
 \big) . 
\hspace{0.4\columnwidth}\hspace{-0.4\twocolumnwidth}%
\label{eq:93CL} 
\end{multline}%
The ``output'' exponent here acts only on \mbox{$
\Phi ^{\mathrm{I}}_{\mathrm{dev}B}
$}; Eqs.\ (\ref{eq:96FN})--(\ref{eq:98FQ}) should therefore be applied to the ``input'' exponent (with \mbox{$
\zeta \to\zeta _{\mathrm{i}}
$} and \mbox{$
a_{\mathrm{e}}\to A_{{\mathrm{i}}}
$}). 
The dressing relations (\ref{eq:65BJ}) now read, 
\begin{align} 
\begin{aligned} 
 &\exp\bigg(
-i\frac{\delta }{\delta A_{{\mathrm{i}}}} G_{\text{R}}
\frac{\delta }{\delta \zeta_{\mathrm{i}}}
 \bigg) 
\Phi_{\mathrm{dev}A}^{\mathrm{I}}\big(
\zeta_{\mathrm{i}}\big | A_{{\mathrm{i}}}
 \big) 
= 
\Phi_{\mathrm{dev}A}\big(
\zeta_{\mathrm{i}}\big | _{\text{A}}
 \big) 
, \\ 
 &\exp\bigg(
-i\frac{\delta }{\delta A_{{\mathrm{i}}}'} G_{\text{R}}
\frac{\delta }{\delta \zeta_{\mathrm{i}}' }
 \bigg)\exp\bigg(
-i\frac{\delta}{\delta A_{{\mathrm{o}}}}G_{\mathrm{R}{\mathrm{o}}}\frac{\delta}{\delta \zeta_{\mathrm{o}}} 
 \bigg) 
\\ &\qquad \times 
\Phi_{\mathrm{dev}B}^{\mathrm{I}}\big(
\zeta_{\mathrm{i}}' ,\zeta _{\mathrm{o}}\big | A_{{\mathrm{i}}}',A_{{\mathrm{o}}}
 \big) 
= 
\Phi_{\mathrm{dev}B}\big(
\zeta_{\mathrm{i}}',\zeta _{\mathrm{o}}\big | A_{{\mathrm{i}}}',A_{{\mathrm{o}}}
 \big) 
, 
\end{aligned} 
\label{eq:94CM} 
\end{align}%
while Eq.\ (\ref{eq:66BK}) expressing interactions of dressed devices becomes, 
\begin{multline} 
\hspace{0.4\columnwidth}\hspace{-0.4\twocolumnwidth}
\Phi_{\mathrm{dev}}\big(
\zeta _{\mathrm{i}},\zeta _{\mathrm{o}}\big|A_{{\mathrm{i}}},A_{{\mathrm{o}}}
 \big) = \\ 
\exp\bigg(
-i\frac{\delta }{\delta A_{{\mathrm{i}}}'}G_{\mathrm{R}{\mathrm{i}}}
\frac{\delta }{\delta \zeta_{\mathrm{i}}} 
-i\frac{\delta }{\delta A_{{\mathrm{i}}}}G_{\mathrm{R}{\mathrm{i}}}
\frac{\delta }{\delta \zeta_{\mathrm{i}}' }
 \bigg)
\\ \times 
\Phi_{\mathrm{dev}A}\big(
\zeta_{\mathrm{i}}'\big|A_{{\mathrm{i}}}'
 \big)\Phi_{\mathrm{dev}B}\big(
\zeta_{\mathrm{i}},\zeta _{\mathrm{o}}\big|A_{{\mathrm{i}}},A_{{\mathrm{o}}}
 \big)\settoheight{\auxlv}{$|$}%
\raisebox{-0.3\auxlv}{$|_{\zeta_{\mathrm{i}}'=\zeta_{\mathrm{i}},A_{{\mathrm{i}}}'=A_{{\mathrm{i}}}}$} . 
\hspace{0.4\columnwidth}\hspace{-0.4\twocolumnwidth}%
\label{eq:61BA} 
\end{multline}%
\subsection{Approximations and solution to the photodetection problem}\label{ch:CBA}
Equation (\ref{eq:61BA}) is not yet a photodetection formula. Firstly, the source-detector interaction in it is bi-directional. Secondly, it contains a lot of irrelevant information, in particular, full quantum (response) properties of the input and output fields. So, it ``knows'' how an attempt to measure the input field would affect the detection, and how simultaneous measurements of the input field and output current would be correlated. Subject to valid quantum models of the devices, it also ``knows'' about all limitations imposed by quantum mechanics on such simultaneous measurements. 

By definition, of interest to us is the {\em output signal\/} of the detector under the condition of {\em avoided macroscopic back-action\/} of the detector on the source. The said signal is formally expressed by the joint time-normal averages of the field \mbox{$
{\hat{\mathcal A}}_{{\mathrm{o}}}(t)
$} and current \mbox{$
{\hat{\mathcal J}}_{{\mathrm{o}}}(t)
$}, defined in the absence of external sources. These averages are accessible through their characteristic functional, 
\begin{multline} 
\hspace{0.4\columnwidth}\hspace{-0.4\twocolumnwidth}
\big \langle {\mathcal T}{\mbox{\rm\boldmath$:$}} 
\exp\big(
i\eta_{\mathrm{o}}{\hat{\mathcal A}}_{{\mathrm{o}}}+
i\zeta_{\mathrm{o}}{\hat{\mathcal J}}_{{\mathrm{o}}}
 \big) 
{\mbox{\rm\boldmath$:$}} \big \rangle_{\mathrm{o}} \\ 
= \Phi_{\mathrm{dev}}{\big( 
0,\zeta _{\mathrm{o}}+ \eta _{\mathrm{o}}G_{\mathrm{R}{\mathrm{o}}}\big| 0,0
 \big)} \equiv \Phi_{\mathrm{o}}\big( 
\zeta _{\mathrm{o}}+ \eta _{\mathrm{o}}G_{\mathrm{R}{\mathrm{o}}}
 \big) , 
\hspace{0.4\columnwidth}\hspace{-0.4\twocolumnwidth}%
\label{eq:95CN} 
\end{multline}%
cf.\ Eq.\ (\ref{eq:24KA}). The symbol \mbox{$
 \langle 
\cdots
 \rangle_{\mathrm{o}} 
$} was defined in Sec.\ \ref{ch:TFM} (see table \ref{T3}). 

Equation (\ref{eq:95CN}) suppresses the aforementioned irrelevant information. The no-macroscopic-back-action approximation is imposed dropping the signals propagating from device $B$ to device $A$. Formally, this means a replacement in Eq.\ (\ref{eq:61BA}),
\begin{multline} 
\hspace{0.4\columnwidth}\hspace{-0.4\twocolumnwidth}
\exp\bigg(
-i\frac{\delta }{\delta A_{{\mathrm{i}}}'}G_{\mathrm{R}{\mathrm{i}}}
\frac{\delta }{\delta \zeta_{\mathrm{i}}} 
-i\frac{\delta }{\delta A_{{\mathrm{i}}}}G_{\mathrm{R}{\mathrm{i}}}
\frac{\delta }{\delta \zeta_{\mathrm{i}}' }
 \bigg) \\ 
\to\exp\bigg(
-i\frac{\delta }{\delta A_{{\mathrm{i}}}}G_{\mathrm{R}{\mathrm{i}}}
\frac{\delta }{\delta \zeta_{\mathrm{i}}' }
 \bigg) . 
\hspace{0.4\columnwidth}\hspace{-0.4\twocolumnwidth}%
\label{eq:96CP} 
\end{multline}%
Under this assumption Eq.\ (\ref{eq:61BA}) yields, 
\begin{multline} 
\hspace{0.4\columnwidth}\hspace{-0.4\twocolumnwidth}
\Phi_{\mathrm{o}}\big(
\zeta _{\mathrm{o}}
 \big) = 
\exp\bigg(
-i\frac{\delta }{\delta A_{{\mathrm{i}}}}G_{\mathrm{R}{\mathrm{i}}}
\frac{\delta }{\delta \zeta_{\mathrm{i}}}
 \bigg)
\\ \times 
\Phi_{\mathrm{dev}A}\big(
\zeta_{\mathrm{i}}\big|0
 \big)\Phi_{\mathrm{dev}B}\big(
0,\zeta _{\mathrm{o}}\big|A_{{\mathrm{i}}},0
 \big)\settoheight{\auxlv}{$|$}%
\raisebox{-0.3\auxlv}{$|_{\zeta_{\mathrm{i}}=0 ,A_{{\mathrm{i}}}=0}$} . 
\hspace{0.4\columnwidth}\hspace{-0.4\twocolumnwidth}%
\label{eq:97CQ} 
\end{multline}%
We dropped the prime at \mbox{$
\zeta_{\mathrm{i}}'(t)
$} which after setting \mbox{$
\zeta_{\mathrm{i}}(t)=0
$} became redundant. 

Confining out interest to the output signal reduces the information required about the devices. The source is described by the time-normal current averages generated by the functional, 
\begin{align} 
\begin{aligned} 
\Phi_{\mathrm{dev}A}\big(
\zeta_{\mathrm{i}}\big|0
 \big) = 
\big \langle {\mathcal T}{\mbox{\rm\boldmath$:$}} 
\exp\big(
i\zeta _{\mathrm{i}}{\hat{\mathcal J}}_A
 \big) 
{\mbox{\rm\boldmath$:$}} \big \rangle _{\mathrm{s}} \equiv \Phi_{\mathrm{s}}\big(
\zeta_{\mathrm{i}}
 \big) , 
\end{aligned} 
\label{eq:98CR} 
\end{align}%
while the detector is characterised by the time-normal current averages conditional on the input field, 
\begin{multline} 
\hspace{0.4\columnwidth}\hspace{-0.4\twocolumnwidth}
\Phi_{\mathrm{dev}B}\big(
0,\zeta _{\mathrm{o}}\big|A_{{\mathrm{i}}},0
 \big) 
\\ 
= 
\big \langle {\mathcal T}{\mbox{\rm\boldmath$:$}} 
\exp\big(
i\zeta _{\mathrm{o}}{\hat{\mathcal J}}_{B{\mathrm{o}}}
 \big) 
{\mbox{\rm\boldmath$:$}} \big \rangle _{\mathrm{d}} \equiv \Phi_{\mathrm{d}}{\big( 
\zeta_{\mathrm{o}}\big| A_{{\mathrm{i}}}
 \big)} . 
\hspace{0.4\columnwidth}\hspace{-0.4\twocolumnwidth}%
\label{eq:99CS} 
\end{multline}%
The symbols \mbox{$
 \langle 
\cdots
 \rangle_{\mathrm{s}} 
$} and \mbox{$
 \langle 
\cdots
 \rangle_{\mathrm{d}} 
$} were defined in Sec.\ \ref{ch:TFM} (see table \ref{T3}). With irrelevant information hidden from view we find the photodetection formula, 
\begin{multline} 
\hspace{0.4\columnwidth}\hspace{-0.4\twocolumnwidth}
\Phi_{\mathrm{o}}\big(
\zeta _{\mathrm{o}}
 \big) = 
\exp\bigg(
-i\frac{\delta }{\delta A_{{\mathrm{i}}}}G_{\mathrm{R}{\mathrm{i}}}
\frac{\delta }{\delta \zeta_{\mathrm{i}}}
 \bigg)
\\ \times 
\Phi_{\mathrm{s}}\big(
\zeta_{\mathrm{i}}
 \big)\Phi_{\mathrm{d}}{\big( 
\zeta_{\mathrm{o}}\big| A_{{\mathrm{i}}}
 \big)}\settoheight{\auxlv}{$|$}%
\raisebox{-0.3\auxlv}{$|_{\zeta_{\mathrm{i}}=0 ,A_{{\mathrm{i}}}=0}$} . 
\hspace{0.4\columnwidth}\hspace{-0.4\twocolumnwidth}%
\label{eq:1CT} 
\end{multline}%

It is worth empasising that, to obtain Eq.\ (\ref{eq:1CT}), the ``one-way'' assumption does not suffice. We had also to assume what we {\em do not do\/} certain things in the experiment: do not attempt to perform any additional measurement on the optical mode, and do not attempt to probe the detector by radiation in the photocurrent mode. 
\subsection{Photodetection statistics as a quantum average over the detected field}\label{ch:CBC}
It is straightforward to rewrite Eq.\ (\ref{eq:1CT}) in terms of the detected field rather than the source current (which is a customary viewpoint). Time-normal averages of the field radiated by the source are generated by the functional, 
\begin{align} 
\begin{aligned} 
\Gamma _{\mathrm{s}}\big(
\eta_{\mathrm{i}}
 \big) = 
\big \langle {\mathcal T}{\mbox{\rm\boldmath$:$}} 
\exp\big(
i\eta _{\mathrm{i}}{\hat{\mathcal A}}_{{\mathrm{i}}}
 \big) 
{\mbox{\rm\boldmath$:$}} \big \rangle _{\mathrm{s}} = \Phi_{\mathrm{s}}\big(
\eta_{\mathrm{i}}G_{\mathrm{R}{\mathrm{i}}}
 \big) , 
\end{aligned} 
\label{eq:2CU} 
\end{align}%
where use was again made of Eq.\ (\ref{eq:97YH}). Applying the obvious formula, 
\begin{align} 
\begin{aligned} 
\Phi_{\mathrm{s}}\big(
\eta_{\mathrm{i}}G_{\mathrm{R}{\mathrm{i}}}
 \big) = \exp\bigg(
\eta_{\mathrm{i}}G_{\mathrm{R}{\mathrm{i}}}
\frac{\delta }{\delta \zeta_{\mathrm{i}}}
 \bigg)\Phi_{\mathrm{s}}\big(
\zeta_{\mathrm{i}}
 \big)\settoheight{\auxlv}{$|$}%
\raisebox{-0.3\auxlv}{$|_{\zeta_{\mathrm{i}}=0}$}, 
\end{aligned} 
\label{eq:3CV} 
\end{align}%
we can rewrire (\ref{eq:1CT}) as, 
\begin{align} 
\begin{aligned} 
\Phi_{\mathrm{o}}\big(
\zeta _{\mathrm{o}}
 \big) = 
\Gamma _{\mathrm{s}}\bigg(
-i\frac{\delta }{\delta A_{{\mathrm{i}}}} 
 \bigg)\Phi_{\mathrm{d}}{\big( 
\zeta_{\mathrm{o}}\big| A_{{\mathrm{i}}}
 \big)}\settoheight{\auxlv}{$|$}%
\raisebox{-0.3\auxlv}{$|_{A_{{\mathrm{i}}}=0}$} . 
\end{aligned} 
\label{eq:4CW} 
\end{align}%
Application of the differential operator here cannot be anything but a fancy way of expressing quantum averaging over the detected field. Indeed, by definition
\begin{align} 
\begin{aligned} 
\Gamma _{\mathrm{s}}\bigg(
-i\frac{\delta }{\delta A_{{\mathrm{i}}}} 
 \bigg) = \bigg \langle {\mathcal T}{\mbox{\rm\boldmath$:$}} 
\exp\bigg(
{\hat{\mathcal A}}_{{\mathrm{i}}}\frac{\delta }{\delta A_{{\mathrm{i}}}}
 \bigg){\mbox{\rm\boldmath$:$}} \bigg \rangle_{\mathrm{s}} , 
\end{aligned} 
\label{eq:6CY} 
\end{align}%
cf.\ Eq.\ (\ref{eq:2CU}). Under orderings, operators behave as c-numbers, and the exponent in (\ref{eq:6CY}) may be interpreted as a functional shift operator. 
Hence for any c-number functional \mbox{$
\mathcal{F}(A_{{\mathrm{i}}})
$}, 
\begin{multline} 
\hspace{0.4\columnwidth}\hspace{-0.4\twocolumnwidth}
\Gamma _{\mathrm{s}}\bigg(
-i\frac{\delta }{\delta A_{{\mathrm{i}}}} 
 \bigg)\mathcal{F}(A_{{\mathrm{i}}})\settoheight{\auxlv}{$|$}%
\raisebox{-0.3\auxlv}{$|_{A_{{\mathrm{i}}}=0}$} \\ 
= \big [ 
\big \langle {\mathcal T}{\mbox{\rm\boldmath$:$}} 
\mathcal{F}(A_{{\mathrm{i}}}+{\hat{\mathcal A}}_{{\mathrm{i}}})
{\mbox{\rm\boldmath$:$}} \big \rangle_{\mathrm{s}}
 \big ]\settoheight{\auxlv}{$|$}%
\raisebox{-0.3\auxlv}{$|_{A_{{\mathrm{i}}}=0}$} = \big \langle {\mathcal T}{\mbox{\rm\boldmath$:$}} 
\mathcal{F}({\hat{\mathcal A}}_{{\mathrm{i}}})
{\mbox{\rm\boldmath$:$}} \big \rangle_{\mathrm{s}} . 
\hspace{0.4\columnwidth}\hspace{-0.4\twocolumnwidth}%
\label{eq:5CX} 
\end{multline}%
Using this formula we can rewrite Eq.\ (\ref{eq:4CW}) explicitly as a quantum average over the detected field, 
\begin{align} 
\begin{aligned} 
\Phi_{\mathrm{o}}\big(
\zeta _{\mathrm{o}}
 \big) 
= \big \langle {\mathcal T}{\mbox{\rm\boldmath$:$}} 
\Phi_{\mathrm{d}}{\big( 
\zeta_{\mathrm{o}}\big| {\hat{\mathcal A}}_{{\mathrm{i}}}
 \big)}
{\mbox{\rm\boldmath$:$}} \big \rangle_{\mathrm{s}} . 
\end{aligned} 
\label{eq:7CZ} 
\end{align}%
Equation (\ref{eq:21DQ}) readily follows from this formula, which in turn gives justification to Eq.\ (\ref{eq:91HR}). 
\section{Quantum theory of a cascaded system}\label{ch:A}
\subsection{The model}\label{ch:TM}
The high-frequency (HF) optical fields 0 and 1 in Fig.\ \ref{fig:CascadeF}b are treated under the rotating wave approximation, while the photovoltage --- which is a low-frequency (LF) field --- without the rotating wave approximation. 
The corresponding field operators are, 
\begin{align} 
\begin{aligned} 
 &%
\begin{aligned} &\hat E_0(t) = i\sqrt{\frac{2\pi \hbar \omega_0}{V_0}}\,\hat a_0 , 
 &\hat E_1(t) = i\sqrt{\frac{2\pi \hbar \omega_1}{V_1}}\,\hat a_1 , 
\end{aligned} \\ 
 &\hat A_{\mathrm{o}}(t) = \sqrt{\frac{2\pi \hbar}{\omega_2 V_2}}
\,\hat a_2\mathrm{e}^{-i\omega _2 t} + {\mathrm{H.c.}}
\, , 
\end{aligned} 
\label{eq:29EB} 
\end{align}%
where $\hat a_{0,1,2}$ are the standard annihilation operators, $\omega _{0,1,2}$ are the frequencies and $V_{0,1,2}$ are the so-called mode volumes. Recall that $\hat E_{0,1}(t)$ are slow amplitudes; they lack time exponents. The Hamiltonian of the electromagnetic\ field is a sum of three oscillator Hamiltonians, 
\begin{align} 
\begin{aligned} 
\hat H^{\mathrm{f}} = \hbar \sum_{\mathrm{\kappa =0}}^2 \omega _{\kappa }
\hat a^{\dag}_{\kappa }\hat a_{\kappa }
\end{aligned} 
\label{eq:58FE} 
\end{align}%
The Hamiltonians of bare devices are $\hat H_{\mathrm{dev}A,B,C}(t)$ and their states are $\hat\rho_{\mathrm{dev}A,B,C}$. Device $A$ interacts with mode 0 by means of the dipole operator $\hat D_{A0}(t)$. Device $B$ interacts with mode 0 by means of $\hat D_{B0}(t)$ and with mode 1 by means of $\hat D_{B1}(t)$. Device $c$ interacts with mode 1 by means of $\hat D_{C1}(t)$ and with mode 1 by means of the current $\hat J_{C2}(t)$. The state of all oscillators is vacuum, so that the state of the full system is, 
\begin{align} 
\begin{aligned} 
\hat \rho =  | 0 \rangle  \langle 0 | 
\otimes \hat \rho _{\mathrm{dev}A} 
\otimes \hat \rho _{\mathrm{dev}B} 
\otimes \hat \rho _{\mathrm{dev}C} 
. 
\end{aligned} 
\label{eq:38AA} 
\end{align}%

\begin{table*}
\begin{tabular}{l@{\hspace{12pt}}l@{\hspace{12pt}}l@{\hspace{12pt}}l@{\hspace{12pt}}l}
\hline
Problem & ``Jumper'' settings & Relevant & \multicolumn{2}{l}{Notation for averages} \\ 
&&ext.\ souces&``Raw''&``Physical''\\
\hline
Light source & ${s_A}=1,{s_B}=0,{s_C}=0$ & $(E_0)$ & $ \langle \cdots \rangle_A$ & $ \langle \cdots \rangle_{\mathrm{s}}=(
 \langle \cdots \rangle_A
 )\settoheight{\auxlv}{$|$}%
\raisebox{-0.3\auxlv}{$|_{E_0=0}$}$ \\ 
Amplifier & ${s_A}=0,{s_B}=1,{s_C}=0$ & $E_0,(E_1)$ & $ \langle \cdots \rangle_B$ & $ \langle \cdots \rangle_{\mathrm{a}}=(
 \langle \cdots \rangle_B
 )\settoheight{\auxlv}{$|$}%
\raisebox{-0.3\auxlv}{$|_{E_1=0}$}$ \\ 
Detector & ${s_A}=0,{s_B}=0,{s_C}=1$ & $E_1,(A_{{\mathrm{o}}})$ & $ \langle \cdots \rangle_C$ & $ \langle \cdots \rangle_{\mathrm{d}}=(
 \langle \cdots \rangle_C
 )\settoheight{\auxlv}{$|$}%
\raisebox{-0.3\auxlv}{$|_{A_{{\mathrm{o}}}=0}$}$ \\ 
{Composite source}& ${s_A}=1,{s_B}=1,{s_C}=0$ & $(E_0,E_1)$ & $ \langle \cdots \rangle_{AB}$ & $ \langle \cdots \rangle_{\mathrm{cs}}=(
 \langle \cdots \rangle_{AB}
 )\settoheight{\auxlv}{$|$}%
\raisebox{-0.3\auxlv}{$|_{E_0=E_1=0}$}$ \\ 
{Composite detector}& ${s_A}=0,{s_B}=1,{s_C}=1$ & $E_0,(E_1,A_{{\mathrm{o}}})$ & $ \langle \cdots \rangle_{BC}$ & $ \langle \cdots \rangle_{\mathrm{cd}}=(
 \langle \cdots \rangle_{BC}
 )\settoheight{\auxlv}{$|$}%
\raisebox{-0.3\auxlv}{$|_{E_1=0,A_{{\mathrm{o}}}=0}$}$ \\ 
Full system & ${s_A}=1,{s_B}=1,{s_C}=1$ & $(E_0,E_1,A_{{\mathrm{o}}})$ & $ \langle \cdots \rangle$ & $ \langle \cdots \rangle_{\mathrm{o}}=
{
 \langle \cdots \rangle
}\settoheight{\auxlv}{$|$}%
\raisebox{-0.3\auxlv}{$|_{E_0=E_1=0,A_{{\mathrm{o}}}=0}$}$ \\
\hline
\end{tabular}
\caption{Six problems relevant to the arrangement in Fig.\ \ref{fig:CascadeF}b. ``Raw'' averages imply the density matrix (\ref{eq:38AA}) and Hamiltonian (\ref{eq:98HY}), the latter with ``jumpers'' set to listed values. ``Physical'' averages follow by setting some or all c-number sources to zero. The table also lists the c-number sources on which the ``raw'' averages depend; those shown in brackets are set to zero in ``physical'' averages.}
\label{T6}
\end{table*}
Formally, we have to consider six physical problems: three of solitary devices, one of the composite source, one of the composite detector, and one of the whole system (cf.\ Fig.\ \ref{fig:CascadeF}b). To unify the bookkeeping we postulate the Hamiltonian in the form, 
\begin{multline} 
\hspace{0.4\columnwidth}\hspace{-0.4\twocolumnwidth}
\hat H(t) = \hat H^{\mathrm{f}} + \hat H_{\mathrm{dev}}(t) 
- \big [ 
\hat A_{{\mathrm{o}}}(t)+A_{{\mathrm{o}}}(t)
 \big ]\hat J_{\mathrm{o}}(t) \\ 
- \big \{ 
\big [ 
\hat E_0(t) + E_0(t)
 \big ]\hat D_{0}^{\dag}(t) 
+\big [ 
\hat E_1(t) + E_1(t)
 \big ]\hat D_{1}^{\dag}(t) + {\mathrm{H.c.}}
 \big \} 
 , 
\hspace{0.4\columnwidth}\hspace{-0.4\twocolumnwidth}%
\label{eq:98HY} 
\end{multline}%
where 
\begin{align} 
\begin{aligned} 
 &H_{\mathrm{dev}}(t) = {s_A}H_{\mathrm{dev}A}(t) 
+ {s_B}H_{\mathrm{dev}B}(t) + {s_C}H_{\mathrm{dev}C}(t) , \\ 
 &\hat D_0(t) = {s_A}\hat D_{A0}(t) + {s_B}\hat D_{B0}(t) , \\
 &\hat D_1(t) = {s_B}\hat D_{B1}(t) + {s_C}\hat D_{C1}(t) , \\
 &\hat J_{\mathrm{o}}(t) = {s_C}\hat J_{C2}(t) . 
\end{aligned} 
\label{eq:99HZ} 
\end{align}%
The ``jumpers'' $s_{A,B,C}=0,1$ serve to ``commute'' the problems. For example, with ${s_A}=1,{s_B}={s_C}=0$ we recover the problem of a solitary source. More precisely speaking, we have to distinguish the problem of device $A$ and that of the light source, which differ in whether the c-number source \mbox{$
E_0(t)
$} is nonzero or zero. Correspondingly we have to define two types of quantities (averages): with nonzero \mbox{$
E_0(t)
$}, denoted \mbox{$
 \langle 
\cdots
 \rangle_A
$}, and with zero \mbox{$
E_0(t)
$}, denoted \mbox{$
 \langle 
\cdots
 \rangle_{\mathrm{s}}
$}. For a summary of all definitions see table \ref{T6}. 

``Dressed'' devices are characterised by the averages, 
\begin{widetext} 
\begin{align} 
\begin{aligned} 
\Phi_{\mathrm{dev}A}{\big( 
\nu_0,\nu_0^* \big| E_0,E_0^*
 \big)} &= \big \langle {\mathcal T}{\mbox{\rm\boldmath$:$}} 
\exp\big(
i\nu^*_0 {\hat{\mathcal D}}_0 - i\nu_0 {\hat{\mathcal D}}_0^{\dagger}
 \big) 
{\mbox{\rm\boldmath$:$}} \big \rangle_A, \\
\Phi_{\mathrm{dev}B}{\big( 
\nu_0,\nu_0^*, 
\nu_1,\nu_1^* 
 \big| 
E_0,E_0^*,
E_1,E_1^*
 \big)} &= \big \langle {\mathcal T}{\mbox{\rm\boldmath$:$}} 
\exp\big(
i\nu^*_0 {\hat{\mathcal D}}_0 - i\nu_0 {\hat{\mathcal D}}_0^{\dagger}
+i\nu^*_1 {\hat{\mathcal D}}_1 - i\nu_1 {\hat{\mathcal D}}_1^{\dagger}
 \big) 
{\mbox{\rm\boldmath$:$}} \big \rangle_B, \\
\Phi_{\mathrm{dev}C}{\big( 
\nu_1,\nu_1^*,\zeta _{{\mathrm{o}}} 
 \big| 
E_1,E_1^*,A_{{\mathrm{o}}}
 \big)} &= \big \langle {\mathcal T}{\mbox{\rm\boldmath$:$}} 
\exp\big(
i\nu^*_1 {\hat{\mathcal D}}_1 - i\nu_1 {\hat{\mathcal D}}_1^{\dagger} + i\zeta _{{\mathrm{o}}}{\hat{\mathcal J}}_{{\mathrm{o}}}
 \big) 
{\mbox{\rm\boldmath$:$}} \big \rangle_C, \\
\Phi_{\mathrm{dev}BC}{\big( 
\nu_0,\nu_0^*, 
\nu_1,\nu_1^*,\zeta _{{\mathrm{o}}} 
 \big| 
E_0,E_0^*,
E_1,E_1^*,A_{{\mathrm{o}}}
 \big)} &= \big \langle {\mathcal T}{\mbox{\rm\boldmath$:$}} 
\exp\big(
i\nu^*_0 {\hat{\mathcal D}}_0 - i\nu_0 {\hat{\mathcal D}}_0^{\dagger}
\\ &\quad\qquad\qquad\qquad 
+i\nu^*_1 {\hat{\mathcal D}}_1 - i\nu_1 {\hat{\mathcal D}}_1^{\dagger} + i\zeta _{{\mathrm{o}}}{\hat{\mathcal J}}_{{\mathrm{o}}}
 \big) 
{\mbox{\rm\boldmath$:$}} \big \rangle_{BC}, \\
\Phi_{\mathrm{dev}}{\big( 
\nu_0,\nu_0^*, 
\nu_1,\nu_1^*,\zeta _{{\mathrm{o}}} 
 \big| 
E_0,E_0^*,
E_1,E_1^*,A_{{\mathrm{o}}}
 \big)} &= \big \langle {\mathcal T}{\mbox{\rm\boldmath$:$}} 
\exp\big(
i\nu^*_0 {\hat{\mathcal D}}_0 - i\nu_0 {\hat{\mathcal D}}_0^{\dagger}
\\ &\quad\qquad\qquad\qquad 
+i\nu^*_1 {\hat{\mathcal D}}_1 - i\nu_1 {\hat{\mathcal D}}_1^{\dagger} + i\zeta _{{\mathrm{o}}}{\hat{\mathcal J}}_{{\mathrm{o}}}
 \big) 
{\mbox{\rm\boldmath$:$}} \big \rangle . 
\end{aligned} 
\label{eq:1JA} 
\end{align}%
Calligraphic letters are as always for Heisenberg\ operators. 
For the ``bare'' devices, 
\begin{align} 
\begin{aligned} 
\Phi^{\mathrm{I}}_{\mathrm{dev}A}{\big( 
\nu_0,\nu_0^* \big| E_0,E_0^*
 \big)} &= \big \langle {\mathcal T}{\mbox{\rm\boldmath$:$}} 
\exp\big(
i\nu^*_0 D_0' - i\nu_0 D_0^{\prime\dagger}
 \big) 
{\mbox{\rm\boldmath$:$}} \big \rangle_A, \\
\Phi^{\mathrm{I}}_{\mathrm{dev}B}{\big( 
\nu_0,\nu_0^*, 
\nu_1,\nu_1^* 
 \big| 
E_0,E_0^*,
E_1,E_1^*
 \big)} &= \big \langle {\mathcal T}{\mbox{\rm\boldmath$:$}} 
\exp\big(
i\nu^*_0 D_0' - i\nu_0 D_0^{\prime\dagger}
+i\nu^*_1 D_1' - i\nu_1 D_1^{\prime\dagger}
 \big) 
{\mbox{\rm\boldmath$:$}} \big \rangle_B, \\
\Phi^{\mathrm{I}}_{\mathrm{dev}C}{\big( 
\nu_1,\nu_1^*,\zeta _{{\mathrm{o}}} 
 \big| 
E_1,E_1^*,A_{{\mathrm{o}}}
 \big)} &= \big \langle {\mathcal T}{\mbox{\rm\boldmath$:$}} 
\exp\big(
i\nu^*_1 D_1' - i\nu_1 D_1^{\prime\dagger} + i\zeta _{{\mathrm{o}}}\hat J_{{\mathrm{o}}}'
 \big) 
{\mbox{\rm\boldmath$:$}} \big \rangle_C, \\
\Phi^{\mathrm{I}}_{\mathrm{dev}BC}{\big( 
\nu_0,\nu_0^*, 
\nu_1,\nu_1^*,\zeta _{{\mathrm{o}}} 
 \big| 
E_0,E_0^*,
E_1,E_1^*,A_{{\mathrm{o}}}
 \big)} &= \big \langle {\mathcal T}{\mbox{\rm\boldmath$:$}} 
\exp\big(
i\nu^*_0 D_0' - i\nu_0 D_0^{\prime\dagger}
\\ &\quad\qquad\qquad\qquad 
+i\nu^*_1 D_1' - i\nu_1 D_1^{\prime\dagger} + i\zeta _{{\mathrm{o}}}\hat J_{{\mathrm{o}}}'
 \big) 
{\mbox{\rm\boldmath$:$}} \big \rangle_{BC}, \\
\Phi^{\mathrm{I}}_{\mathrm{dev}}{\big( 
\nu_0,\nu_0^*, 
\nu_1,\nu_1^*,\zeta _{{\mathrm{o}}} 
 \big| 
E_0,E_0^*,
E_1,E_1^*,A_{{\mathrm{o}}}
 \big)} &= \big \langle {\mathcal T}{\mbox{\rm\boldmath$:$}} 
\exp\big(
i\nu^*_0 D_0' - i\nu_0 D_0^{\prime\dagger}
\\ &\quad\qquad\qquad\qquad 
+i\nu^*_1 D_1' - i\nu_1 D_1^{\prime\dagger} + i\zeta _{{\mathrm{o}}}\hat J_{{\mathrm{o}}}'
 \big) 
{\mbox{\rm\boldmath$:$}} \big \rangle. 
\end{aligned} 
\label{eq:5JE} 
\end{align}%
The primed operators are the Heisenberg\ ones with respect to the Hamiltonian, 
\begin{align} 
\begin{aligned} 
\hat H(t) = \hat H^{\mathrm{f}} + \hat H_{\mathrm{dev}}(t) 
- {
A_{{\mathrm{o}}}(t)
}\hat J_{\mathrm{o}}(t) 
- \big [ 
{
E_0(t)
}\hat D_{0}^{\dag}(t) 
+{
E_1(t)
}\hat D_{1}^{\dag}(t) + {\mathrm{H.c.}}
 \big ] 
 . 
\end{aligned} 
\label{eq:2JB} 
\end{align}%
In (\ref{eq:1JA}) and (\ref{eq:5JE}), specifications at the averages apply in fact to averaged operators, while quantum averaging as such is always over the $\rho $-matrix (\ref{eq:38AA}). Redundant degrees of freedom are traced out automatically. 
\subsection{Formal solution}\label{ch:TZ}
Adapting the general dressing formula (\ref{eq:91FH}) to the six problems at hand we have, 
\begin{align} 
\begin{aligned} 
 &\Phi_{\mathrm{dev}A}{\big( 
\nu_0,\nu_0^* 
 \big| 
E_0,E_0^* 
 \big)} 
= \exp
Z_0\bigg(
\frac{\delta}{\delta E_0},
\frac{\delta}{\delta \nu _0}
 \bigg) 
\Phi^{\mathrm{I}}_{\mathrm{dev}A}{\big( 
\nu_0,\nu_0^* 
 \big| 
E_0,E_0^* 
 \big)} , 
\\ 
 &\Phi_{\mathrm{dev}B}{\big( 
\nu_0,\nu_0^*, 
\nu_1,\nu_1^* 
 \big| 
E_0,E_0^*,
E_1,E_1^*
 \big)} \\ &\quad\qquad 
= \exp\bigg [ 
Z_0\bigg(
\frac{\delta}{\delta E_0},
\frac{\delta}{\delta \nu _0}
 \bigg) 
+ 
Z_1\bigg(
\frac{\delta}{\delta E_1},
\frac{\delta}{\delta \nu _1}
 \bigg) 
 \bigg ] 
\Phi^{\mathrm{I}}_{\mathrm{dev}B}{\big( 
\nu_0,\nu_0^*, 
\nu_1,\nu_1^* 
 \big| 
E_0,E_0^*,
E_1,E_1^*
 \big)} , 
\\ 
 &\Phi_{\mathrm{dev}C}{\big( 
\nu_0,\nu_0^*, 
\nu_1,\nu_1^*,\zeta _{{\mathrm{o}}} 
 \big| 
E_1,E_1^*,A_{{\mathrm{o}}}
 \big)} \\ &\quad\qquad 
= \exp\bigg [ 
Z_1\bigg(
\frac{\delta}{\delta E_1},
\frac{\delta}{\delta \nu _1}
 \bigg) 
+ 
\mathcal{Z}_{{\mathrm{o}}}\bigg(
\frac{\delta}{\delta A_{{\mathrm{o}}}},
\frac{\delta}{\delta \zeta_{{\mathrm{o}}}}
 \bigg) 
 \bigg ] 
\Phi^{\mathrm{I}}_{\mathrm{dev}C}{\big( 
\nu_1,\nu_1^*,\zeta _{{\mathrm{o}}} 
 \big| 
E_1,E_1^*,A_{{\mathrm{o}}}
 \big)} , 
\\ 
 &\Phi_{\mathrm{dev}BC}{\big( 
\nu_0,\nu_0^*, 
\nu_1,\nu_1^*,\zeta _{{\mathrm{o}}} 
 \big| 
E_0,E_0^*,
E_1,E_1^*,A_{{\mathrm{o}}}
 \big)} \\ &\quad\qquad 
= \exp\bigg [ 
Z_0\bigg(
\frac{\delta}{\delta E_0},
\frac{\delta}{\delta \nu _0}
 \bigg) 
+ 
Z_1\bigg(
\frac{\delta}{\delta E_1},
\frac{\delta}{\delta \nu _1}
 \bigg) 
+ 
\mathcal{Z}_{{\mathrm{o}}}\bigg(
\frac{\delta}{\delta A_{{\mathrm{o}}}},
\frac{\delta}{\delta \zeta_{{\mathrm{o}}}}
 \bigg) 
 \bigg ] 
\\ &\quad\qquad\qquad\qquad\qquad\qquad\qquad\qquad\times 
\Phi^{\mathrm{I}}_{\mathrm{dev}BC}{\big( 
\nu_0,\nu_0^*, 
\nu_1,\nu_1^*,\zeta _{{\mathrm{o}}} 
 \big| 
E_0,E_0^*,
E_1,E_1^*,A_{{\mathrm{o}}}
 \big)} , 
\\ 
 &\Phi_{\mathrm{dev}}{\big( 
\nu_0,\nu_0^*, 
\nu_1,\nu_1^*,\zeta _{{\mathrm{o}}} 
 \big| 
E_0,E_0^*,
E_1,E_1^*,A_{{\mathrm{o}}}
 \big)} \\ &\quad\qquad 
= \exp\bigg [ 
Z_0\bigg(
\frac{\delta}{\delta E_0},
\frac{\delta}{\delta \nu _0}
 \bigg) 
+ 
Z_1\bigg(
\frac{\delta}{\delta E_1},
\frac{\delta}{\delta \nu _1}
 \bigg) 
+ 
\mathcal{Z}_{{\mathrm{o}}}\bigg(
\frac{\delta}{\delta A_{{\mathrm{o}}}},
\frac{\delta}{\delta \zeta_{{\mathrm{o}}}}
 \bigg) 
 \bigg ] 
\\ &\quad\qquad\qquad\qquad\qquad\qquad\qquad\qquad\times 
\Phi^{\mathrm{I}}_{\mathrm{dev}}{\big( 
\nu_0,\nu_0^*, 
\nu_1,\nu_1^*,\zeta _{{\mathrm{o}}} 
 \big| 
E_0,E_0^*,
E_1,E_1^*,A_{{\mathrm{o}}}
 \big)} , 
\end{aligned} 
\label{eq:6JF} 
\end{align}%
where 
\begin{align} 
\begin{aligned} 
Z_0\bigg(
\frac{\delta}{\delta E_0},
\frac{\delta}{\delta \nu _0}
 \bigg) &= -i 
\frac{\delta}{\delta E_0}
\Delta_{\mathrm{R}0}
\frac{\delta}{\delta \nu _0^*}
+i\frac{\delta}{\delta E_0^*}
\Delta_{\mathrm{R}0}^*
\frac{\delta}{\delta \nu _0}
\\ 
Z_1\bigg(
\frac{\delta}{\delta E_1},
\frac{\delta}{\delta \nu _1}
 \bigg) &= -i 
\frac{\delta}{\delta E_1}
\Delta_{\mathrm{R}1}
\frac{\delta}{\delta \nu _1^*}
+i\frac{\delta}{\delta E_1^*}
\Delta_{\mathrm{R}1}^*
\frac{\delta}{\delta \nu _1}
\\ 
\mathcal{Z}_{{\mathrm{o}}}\bigg(
\frac{\delta}{\delta A_{{\mathrm{o}}}},
\frac{\delta}{\delta \zeta_{{\mathrm{o}}}}
 \bigg) &= -i 
\frac{\delta}{\delta A_{{\mathrm{o}}}} 
G_{\mathrm{R}{\mathrm{o}}}
\frac{\delta}{\delta \zeta_{{\mathrm{o}}}} . 
\end{aligned} 
\label{eq:7JH} 
\end{align}%
The kernels are given by the formulae, 
\begin{align} 
\begin{aligned} 
\Delta_{\mathrm{R}0}(t-t') = \frac{i}{\hbar }\theta(t-t')\big [ 
\hat E_0(t),\hat E_0^{\dag}(t')
 \big ] , \\ 
\Delta_{\mathrm{R}1}(t-t') = \frac{i}{\hbar }\theta(t-t')\big [ 
\hat E_1(t),\hat E_1^{\dag}(t')
 \big ] , \\ 
G_{\mathrm{R}{\mathrm{o}}}(t-t') = \frac{i}{\hbar }\theta(t-t')\big [ 
\hat A_{{\mathrm{o}}}(t),\hat A_{{\mathrm{o}}}(t')
 \big ] . 
\end{aligned} 
\label{eq:8JJ} 
\end{align}%

Functionals (\ref{eq:1JA}) may be redefined in terms of ``bare'' dipole and current operators without sources in the manner of Eqs.\ (\ref{eq:92FJ}), (\ref{eq:21FB}). This leads to the factorisation properties, 
\begin{align} 
\begin{aligned} 
 &%
\begin{aligned}\Phi^{\mathrm{I}}_{\mathrm{dev}BC}{\big( 
\nu_0,\nu_0^*, 
\nu_1,\nu_1^*,\zeta _{{\mathrm{o}}} 
 \big| 
E_0,E_0^*,
E_1,E_1^*,A_{{\mathrm{o}}}
 \big)} &= \Phi^{\mathrm{I}}_{\mathrm{dev}B}{\big( 
\nu_0,\nu_0^*, 
\nu_1,\nu_1^* 
 \big| 
E_0,E_0^*,
E_1,E_1^*
 \big)} 
\\ &\quad\times 
\Phi^{\mathrm{I}}_{\mathrm{dev}C}{\big( 
\nu_1,\nu_1^*,\zeta _{{\mathrm{o}}} 
 \big| 
E_1,E_1^*,A_{{\mathrm{o}}}
 \big)}, 
\end{aligned} \\ 
 &%
\begin{aligned}\Phi^{\mathrm{I}}_{\mathrm{dev}}{\big( 
\nu_0,\nu_0^*, 
\nu_1,\nu_1^*,\zeta _{{\mathrm{o}}} 
 \big| 
E_0,E_0^*,
E_1,E_1^*,A_{{\mathrm{o}}}
 \big)} &= \Phi^{\mathrm{I}}_{\mathrm{dev}A}{\big( 
\nu_0,\nu_0^* \big| E_0,E_0^*
 \big)} 
\\ &\quad\times 
\Phi^{\mathrm{I}}_{\mathrm{dev}BC}{\big( 
\nu_0,\nu_0^*, 
\nu_1,\nu_1^*,\zeta _{{\mathrm{o}}} 
 \big| 
E_0,E_0^*,
E_1,E_1^*,A_{{\mathrm{o}}}
 \big)} 
\end{aligned}
\\ &\quad
= \Phi^{\mathrm{I}}_{\mathrm{dev}A}{\big( 
\nu_0,\nu_0^* \big| E_0,E_0^*
 \big)}\Phi^{\mathrm{I}}_{\mathrm{dev}B}{\big( 
\nu_0,\nu_0^*, 
\nu_1,\nu_1^* 
 \big| 
E_0,E_0^*,
E_1,E_1^*
 \big)}\Phi^{\mathrm{I}}_{\mathrm{dev}C}{\big( 
\nu_1,\nu_1^*,\zeta _{{\mathrm{o}}} 
 \big| 
E_1,E_1^*,A_{{\mathrm{o}}}
 \big)} . 
\end{aligned} 
\label{eq:4JD} 
\end{align}%
Combining these formulae with the dressing ones (\ref{eq:6JF}) we find the relations among properties of the ``dressed'' devices, 
\begin{align} 
\begin{aligned} 
 &\Phi_{\mathrm{dev}BC}{\big( 
\nu_0,\nu_0^*, 
\nu_1,\nu_1^*,\zeta_{{\mathrm{o}}} 
 \big| 
E_0,E_0^*,
E_1,E_1^*,A_{{\mathrm{o}}}
 \big)} 
= \exp\bigg [ 
Z_1\bigg(
\frac{\delta}{\delta E_1},
\frac{\delta}{\delta \nu _1'}
 \bigg) 
+ 
Z_1\bigg(
\frac{\delta}{\delta E_1'},
\frac{\delta}{\delta \nu _1}
 \bigg) 
 \bigg ] 
\\ &\quad\qquad\times 
\Phi_{\mathrm{dev}B}{\big( 
\nu_0,\nu_0^*, 
\nu_1',\nu_1^{\prime *} 
 \big| 
E_0,E_0^*,
E_1',E_1^{\prime *}
 \big)} 
\Phi_{\mathrm{dev}C}{\big( 
\nu_1,\nu_1^*,\zeta _{{\mathrm{o}}} 
 \big| 
E_1,E_1^*,A_{{\mathrm{o}}}
 \big)}\settoheight{\auxlv}{$|$}%
\raisebox{-0.3\auxlv}{$|_{\nu _1'=\nu _1,E_1'=E_1}$} , 
\\ 
 &\Phi_{\mathrm{dev}}{\big( 
\nu_0,\nu_0^*, 
\nu_1,\nu_1^*,\zeta_{{\mathrm{o}}} 
 \big| 
E_0,E_0^*,
E_1,E_1^*,A_{{\mathrm{o}}}
 \big)} 
= \exp\bigg [ 
Z_0\bigg(
\frac{\delta}{\delta E_0},
\frac{\delta}{\delta \nu _0'}
 \bigg) 
+ 
Z_0\bigg(
\frac{\delta}{\delta E_0'},
\frac{\delta}{\delta \nu _0}
 \bigg) 
 \bigg ] 
\\ &\quad\qquad\times 
\Phi_{\mathrm{dev}A}{\big( 
\nu_0',\nu_0^{\prime *} 
 \big| 
E_0',E_0^{\prime *}
 \big)} 
\Phi_{\mathrm{dev}BC}{\big( 
\nu_0,\nu_0^*, 
\nu_1,\nu_1^*,\zeta_{{\mathrm{o}}} 
 \big| 
E_0,E_0^*,
E_1,E_1^*,A_{{\mathrm{o}}}
 \big)}\settoheight{\auxlv}{$|$}%
\raisebox{-0.3\auxlv}{$|_{\nu _0'=\nu _0,E_0'=E_0}$} 
\\ &\quad
= \exp\bigg [ 
Z_0\bigg(
\frac{\delta}{\delta E_0},
\frac{\delta}{\delta \nu _0'}
 \bigg) 
+ 
Z_0\bigg(
\frac{\delta}{\delta E_0'},
\frac{\delta}{\delta \nu _0}
 \bigg) 
+ 
Z_1\bigg(
\frac{\delta}{\delta E_1},
\frac{\delta}{\delta \nu _1'}
 \bigg) 
+ 
Z_1\bigg(
\frac{\delta}{\delta E_1'},
\frac{\delta}{\delta \nu _1}
 \bigg) 
 \bigg ] 
\\ &\quad\qquad\times 
\Phi_{\mathrm{dev}A}{\big( 
\nu_0',\nu_0^{\prime *} 
 \big| 
E_0',E_0^{\prime *}
 \big)} 
\\ &\quad\qquad\times 
\Phi_{\mathrm{dev}B}{\big( 
\nu_0,\nu_0^*, 
\nu_1',\nu_1^{\prime *} 
 \big| 
E_0,E_0^*,
E_1',E_1^{\prime *}
 \big)} 
\Phi_{\mathrm{dev}C}{\big( 
\nu_1,\nu_1^*,\zeta _{{\mathrm{o}}} 
 \big| 
E_1,E_1^*,A_{{\mathrm{o}}}
 \big)}\settoheight{\auxlv}{$|$}%
\raisebox{-0.3\auxlv}{$|_{\nu _{0,1}'=\nu _{0,1},E_{0,1}'=E_{0,1}}$} . 
\\ 
\end{aligned} 
\label{eq:10JL} 
\end{align}%
\end{widetext}%
In the formula for $\Phi_{\mathrm{dev}}$, we have to suppress all information about the HF modes and that about response properties of the LF mode. Preserved is only the information about the output current, 
\begin{multline} 
\hspace{0.4\columnwidth}\hspace{-0.4\twocolumnwidth}
\Phi_{\mathrm{o}}\big(
\zeta_{\mathrm{o}}
 \big) = \Phi_{\mathrm{dev}}{\big( 
0,0,0,0,\zeta_{{\mathrm{o}}} 
 \big| 
0,0,0,0,0
 \big)} \\ 
= \big \langle {\mathcal T}{\mbox{\rm\boldmath$:$}} 
\exp\big(
i\zeta _{\mathrm{o}}{\hat{\mathcal J}}_{{\mathrm{o}}}
 \big) 
{\mbox{\rm\boldmath$:$}} \big \rangle_{\mathrm{o}}. 
\hspace{0.4\columnwidth}\hspace{-0.4\twocolumnwidth}%
\label{eq:11JM} 
\end{multline}%
We also have to suppress the back-action of the detector on the amplifier and of the amplifier on the light source, which means dropping the corresponding differential operators, 
\begin{align} 
\begin{aligned} 
Z_0\bigg(
\frac{\delta}{\delta E_0'},
\frac{\delta}{\delta \nu _0}
 \bigg) 
, 
Z_1\bigg(
\frac{\delta}{\delta E_1'},
\frac{\delta}{\delta \nu _1}
 \bigg) \to 0 .
\end{aligned} 
\label{eq:12JN} 
\end{align}%
Consider firstly a photodetection formula in terms of the light source and composite detector. The corresponding part of Eq.\ (\ref{eq:10JL}) reduces to, 
\begin{multline} 
\hspace{0.4\columnwidth}\hspace{-0.4\twocolumnwidth}
\Phi_{\mathrm{o}}\big(
\zeta _{\mathrm{o}}
 \big) = \Phi_{\mathrm{s}}\bigg(
i\frac{\delta}{\delta E_0^*} \Delta_{\mathrm{R}0}^*, 
-i\frac{\delta}{\delta E_0} \Delta_{\mathrm{R}0} 
 \bigg) 
\\ \times 
\Phi_{\mathrm{cd}}{\big( 
\zeta _{\mathrm{o}}\big| E_0,E_0^* \big)}\settoheight{\auxlv}{$|$}%
\raisebox{-0.3\auxlv}{$|_{E_0=0}$} , 
\hspace{0.4\columnwidth}\hspace{-0.4\twocolumnwidth}%
\label{eq:13JP} 
\end{multline}%
where 
\begin{align} 
\begin{aligned} 
 &%
\begin{aligned}
\Phi_{\mathrm{s}}\big(
\nu_0 ,\nu_0 ^* 
 \big) &= \Phi_{\mathrm{dev}A}{\big( 
\nu_0,\nu_0^{*} 
 \big| 
0,0
 \big)} \\ 
 &= \big \langle {\mathcal T}{\mbox{\rm\boldmath$:$}} 
\exp\big(
i\nu_0^* {\hat{\mathcal D}}_0 - i\nu _0 {\hat{\mathcal D}}_0^{\dag}
 \big) 
{\mbox{\rm\boldmath$:$}} \big \rangle_{\mathrm{s}} , 
\end{aligned}
\\ 
 &%
\begin{aligned}
\Phi_{\mathrm{cd}}{\big( 
\zeta _{\mathrm{o}}\big| E_0,E_0^* \big)} &= \Phi_{\mathrm{dev}BC}{\big( 
0,0,0,0,\zeta_{{\mathrm{o}}} 
 \big| 
E_0,E_0^*,0,0,0
 \big)} \\ 
 &= \big \langle {\mathcal T}{\mbox{\rm\boldmath$:$}} 
\exp\big(
i\zeta_{\mathrm{o}}{\hat{\mathcal J}}_{{\mathrm{o}}}
 \big) 
{\mbox{\rm\boldmath$:$}} \big \rangle_{\mathrm{cd}} . 
\end{aligned}
\end{aligned} 
\label{eq:14JQ} 
\end{align}%
The differential operator in (\ref{eq:13JP}) may be rewritten as a quantum average, 
\begin{multline} 
\hspace{0.4\columnwidth}\hspace{-0.4\twocolumnwidth}
\Phi_{\mathrm{s}}\bigg(
i\frac{\delta}{\delta E_0^*} \Delta_{\mathrm{R}0}^*, 
-i\frac{\delta}{\delta E_0} \Delta_{\mathrm{R}0} 
 \bigg) \\ 
= \bigg \langle {\mathcal T}{\mbox{\rm\boldmath$:$}} 
\exp\bigg(
\frac{\delta}{\delta E_0} \Delta_{\mathrm{R}0}{\hat{\mathcal D}}_0 +\frac{\delta}{\delta E_0^*} \Delta_{\mathrm{R}0}^*{\hat{\mathcal D}}_0^{\dag}
 \bigg) 
{\mbox{\rm\boldmath$:$}} \bigg \rangle_{\mathrm{s}} \\ 
= \bigg \langle {\mathcal T}{\mbox{\rm\boldmath$:$}} 
\exp\bigg(
\frac{\delta}{\delta E_0} {\hat{\mathcal E}}_0 +\frac{\delta}{\delta E_0^*} {\hat{\mathcal E}}_0^{\dag}
 \bigg) 
{\mbox{\rm\boldmath$:$}} \bigg \rangle_{\mathrm{s}} . 
\hspace{0.4\columnwidth}\hspace{-0.4\twocolumnwidth}%
\label{eq:15JR} 
\end{multline}%
We used here the fact that, under the time-normal ordering, classical radiation laws apply directly to operators, so that we could write, 
\begin{align} 
\begin{aligned} 
{\hat{\mathcal E}}_0(t) = \int dt' \Delta _{\mathrm{R}0}(t-t'){\hat{\mathcal D}}_0(t') . 
\end{aligned} 
\label{eq:16JS} 
\end{align}%
Similar to Eq.\ (\ref{eq:5CX}), 
\begin{multline} 
\hspace{0.4\columnwidth}\hspace{-0.4\twocolumnwidth}
\bigg \langle {\mathcal T}{\mbox{\rm\boldmath$:$}} 
\exp\bigg(
\frac{\delta}{\delta E_0} {\hat{\mathcal E}}_0 +\frac{\delta}{\delta E_0^*} {\hat{\mathcal E}}_0^{\dag}
 \bigg) 
{\mbox{\rm\boldmath$:$}} \bigg \rangle_{\mathrm{s}}\mathcal{F}\big(
E_0,E_0^*
 \big)\settoheight{\auxlv}{$|$}%
\raisebox{-0.3\auxlv}{$|_{E_0=0}$} \\ 
= \big \langle {\mathcal T}{\mbox{\rm\boldmath$:$}} 
\mathcal{F}\big(
{\hat{\mathcal E}}_0,{\hat{\mathcal E}}_0^{\dag}
 \big) 
{\mbox{\rm\boldmath$:$}} \big \rangle_{\mathrm{s}} . 
\hspace{0.4\columnwidth}\hspace{-0.4\twocolumnwidth}%
\label{eq:18JU} 
\end{multline}%
Equation (\ref{eq:13JP}) may therefore be written as, 
\begin{align} 
\begin{aligned} 
\Phi_{\mathrm{o}}\big(
\zeta _{\mathrm{o}}
 \big) = \big \langle {\mathcal T}{\mbox{\rm\boldmath$:$}} 
\Phi_{\mathrm{cd}}{\big( 
\zeta _{\mathrm{o}}\big| 
{\hat{\mathcal E}}_0,{\hat{\mathcal E}}_0^{\dag}
 \big)} 
{\mbox{\rm\boldmath$:$}} \big \rangle_{\mathrm{s}}. 
\end{aligned} 
\label{eq:19JV} 
\end{align}%
Remembering the definitions of $\Phi_{\mathrm{o}}$ and $\Phi_{\mathrm{cd}}$, 
this is equivalent to the two-layer average, 
\begin{multline} 
\hspace{0.4\columnwidth}\hspace{-0.4\twocolumnwidth}
\big \langle {\mathcal T}{\mbox{\rm\boldmath$:$}} 
\exp\big(
i\zeta _{\mathrm{o}}{\hat{\mathcal J}}_{{\mathrm{o}}}
 \big) 
{\mbox{\rm\boldmath$:$}} \big \rangle _{\mathrm{o}}\\ 
= \big \langle {\mathcal T}{\mbox{\rm\boldmath$:$}} 
\big [ 
\big \langle {\mathcal T}{\mbox{\rm\boldmath$:$}} 
\exp\big(
i\zeta _{\mathrm{o}}{\hat{\mathcal J}}_{{\mathrm{o}}}
 \big) 
{\mbox{\rm\boldmath$:$}} \big \rangle_{\mathrm{cd}} 
 \big ] \settoheight{\auxlv}{$|$}%
\raisebox{-0.3\auxlv}{$|_{E_0,E_0^*\to{\hat{\mathcal E}}_0,{\hat{\mathcal E}}_0^{\dag}}$}
{\mbox{\rm\boldmath$:$}} \big \rangle_{\mathrm{s}} . 
\hspace{0.4\columnwidth}\hspace{-0.4\twocolumnwidth}%
\label{eq:21JX} 
\end{multline}%
To decifer this relation, recall that averages denoted \mbox{$
\big \langle {\mathcal T}{\mbox{\rm\boldmath$:$}} 
\cdots 
{\mbox{\rm\boldmath$:$}} \big \rangle_{\mathrm{cd}}
$} are by definition conditional on $E_0(t)$. 
In terms of current averages, 
\begin{multline} 
\hspace{0.4\columnwidth}\hspace{-0.4\twocolumnwidth}
\big \langle {\mathcal T}{\mbox{\rm\boldmath$:$}} 
{\hat{\mathcal J}}_{{\mathrm{o}}}(t_1)\cdots{\hat{\mathcal J}}_{{\mathrm{o}}}(t_m)
{\mbox{\rm\boldmath$:$}} \big \rangle _{\mathrm{o}}\\ 
= \big \langle {\mathcal T}{\mbox{\rm\boldmath$:$}} 
\big [ 
\big \langle {\mathcal T}{\mbox{\rm\boldmath$:$}} 
{\hat{\mathcal J}}_{{\mathrm{o}}}(t_1)\cdots{\hat{\mathcal J}}_{{\mathrm{o}}}(t_m)
{\mbox{\rm\boldmath$:$}} \big \rangle_{\mathrm{cd}} 
 \big ] \settoheight{\auxlv}{$|$}%
\raisebox{-0.3\auxlv}{$|_{E_0,E_0^*\to{\hat{\mathcal E}}_0,{\hat{\mathcal E}}_0^{\dag}}$}
{\mbox{\rm\boldmath$:$}} \big \rangle_{\mathrm{s}} , 
\hspace{0.4\columnwidth}\hspace{-0.4\twocolumnwidth}%
\label{eq:20JW} 
\end{multline}%
which is an under-the-RWA counterpart of Eq.\ (\ref{eq:21DQ}). 

The rest of Eq.\ (\ref{eq:10JL}) under approximations (\ref{eq:11JM}), (\ref{eq:12JN}) is manupulated similarly. So, photodetection statistics in terms of the properties of the three devices reads, 
\begin{multline} 
\hspace{0.4\columnwidth}\hspace{-0.4\twocolumnwidth}
\big \langle {\mathcal T}{\mbox{\rm\boldmath$:$}} 
\exp\big(i\zeta _{\mathrm{o}}{\hat{\mathcal J}}_{{\mathrm{o}}}\big)
{\mbox{\rm\boldmath$:$}} \big \rangle_{\mathrm{o}} = \bigg [ 
 \bigg \langle {\mathcal T}{\mbox{\rm\boldmath$:$}} 
\exp\bigg(
\frac{\delta}{\delta E_0} {\hat{\mathcal E}}_0 +\frac{\delta}{\delta E_0^*} {\hat{\mathcal E}}_0^{\dag}
 \bigg) 
{\mbox{\rm\boldmath$:$}} \bigg \rangle_{\mathrm{s}} 
\\ \times
\bigg \langle {\mathcal T}{\mbox{\rm\boldmath$:$}} 
\exp\bigg(
\frac{\delta}{\delta E_1} {\hat{\mathcal E}}_1 +\frac{\delta}{\delta E_1^*} {\hat{\mathcal E}}_1^{\dag}
 \bigg) 
{\mbox{\rm\boldmath$:$}} \bigg \rangle_{\mathrm{a}} 
\\ \times
\big \langle {\mathcal T}{\mbox{\rm\boldmath$:$}} 
\exp\big(
i\zeta _{\mathrm{o}}{\hat{\mathcal J}}_{{\mathrm{o}}}
 \big) 
{\mbox{\rm\boldmath$:$}} \big \rangle_{\mathrm{d}}
 \bigg ] \settoheight{\auxlv}{$\Big|$}%
\raisebox{-0.3\auxlv}{$\Big|_{E_0=E_1=0}$} . 
\hspace{0.4\columnwidth}\hspace{-0.4\twocolumnwidth}%
\label{eq:22JY} 
\end{multline}%
Again, recall that averages \mbox{$
 \langle {\mathcal T}{\mbox{\rm\boldmath$:$}} 
\cdots 
{\mbox{\rm\boldmath$:$}} \rangle_{\mathrm{a}}
$} and \mbox{$
 \langle {\mathcal T}{\mbox{\rm\boldmath$:$}} 
\cdots 
{\mbox{\rm\boldmath$:$}} \rangle_{\mathrm{d}}
$} are by definition conditional on, respectively, $E_0(t)$ and $E_1(t)$. Manipulations similar to those leading to Eq.\ (\ref{eq:20JW}) allow one to write this relation as a three-layer average, 
\begin{widetext} 
\begin{align} 
\begin{aligned} 
\big \langle {\mathcal T}{\mbox{\rm\boldmath$:$}} 
{\hat{\mathcal J}}_{{\mathrm{o}}}(t_1)\cdots{\hat{\mathcal J}}_{{\mathrm{o}}}(t_m)
{\mbox{\rm\boldmath$:$}} \big \rangle _{\mathrm{o}}
= 
\big \langle {\mathcal T}{\mbox{\rm\boldmath$:$}} 
\big \{ 
\big \langle {\mathcal T}{\mbox{\rm\boldmath$:$}} 
\big [ 
\big \langle {\mathcal T}{\mbox{\rm\boldmath$:$}} 
{\hat{\mathcal J}}_{{\mathrm{o}}}(t_1)\cdots{\hat{\mathcal J}}_{{\mathrm{o}}}(t_m)
{\mbox{\rm\boldmath$:$}} \big \rangle_{\mathrm{d}} 
 \big ]\settoheight{\auxlv}{$|$}%
\raisebox{-0.3\auxlv}{$|_{E_1,E_1^*\to{\hat{\mathcal E}}_1,{\hat{\mathcal E}}_1^{\dag}}$}
{\mbox{\rm\boldmath$:$}} \big \rangle_{\mathrm{a}} 
 \big \}\settoheight{\auxlv}{$|$}%
\raisebox{-0.3\auxlv}{$|_{E_0,E_0^*\to{\hat{\mathcal E}}_0,{\hat{\mathcal E}}_0^{\dag}}$} 
{\mbox{\rm\boldmath$:$}} \big \rangle_{\mathrm{s}} . 
\end{aligned} 
\label{eq:23JZ} 
\end{align}%
\end{widetext}%
This relation fully justifies ``doing quantum electrodynamics\ while thinking classically'' in Sec.\ \ref{ch:TC} in the main body of the paper, where quantum averagings \mbox{$
 \langle {\mathcal T}{\mbox{\rm\boldmath$:$}} 
\cdots 
{\mbox{\rm\boldmath$:$}} \rangle_{\mathrm{a}}
$} and \mbox{$
 \langle {\mathcal T}{\mbox{\rm\boldmath$:$}} 
\cdots 
{\mbox{\rm\boldmath$:$}} \rangle_{\mathrm{d}}
$} were subject to semiclassical models.


\begin{thebibliography}{26}
\bibitem[{end({\natexlab{c}})}]{endLin}
\bibinfo{note}{We assumed that the amplifier remains linear for arbitrarily
  strong input signal. Equation (\ref{eq:47ET}) is therefore a consistency
  condition for the model (\ref{eq:25DU}) rather than a restriction on a real
  device.}
\end{thebibliography}
\end{document}